\documentclass[aop,preprint]{imsart}

\RequirePackage{amsthm,amsmath,amsfonts,amssymb}
\RequirePackage[numbers]{natbib}
\RequirePackage[colorlinks,citecolor=blue,urlcolor=blue]{hyperref}
\RequirePackage{graphicx,booktabs,array,needspace}

\startlocaldefs
\numberwithin{equation}{section}
\theoremstyle{plain}
\newtheorem{theorem}{Theorem}[section]
\newtheorem{proposition}[theorem]{Proposition}
\newtheorem{lemma}[theorem]{Lemma}
\newtheorem{corollary}[theorem]{Corollary}
\theoremstyle{definition}

\newtheorem{remark}[theorem]{Remark}
\newcommand{\Pp}{\mathbb P}
\newcommand{\E}{\mathbb E}
\newcommand{\Var}{\operatorname{Var}}
\newcommand{\clip}{\operatorname{clip}_{[0,n]}}
\newcommand{\Ber}{\operatorname{Bernoulli}}
\newcommand{\one}{\boldsymbol 1}

\renewcommand{\theenumi}{\roman{enumi}}

\endlocaldefs

\begin{document}

\begin{frontmatter}
\pdfsubject{Finite-state prediction, generation, and sliding-window estimation}
\title{Recursive memory for prediction and generation}
\runtitle{Recursive memory for prediction and generation}
\runauthor{P. Feng and C. Meng}
\begin{aug}
\author[A]{\fnms{Pengbin}~\snm{Feng}\ead[label=e1]{pengbinf@alumni.usc.edu}}
\author[B]{\fnms{Chunlei}~\snm{Meng}\ead[label=e2]{clmeng23@m.fudan.edu.cn}}
\address[A]{University of Southern California\printead[presep={,\ }]{e1}}
\address[B]{Fudan University\printead[presep={,\ }]{e2}}
\end{aug}

\begin{abstract}
Recurrent and state-space language models compress the past into a state of fixed size. The state must answer questions about the future now, and it must remain useful as new tokens arrive, including tokens the model samples itself. We show that the second requirement can cost far more memory than the first. For a known hidden Markov source with positive transition probabilities, a state of $O(\log(1/\varepsilon))$ bits answers every question about the future to accuracy $\varepsilon$, however far ahead it looks, because the Bayesian filter forgets its rounding errors. Generating the future is different. When a model feeds each sampled token back into its state, the small errors in its next-token distributions add up, and $T$ tokens can require a state of order $\log(T/\varepsilon)$ bits. The lower bound comes from the KL chain rule: at every step, a model with $K$ states pays for approximating the true next-token distribution by one of only $K$ distributions. The argument applies to any source, including a language model continuing a prompt, and a small-ball estimate makes it tight for two-state sources. The extra $\log T$ disappears if the model keeps private randomness, or if error is budgeted per token, as in perplexity. Statistics of the recent past show the same gap between describing and maintaining. The number of ones among the last $n$ bits takes $\log n$ bits to write down. But if each bit is a noisy copy of the one $n$ steps earlier, tracking that count to a mean-square error $E^2$ well below its variance costs order $n/(1+E^2)+\log n$ bits. Information about each bit must be kept until the bit leaves the window, and a queue holding the counts of recent blocks keeps just enough of it.
\end{abstract}

\end{frontmatter}

\setcounter{tocdepth}{1}
\begingroup
\small
\tableofcontents
\endgroup
\medskip

\section{Introduction}\label{sec:introduction}

A recurrent or state-space language model reads a long context but carries only a fixed-size state from one token to the next. That state serves two purposes. It must support the model's predictions at the current position, and it must remain a good starting point for every later update, since anything it drops is gone for the rest of the sequence. This paper asks how much memory each purpose requires, and shows that the second can be far more expensive than the first.

Two simple tasks make the difference concrete. The number of ones among the last $n$ bits can be written down with about $\log_2n$ bits, but a state that holds only this number cannot keep it current, because the next update must subtract the bit that is leaving the window. Prediction shows a similar gap. A state can answer every question about the future of a hidden Markov source to high accuracy and still be too small to generate that future one symbol at a time, as an autoregressive model does when it feeds its own samples back into its state. We measure both gaps for finite-state machines that observe a known stationary source.

The machine starts from a fixed state, knows the current time, and after each observation keeps one of $K$ labels, at a cost of $\lceil\log_2K\rceil$ bits. Giving two histories the same label can cost accuracy at the present time. It also constrains the future, because the two histories stay identified after every common continuation, even when a later question would need to tell them apart. Proposition~\ref{prop:common-risk} makes both effects precise. For hidden Markov models that are only queried, both can be controlled with memory independent of the horizon. The horizon enters when the machine must generate the future and so reuse an approximate predictor at every step. For moving-window counts, the constraint on the future dominates, since memory is needed to form an estimate from arriving observations and to keep information about each observation until it leaves the window.

\subsection{Main results}

For a hidden Markov model with a positive transition matrix and a finite output alphabet, rounding the filter after each observation gives a finite-state observer that answers every question about future events to within log-probability error $\varepsilon$, however far ahead the question looks (Theorem~\ref{thm:hmm-memory}). The conditional probability of any future event is a nonnegative linear functional of the predictive belief, so a relative approximation of the belief controls all of them at once, and filter contraction keeps the rounding errors from accumulating. For a stationary, strictly positive model with two hidden states and distinct transition rows and emission laws, the optimal memory is $\Theta(\log(1/\varepsilon))$ bits as $\varepsilon\downarrow0$ (Theorem~\ref{thm:query-memory}).

Generating the future one symbol at a time costs more. Suppose the generator starts in a fixed state and updates deterministically from the symbols it emits, so that its output determines its state; we call such a generator \emph{unifilar}. Its $K$ states supply at most $K$ next-symbol distributions at each time, and by the KL chain rule their approximation errors add up along the continuation. For the two-state models above,
\begin{equation}\label{eq:intro-generation}
b_{\mathrm{gen}}(T,\varepsilon),\ b_{\mathrm{KL}}(T,\varepsilon)
=\Theta\!\left(\log\frac{T}{\varepsilon}\right),
\qquad T/\varepsilon\longrightarrow\infty,
\end{equation}
where $b_{\mathrm{gen}}$ is the memory needed for log error at most $\varepsilon$ on every word of length at most $T$, and $b_{\mathrm{KL}}$ the memory needed for expected total forward KL at most $\varepsilon$ (Theorem~\ref{thm:gen-binary-classification}). The lower bound combines this accumulation with a small-ball estimate for the stationary predictor law; the matching constructions round the filter to precision of order $\varepsilon/T$, or $\sqrt{\varepsilon/T}$ for total KL.

Two changes remove the dependence on $T$. A generator that may keep a private hidden state can draw it from the retained belief when a forecast is requested and then run the original hidden dynamics. The approximate belief is then used only once. With the labels of both phases counted, the optimal memory is again $\Theta(\log(1/\varepsilon))$ bits, uniformly in $T$ (Corollary~\ref{cor:gen-private-memory}). Alternatively, if the allowed excess log loss is $a$ per symbol, as in comparisons by perplexity, the total KL budget is $aT$ and $b_{\mathrm{KL}}(T,aT)=\Theta(\log(1/a))$ uniformly in $T$; see \eqref{eq:per-token-bit-law}. The $\log T$ in \eqref{eq:intro-generation} is thus the price of a state determined by the visible output combined with an error budget fixed in total.

The converse behind the lower bound holds for every finite-alphabet source and every forecast start, and the state count may vary with time (Theorem~\ref{thm:gen-quantization}). For an autoregressive language model continuing a fixed prompt, it gives a lower bound on the total forward KL of any student whose state is determined by the prompt and the emitted tokens, such as a recurrent or state-space model distilled from the teacher. The bound is twice the summed error of quantizing the teacher's next-token laws with as many codewords as the student has states (Section~\ref{sec:sequence-models}).

The second setting concerns a statistic of the recent past. Consider a stationary binary source in which each new bit copies the bit $n$ positions earlier, flipped independently with probability $(1-\delta)/2$, where $0<\delta\le1/2$. Every window of $n$ consecutive bits consists of independent fair bits, but the next $n$ bits are noisy copies of the current window. The window count is therefore a forecast: it determines the conditional log probability that the next $n$ bits are all ones, an event of unconditional probability $2^{-n}$. For fixed $c\in(0,1/4)$ and mean-square count error at most $E^2$ at each query time, the optimal memory is
\begin{equation}\label{eq:intro-main}
\Theta_c\!\left(\frac{n}{1+E^2}+\log_2n\right),
\qquad E^2\le cn,\quad 0<c<\frac14
\end{equation}
(Theorem~\ref{thm:main}). At fixed $E$ this is order $n$ bits, for a count that $\lceil\log_2(n+1)\rceil$ bits describe exactly. The first term pays for keeping information about each observation until it leaves the window. Running the machine along a common continuation recovers the counts of the blocks about to expire, and their residual variance given the next window forces the lower bound. A fixed-capacity queue of block counts attains it. The second term is already needed to form the first count, by a streaming lower bound of Braverman, Garg and Woodruff~\cite{bgw}. A single complete window, by contrast, can be encoded to mean-square error $cn$ with a bounded number of labels (Proposition~\ref{prop:offline}).

A constant estimate has mean-square error $n/4$, so no memory is needed once $E^2\ge n/4$ (Proposition~\ref{prop:zero}). Just below this level, bounded memory already helps, and it can reduce the error by amounts that grow like any fixed power of $\log n$ (Section~\ref{sec:critical-memory}). Hence the logarithmic lower bound in Theorem~\ref{thm:main} cannot hold uniformly as $c\uparrow1/4$.

\subsection{Relation to previous work}

Predictive sufficiency groups histories with the same conditional law of the future, and computational mechanics turns these groups into recursively updated causal states~\cite{sc}. Their relation to generative representations, including posterior initialization of hidden-state generators, is developed by L\"ohr and Ay~\cite{lohray2009}, L\"ohr~\cite{lohr2012}, and Ruebeck et al.\ \cite{ruebeck2018}. Approximate sufficiency leads to predictive rate-distortion theory~\cite{mc,mccorrection}. Marzen and Crutchfield~\cite{mc2017} bound the number and entropy of features obtained by coarse-graining mixed states, using conditional mutual information as distortion, and mixed-state geometry distinguishes state cardinality, entropy and dimension~\cite{jurgens2021divergent,jurgens2021ambiguity}. Wieck-Sosa and Shalizi~\cite{wiecksosa2026} relate the statistical convergence of conditional-law estimators to the intrinsic dimension of predictive states. Here accuracy is measured directly on future-event probabilities or on total forecast loss, and the retained label must itself be updated recursively from a fixed initial state.

In the individual-sequence setting, Feder, Merhav and Gutman~\cite{fmg1992} introduced finite-state predictability, and Meron and Feder~\cite{meronfeder2004} studied the asymptotic regret attainable with a fixed number of states. Recent-history prediction~\cite{sharan2018} and minimax prediction of unknown HMMs through universal compression~\cite{han2024} ask how accurately a law can be predicted or learned from observations. In our setting the law is known, the scarce resource is the retained state, and accuracy is required at every time rather than on average over time.

For the HMM constructions, Hilbert-metric stability of the filter~\cite{lgo} controls the effect of repeated rounding; quantized filters~\cite{pagespham2005} and finite-memory beliefs~\cite{karayuksel2022} are related approximations. For generation, we start from the one-step coarse-graining argument of Yang et al.\ \cite[Appendix F]{yang2023}, extend it to finite histories and time-dependent codebooks, add up the loss over the forecast, and then pass to the stationary predictor law. A backward-block estimate gives the required nonconcentration for every nondegenerate positive two-state model, including models whose filter images overlap. The resulting rate is governed by the quantization error of the predictor law, the subject of quantization theory~\cite{graf_luschgy2000,graf_luschgy2002}.

For sliding windows, we build on the counting framework of Datar et al.\ \cite{datar2002}, the uniform additive-error bounds of Ben Basat et al.\ \cite{bb}, and the streaming lower bound of Braverman, Garg and Woodruff~\cite[full version, Theorem~13]{bgw}. For a specified source, we determine the mean-square cost and attain its order $n/(1+E^2)+\log n$ with a finite-capacity queue. Finite-memory hypothesis testing and estimation~\cite{hellman1970,leighton1986,berg2024survey} concern a fixed unknown parameter. In the window problem the target changes with every observation. Near the constant forecast we use alternating canalizing functions~\cite{khs13} and Tribes computations~\cite{bv10,bgz22}. Their response to a bias in the input bits gives the reduction in count variance, and our constructions keep this reduction at every query time. Venkatesh and Franklin~\cite{vf91} studied how much information one bit of memory can retain about a Bernoulli sequence. Our one-bit results concern the mean-square error of a window count at every query time.

Finite-state realization also links these questions to sequence models. Learned belief geometry is studied in~\cite{shai2024,piotrowski2025}, and Heaviside Elman networks realize deterministic probabilistic finite-state automata~\cite{svete2023}. Growing precision allows more general regular language models~\cite{svete2024}, and auxiliary sampled symbols followed by projection onto visible outputs give constant-precision realizations~\cite{nowak2024}; our private-state generator admits a realization of this kind (Appendix~\ref{app:projected-output}). Related bounds concern copying~\cite{jelassi2024}, the attention workspace for arbitrary key--value streams~\cite{harisonak2025}, and the tradeoff between recurrent state size and recall~\cite{arora2024}. These bounds count the information about the input that the model must store. In our setting the extra memory of generation comes instead from adding up one-step approximation errors. Pretrained Transformers can be distilled into state-space models~\cite{bick2024} or linearized with low-rank adaptation~\cite{zhang2025lolcats}; in both cases the student's persistent state is set by its architecture and precision. Section~\ref{sec:sequence-models} applies our converse to such students.

Section~\ref{sec:tasks} sets up the state model and Proposition~\ref{prop:common-risk}. Sections~\ref{sec:hmm} and~\ref{sec:generation} treat queries and generation for hidden Markov models, and Section~\ref{sec:model} treats moving-window counts. Section~\ref{sec:scope} discusses the loss budget and the precision that the generation bounds require, and applies the generation converse to autoregressive sequence models. Proofs and the calculations behind the finite examples are in the appendices.

\section{Tasks, loss, and retained memory}\label{sec:tasks}

At a fixed time, a retained state identifies histories that the machine can no longer distinguish. Accuracy depends on what varies within these groups, while recursion restricts how the groups evolve after a new observation. We formalize both aspects through conditional risk and continuation consistency.

\subsection{Machines and memory}

Fix a known source on a finite alphabet. An observer starts from a fixed state and updates after every incoming symbol:
\begin{equation}\label{eq:common-recursion}
Z_0=z_\star,\qquad Z_{t+1}=F_t(Z_t,Y_{t+1}).
\end{equation}
If $\mathcal Z$ has $K$ elements, the machine uses $\lceil\log_2K\rceil$ bits of memory. A free clock allows the update maps and decoders to depend on $t$, but storing a phase counter or transition table in the state costs memory. Only the state $Z_t$ kept between observations counts as memory. The source law and the maps belong to the description of the machine, and temporary computation by a decoder is not counted. Decoders see the past only through $Z_t$, so a stored record of past symbols, such as a cache, is part of the state. The machine may depend on the requested accuracy and forecast horizon, but its state bound must hold at all observation times.

A \emph{query observer} uses $Z_t$ to answer a supplied future event $B$ through a decoder $G_B$. A \emph{rolling estimator} reports its statistic after each update from the old state and new symbol. Both produce a current answer from the retained history. A generator must also use that history to supply successive outputs. A \emph{unifilar generator} assigns a next-symbol distribution to each state, samples an output, and applies \eqref{eq:common-recursion}; its fixed initial state and emitted word determine the entire state trajectory.

A \emph{private-state generator} observes through the same deterministic recursion and fixed start. At a forecast request, it initializes a simulation state from a law depending on the past only through $(t,Z_t)$. Subsequent transitions may retain latent randomness beyond that revealed by the outputs. Observation and simulation share storage, with disjoint sets of labels, and both sets count toward memory.

For the static comparison, a \emph{single-window encoder} chooses a label after inspecting the complete window. Its label count measures the description length of that one input. Recursive maintenance adds the requirement that each new label be computed from the preceding label and the next observation.

As an example, consider a binary alarm stream driven by a hidden regime that switches rarely, as in Section~\ref{sec:two-regime}. A query observer can keep a rounded predictive belief and answer a question such as ``an alarm within the next $T$ steps?'' by averaging the probability of that event under each regime, weighted by the belief. A unifilar generator also keeps a rounded belief, but it emits the alarms itself and updates the belief from its own outputs, so the rounding errors of successive steps add up over the forecast. A private-state generator instead draws the current regime from the belief once and then runs the hidden chain. For a count over a window of $n$ bits, a single-window encoder sees the whole window and can store just the count, whereas a rolling estimator must update the count as one bit enters and another leaves, and so must remember something about the bits that are about to leave.

\subsection{Conditional risk and recursive consistency}

Conditional variance and conditional relative entropy express the best
accuracy available from a retained state. Let
$\mathcal F_t=\sigma(Y_1,\ldots,Y_t)$, let $W_t$ be a square-integrable
$\mathcal F_t$-measurable scalar target, and let
$\Pi_t(y)=\Pp(Y_{t+1}=y\mid\mathcal F_t)$ be the one-step predictor.
Write
\[
\bar W_t=\E(W_t\mid Z_t),\qquad
\bar\Pi_t=\E(\Pi_t\mid Z_t).
\]

\Needspace{17\baselineskip}
\begin{proposition}[State-conditioned risk and continuation consistency]
\label{prop:common-risk}
For every finite state $Z_t$ obtained from \eqref{eq:common-recursion},
every scalar decoder $a_t$, and every strictly positive probability decoder
$q_t$, one has
\begin{align}
\E(W_t-a_t(Z_t))^2
&=\E\Var(W_t\mid Z_t)
  +\E(\bar W_t-a_t(Z_t))^2,\label{eq:common-variance}\\
\E D_{\mathrm{KL}}(\Pi_t\Vert q_t(Z_t))
&=\E D_{\mathrm{KL}}(\Pi_t\Vert\bar\Pi_t)
  +\E D_{\mathrm{KL}}(\bar\Pi_t\Vert q_t(Z_t)).
\label{eq:common-kl}
\end{align}
If $C$ is any additional random variable and $W=W(Y_{1:t},C)$ is
square-integrable, every decoder $\phi$ satisfies
\begin{equation}\label{eq:common-probe}
\E(W-\phi(Z_t,C))^2\ge\E\Var(W\mid Z_t,C).
\end{equation}
Finally, if histories $h,h'$ of the same length give the same state,
then every common continuation $w$ gives the same subsequent state:
\begin{equation}\label{eq:common-continuation}
Z_t(h)=Z_t(h')\quad\Longrightarrow\quad
Z_{t+s}(hw)=Z_{t+s}(h'w),\qquad |w|=s.
\end{equation}
\end{proposition}

In \eqref{eq:common-variance}, conditioning on the retained state separates the uncertainty left by the encoding from the error of the chosen decoder. For a count of variance $n/4$, the optimal decoder therefore gains $\Var(\E(W_t\mid Z_t))$. Equation~\eqref{eq:common-kl} gives the log-loss analogue, and Section~\ref{sec:generation} sums these one-step losses over the forecast horizon.

Continuation consistency ties the present answer to later updates. Histories with similar current counts can have different expiring bits; once merged, they remain indistinguishable after any common continuation. In the window lower bound, the decoder runs the machine along such a continuation, and its error is still at least the conditional variance in \eqref{eq:common-probe}. Appendix~\ref{app:common-risk} proves the identities.

\subsection{Loss and scale in each comparison}

The three losses treat histories differently. Mean-square estimation controls $\sup_t\E(\widehat W_t-W_t)^2$, so the expectation is taken at each deterministic query time before the supremum over time. Pointwise HMM prediction must meet its relative guarantee after every admissible history. Expected KL instead averages the conditional loss over histories under the source, and the bound must hold at every forecast start time. All logarithms are natural unless a base is displayed, and memory is measured in bits.

The HMM comparisons keep the source fixed while accuracy and forecast horizon vary. The window comparisons vary the delay within a specified stationary family. Table~\ref{tab:tasks} summarizes the resulting costs.
\begin{table}[tb]
\caption{Memory for each task. The HMM is fixed, strictly
stationary, strictly positive and nondegenerate, with two hidden states and a finite
observation alphabet. Query and private-state costs are as $\varepsilon\downarrow0$,
unifilar generation as $T/\varepsilon\to\infty$,
and the window costs as $n\to\infty$ with $E^2=cn$
and fixed $0<c<1/4$.}
\label{tab:tasks}
\centering\small
\setlength{\tabcolsep}{4pt}
\begin{tabular}{@{}p{.29\textwidth}p{.40\textwidth}p{.26\textwidth}@{}}
\toprule
Task & Accuracy and retained state & Memory in bits\\
\midrule
Single-window encoding & One label after reading the full window; MSE $cn$ & $O_c(1)$\\[3pt]
Online count maintenance & Retained state; MSE $cn$ at each $t\ge n$ & $\Theta_c(\log_2n)$\\[3pt]
HMM event queries & Retained history state; uniform event log error $\varepsilon$ & $\Theta(\log_2(1/\varepsilon))$, independent of $T$\\[3pt]
Unifilar generation & State determined by symbols; uniform word log error $\varepsilon$ through $T$ & $\Theta(\log_2(T/\varepsilon))$\\[3pt]
Unifilar generation, total KL & Same state constraint; expected total KL at most $\varepsilon$ & $\Theta(\log_2(T/\varepsilon))$; upper coefficient $1/2$\\[3pt]
Private-state generation & Deterministic observation followed by private simulation; both phases counted; uniform future-event guarantee & $\Theta(\log_2(1/\varepsilon))$, independent of $T$\\
\bottomrule
\end{tabular}
\end{table}

\begin{table}[tb]
\caption{Main notation. Symbols local to a construction are defined where
they are used.}
\label{tab:notation}
\centering\small
\begin{tabular}{@{}lp{.73\textwidth}@{}}
\toprule
Symbol & Meaning\\
\midrule
$K,b$ & Number of retained states and its bit cost $\lceil\log_2K\rceil$\\
$t,T,n$ & Observation time, future generation horizon, and window length\\
$u_t,Z_t$ & Exact predictive belief and retained observer state\\
$P,g_i,D_y$ & Hidden transition matrix, emission law, and emission matrix\\
$\varepsilon$ & Event log-probability tolerance or specified total KL budget\\
$E$ & Root mean-square tolerance in window-count units\\
$\delta$ & Delayed-copy dependence parameter, $0<\delta\le1/2$\\
$\Pi_j,\mu_j$ & One-step predictor after $j$ observations and its law\\
$\mu,\mathcal D_K(\mu)$ & Stationary predictor law and squared TV quantization error\\
$\kappa,d_*$ & Hilbert contraction coefficient and a small-ball exponent\\
$S_t,\Gamma_{n,\delta}(b)$ & Current window count and all-time variance gain with $b$ bits\\
\bottomrule
\end{tabular}
\end{table}

\section{Finite memory for future-event queries}\label{sec:hmm}

\subsection{A two-regime example}\label{sec:two-regime}

Consider an alarm whose rate depends on an unobserved operating
regime. Its binary alarm stream follows the stationary hidden Markov
model
\[
P=\begin{pmatrix}4/5&1/5\\1/5&4/5\end{pmatrix},\qquad
\Pp(Y_t=1\mid H_t=0)=\frac1{10},\qquad
\Pp(Y_t=1\mid H_t=1)=\frac9{10}.
\]
The regimes persist, but neither an alarm nor its absence identifies
the current regime. Past observations enter a forecast through the
predictive belief $u_t=\mathcal L(H_{t+1}\mid Y_{1:t})$. With
$D_0=\operatorname{diag}(9/10,1/10)$, the probability of at least one
alarm in the next $T$ observations is
\[
R_{t,T}=1-u_t(D_0P)^T\one.
\]
Changing the event changes its likelihood vector, while the same
belief supplies the information from the past. We will approximate
this belief by a recursively maintained finite state. The resulting
observer answers event queries uniformly over histories and horizons.

\subsection{Uniform prediction by finite-state filtering}

The construction extends to any fixed, known finite-state hidden
Markov model with positive transitions. Let $(H_t)$ be a Markov chain on $\{1,\ldots,m\}$ with a strictly positive row-stochastic transition matrix $P$. Given the hidden states, observations $(Y_t)$ are independent, take values in a finite alphabet $\mathcal A$, and have probabilities
\[
g_i(y)=\Pp(Y_t=y\mid H_t=i),\qquad
D_y=\operatorname{diag}(g_1(y),\ldots,g_m(y)).
\]
For the predictive belief $u_t=\mathcal L(H_{t+1}\mid Y_{1:t})$, the update is
\begin{equation}\label{eq:filter}
u_{t+1}=\Psi_{Y_{t+1}}(u_t),\qquad
\Psi_y(u)=\frac{uD_y}{uD_y\one}P.
\end{equation}
The image of every defined update lies in the convex hull $\mathcal C$ of the rows of $P$, a compact subset of the positive simplex.

Relative errors in the belief are naturally measured by the Hilbert
distance. For positive probability vectors $u,w$, define
\[
d_H(u,w)=\max_i\log\frac{u_i}{w_i}
-\min_i\log\frac{u_i}{w_i}.
\]
Set
\begin{equation}\label{eq:birkhoff}
\Delta(P)=\log\max_{i,j,k,l}\frac{P_{ik}P_{jl}}{P_{il}P_{jk}},
\qquad \kappa=\tanh\frac{\Delta(P)}4<1.
\end{equation}
Birkhoff's contraction inequality, in the filtering formulation of~\cite{lgo}, gives
\begin{equation}\label{eq:filter-contraction}
d_H(\Psi_y(u),\Psi_y(w))\le\kappa d_H(u,w).
\end{equation}

For a future event $B$, let
\[
v_B(i)=\Pp(B\mid H_{t+1}=i).
\]
We index $B$ relative to the current time. Its conditional probability
is the mixture $uv_B$, so a relative approximation of the mixture
weights will control the answer to every event query.

\begin{theorem}[Finite recursive prediction]\label{thm:hmm-memory}
For every fixed initial predictive belief $u_0$ and every $\varepsilon>0$, there is a deterministic finite-state recursion with the following property. Its state is a probability vector $z$, and the decoder for an event query $B$ is
\begin{equation}\label{eq:hmm-decoder}
G_B(z)=zv_B.
\end{equation}
At every time and every positive-probability observation history,
\begin{equation}\label{eq:hmm-risk-bound}
e^{-\varepsilon}\le
\frac{G_B(Z_t)}{\Pp(B\mid Y_{1:t})}
\le e^\varepsilon
\end{equation}
for all future events $B$ of positive conditional probability. Put
\[
p_{\min}=\min_{i,j}P_{ij},\qquad p_{\max}=\max_{i,j}P_{ij},
\qquad L=\log(p_{\max}/p_{\min}),\qquad
\eta=(1-\kappa)\varepsilon.
\]
If $L>0$, the state space can be chosen with
\begin{equation}\label{eq:hmm-state-count}
K\le1+\left(1+\left\lceil\frac{2L}{\eta}\right\rceil\right)^{m-1}.
\end{equation}
If $L=0$, two states suffice, and one suffices when $u_0$ is uniform.
\end{theorem}

The leading $1$ in \eqref{eq:hmm-state-count} is a state that stores $u_0$ exactly. The initial belief may lie on the boundary of the simplex, but all later predictive beliefs
are positive.
For the two-regime model, the theorem applies both to an alarm-free interval,
whose likelihood vector is $(D_0P)^T\one$, and to its complement,
whose vector is $\one-(D_0P)^T\one$.
Appendix~\ref{app:hmm-queries} gives the rounding construction and its
state count.

Let $K_{\mathrm{query}}(\varepsilon)$ be the smallest number of
states among query observers, with arbitrary decoders, that satisfy
\eqref{eq:hmm-risk-bound} at every time, after every
positive-probability history, and for every future event. Put
\begin{equation}\label{eq:query-memory-cost}
b_{\mathrm{query}}(\varepsilon)
=\left\lceil\log_2K_{\mathrm{query}}(\varepsilon)\right\rceil.
\end{equation}

\begin{theorem}[Optimal order of query memory]\label{thm:query-memory}
Consider a stationary hidden Markov model with two hidden states, a
finite observation alphabet, and strictly positive transition and
emission probabilities. If the transition rows coincide or the two
emission distributions coincide, one state answers every future-event
query exactly. Otherwise there are model-dependent constants
$c_q,C_q,d_q>0$ such that, for $0<\varepsilon\le1$,
\begin{equation}\label{eq:query-state-bounds}
c_q\varepsilon^{-d_q}
\le K_{\mathrm{query}}(\varepsilon)
\le C_q(1+\varepsilon^{-1}).
\end{equation}
Consequently,
\begin{equation}\label{eq:query-bit-law}
b_{\mathrm{query}}(\varepsilon)
=\Theta\!\left(\log\frac1\varepsilon\right)
\qquad\text{as }\varepsilon\downarrow0.
\end{equation}
The lower bound allows time-dependent updates and decoders. The upper
bound is attained by a time-homogeneous observer.
\end{theorem}

Single-symbol queries already force the lower bound. A power-law
small-ball estimate limits how closely a finite codebook can
approximate their stationary predictor. Appendix~\ref{app:hmm-query-lower}
proves the estimate and its transfer to finite histories.

\subsection{Dependence on the query horizon}

The state bound has no horizon term because an event probability is
a mixture over the predictive belief. For positive predictive
beliefs $u,z$ and any nonzero event likelihood vector,
\[
\min_i\frac{z_i}{u_i}\le\frac{zv_B}{uv_B}
\le\max_i\frac{z_i}{u_i}.
\]
Thus the relative error depends on the two beliefs, uniformly over
the duration and probability of the event (Lemma~\ref{lem:mixture}).
Filter contraction keeps those beliefs close throughout observation:
if each rounding step introduces Hilbert error at most $\eta$, then
\[
e_{t+1}\le\kappa e_t+\eta.
\]
Choosing $\eta=(1-\kappa)\varepsilon$ bounds the error by
$\varepsilon$ at every time. Contraction also shows what happens as
the queried horizon grows. The additional log-likelihood discrepancy
between two beliefs decays geometrically.

Let $\mathsf P_u$ denote the future-observation law on $\mathcal A^{\mathbb N}$ with initial predictive belief $u$. For positive $u,w$, define
\begin{align}
d_T(u,w)&=\sup_{B\in\sigma(Y_1,\ldots,Y_T),\,\mathsf P_w(B)>0}
\left|\log\frac{\mathsf P_u(B)}{\mathsf P_w(B)}\right|,\label{eq:horizon-metric}\\
d_{\mathrm{all}}(u,w)&=\sup_{B\in\sigma(Y_1,Y_2,\ldots),\,\mathsf P_w(B)>0}
\left|\log\frac{\mathsf P_u(B)}{\mathsf P_w(B)}\right|.\notag
\end{align}
Because the alphabet is finite, $d_T$ is the largest absolute log likelihood ratio among the possible words of length $T$. The ratio for any union of words is a weighted average of their ratios.

\begin{theorem}[Horizon saturation]\label{thm:saturation}
For every pair of positive predictive beliefs and every $T\ge1$,
\begin{equation}\label{eq:saturation}
0\le d_{\mathrm{all}}(u,w)-d_T(u,w)
\le\kappa^T d_H(u,w).
\end{equation}
In particular, $d_{\mathrm{all}}(u,w)=\sup_Td_T(u,w)\le d_H(u,w)$.
\end{theorem}

A belief that is accurate for events on the next $T$ symbols is therefore accurate for all future events up to an additional $\kappa^Td_H(u,w)$. Appendix~\ref{app:hmm-saturation} gives the proof, and
Appendix~\ref{app:hmm-illustrations} follows the convergence along a
single word.

\Needspace{10\baselineskip}
\section{The memory cost of sequential generation}\label{sec:generation}

\subsection{Deterministic-state generation and its loss criteria}

An event query asks for one conditional probability. A generator must
produce the whole conditional law of the future, one emission at a
time, and each emission leaves a state for the next step. For the
two-regime model, the answers $zv_B$ are the probabilities of the
original HMM started from belief $z$; how much memory it takes to
generate them depends on how the state may be updated. We begin with
states determined by the starting state and visible outputs, then
allow a private hidden regime.

Let $Y=(Y_t)_{t\ge1}$ be a source on a finite alphabet $\mathcal A$.
A deterministic-state generator has a finite state space $\mathcal Z$,
a fixed initial state $z_\star$, transition maps
\[
F_t:\mathcal Z\times\mathcal A\longrightarrow\mathcal Z,
\]
and output distributions $q_{t,z}$ on $\mathcal A$. When observations are
supplied to the machine, its state satisfies
\begin{equation}\label{eq:gen-state-update}
Z_0=z_\star,\qquad Z_{t+1}=F_t(Z_t,Y_{t+1}).
\end{equation}
During generation, the machine samples from $q_{t,Z_t}$ and applies
the same update, so it is a unifilar generator in the sense of Section~\ref{sec:tasks}.

For a word $w=w_1\cdots w_\ell$, define the generated word probability
from state $z$ at time $t$ by
\begin{equation}\label{eq:gen-word-law}
\mathsf Q_{t,z}(w)
=\prod_{j=1}^{\ell}q_{t+j-1,z_{j-1}}(w_j),
\qquad
z_0=z,\quad z_j=F_{t+j-1}(z_{j-1},w_j).
\end{equation}

We require, for every $t\ge0$, every positive-probability history
$h\in\mathcal A^t$, every $1\le\ell\le T$, and every
$w\in\mathcal A^\ell$, that
\begin{equation}\label{eq:gen-criterion}
\begin{split}
e^{-\varepsilon}\Pp(Y_{t+1:t+\ell}=w\mid Y_{1:t}=h)
&\le \mathsf Q_{t,Z_t(h)}(w)\\
&\le e^{\varepsilon}
\Pp(Y_{t+1:t+\ell}=w\mid Y_{1:t}=h).
\end{split}
\end{equation}
This criterion gives equal supports and absolute log-likelihood
error at most $\varepsilon$ on every word of positive conditional
probability. Summing over words gives the equivalent multiplicative
guarantee for every event determined by at most $T$ future symbols.

Let $K_{\mathrm{gen}}(T,\varepsilon)$ be the minimum number of states
among machines satisfying \eqref{eq:gen-criterion}, and set
\begin{equation}\label{eq:gen-bit-cost}
b_{\mathrm{gen}}(T,\varepsilon)
=\left\lceil\log_2 K_{\mathrm{gen}}(T,\varepsilon)\right\rceil.
\end{equation}
Both quantities are infinite when no finite machine satisfies the
criterion.

We also measure source-weighted generation error by total relative
entropy,
\begin{equation}\label{eq:gen-kl-criterion}
\sup_{t\ge0}\E\,D_{\mathrm{KL}}\!\left(
\mathcal L(Y_{t+1:t+T}\mid Y_{1:t})
\,\middle\Vert\,\mathsf Q_{t,Z_t}\right)\le\varepsilon.
\end{equation}
Here histories are averaged under the true source, and relative
entropy has direction $P\Vert Q$. The budget applies to the whole
length-$T$ forecast.
Let $K_{\mathrm{KL}}(T,\varepsilon)$ denote the minimum state count for
\eqref{eq:gen-kl-criterion}, and let
$b_{\mathrm{KL}}=\lceil\log_2K_{\mathrm{KL}}\rceil$. The uniform word
criterion implies \eqref{eq:gen-kl-criterion}, so
\[
K_{\mathrm{KL}}(T,\varepsilon)
\le K_{\mathrm{gen}}(T,\varepsilon).
\]

\subsection{Classification for two hidden states}

The next theorem classifies generation for positive two-state models
under both losses. To distinguish the
cost of a long continuation from the requirement of repeated
forecasting, we also consider a single forecast at time zero. Let
$K_{\mathrm{init}}(T,\varepsilon)$ be the minimum number of states
when only the bounds
\begin{equation}\label{eq:gen-initial-criterion}
e^{-\varepsilon}\Pp(Y_{1:T}=w)
\le\mathsf Q_{0,z_\star}(w)
\le e^{\varepsilon}\Pp(Y_{1:T}=w),
\qquad w\in\mathcal A^T,
\end{equation}
are imposed. Every machine satisfying \eqref{eq:gen-criterion}
satisfies these bounds, so
$K_{\mathrm{init}}(T,\varepsilon)
\le K_{\mathrm{gen}}(T,\varepsilon)$.

Write the transition matrix and emission probabilities as
\begin{equation}\label{eq:gen-general-binary-parameters}
P=\begin{pmatrix}1-\alpha&\alpha\\1-\gamma&\gamma\end{pmatrix},
\qquad
g_i(y)=\Pp(Y_t=y\mid H_t=i)>0,
\quad i\in\{0,1\},\ y\in\mathcal A,
\end{equation}
where $\alpha,\gamma\in(0,1)$ and $\mathcal A$ is finite. Each $g_i$
is a probability distribution. Initialize the hidden chain in its
stationary distribution. For a binary alphabet, write $e_i=g_i(1)$.

\Needspace{7\baselineskip}
\begin{theorem}[Two-state classification]\label{thm:gen-binary-classification}
For the model \eqref{eq:gen-general-binary-parameters}, the following
statements hold.
\begin{enumerate}
\item If $\gamma=\alpha$ or $g_0=g_1$, the observations are independent
and identically distributed, and one state gives exact generation.
\item If $\gamma\ne\alpha$ and $g_0\ne g_1$, there are constants
$c,C,d_*>0$, depending only on the model, such that, for every $T\ge1$
and $\varepsilon>0$, with $x=T/\varepsilon$,
\begin{align}
c x^{d_*/2}
&\le K_{\mathrm{gen}}(T,\varepsilon)\le C(1+x),
\label{eq:gen-classification-state-bounds}\\
c x^{d_*/2}
&\le K_{\mathrm{KL}}(T,\varepsilon)\le C(1+\sqrt{x}).
\label{eq:gen-kl-classification-state-bounds}
\end{align}
The exponent $d_*$ is constructed in
\eqref{eq:gen-general-small-ball-constants}. In particular,
\begin{equation}\label{eq:gen-general-binary-bit-law}
b_{\mathrm{gen}}(T,\varepsilon)
=\Theta_{P,g_0,g_1}\!\left(\log\frac{T}{\varepsilon}\right),
\qquad T/\varepsilon\to\infty,
\end{equation}
and
\begin{equation}\label{eq:gen-kl-bit-law}
b_{\mathrm{KL}}(T,\varepsilon)
=\Theta_{P,g_0,g_1}\!\left(\log\frac{T}{\varepsilon}\right),
\qquad T/\varepsilon\to\infty.
\end{equation}
For a single initial forecast and every fixed $\varepsilon>0$,
\begin{equation}\label{eq:gen-general-binary-initial-law}
\left\lceil\log_2K_{\mathrm{init}}(T,\varepsilon)\right\rceil
=\Theta_{P,g_0,g_1,\varepsilon}(\log T),
\qquad T\to\infty.
\end{equation}
\end{enumerate}
The lower bounds permit a free clock and arbitrary time-dependent
maps. The upper bounds are achieved by time-homogeneous machines.
\end{theorem}

For the nondegenerate models, the horizon costs order $\log T$ bits at fixed tolerance, even for a single initial forecast, because the generator must supply each successive conditional output law from its finite state. When the hidden chain starts in its stationary distribution, the initialization changes the lower bound of Theorem~\ref{thm:gen-quantization} only by an additive constant (Proposition~\ref{prop:gen-initial-forecast}). The proof in Appendix~\ref{app:hmm-binary} bounds the resulting approximation cost.

\subsection{Predictor quantization and accumulated loss}

A $K$-state generator can supply at most $K$ next-symbol
distributions at a given time. Proposition~\ref{prop:common-risk}
bounds the one-step loss caused by identifying distinct predictors. The KL chain rule
then adds this loss at every step of the forecast.

For probability vectors $v,c$ on $\mathcal A$, write
\[
d_{\mathrm{TV}}(v,c)=\frac12\sum_{y\in\mathcal A}|v(y)-c(y)|.
\]
For a probability measure $\mu$ on the simplex
$\Delta_{\mathcal A}$, define its squared quantization error by
\begin{equation}\label{eq:gen-quantization}
\mathcal D_K(\mu)
=\inf_{c_1,\ldots,c_K\in\Delta_{\mathcal A}}
\int \min_{1\le j\le K}d_{\mathrm{TV}}(v,c_j)^2\,\mu(dv).
\end{equation}
Yang et al.~\cite[Appendix F]{yang2023} developed one-step
coarse-graining bounds for stationary unifilar models. We use the
same approximation principle with finite observed histories, a fixed
initial machine state and time-dependent codebooks. For a source
$Y=(Y_t)_{t\ge1}$ on $\mathcal A$, write
\[
\Pi_j(y)=\Pp(Y_{j+1}=y\mid Y_{1:j}),\qquad
\mu_j=\mathcal L(\Pi_j),\qquad j\ge0,
\]
for its one-step predictors and their laws.

\Needspace{14\baselineskip}
\begin{theorem}[Predictor quantization converse]\label{thm:gen-quantization}
Let $Y=(Y_t)_{t\ge1}$ be a source on a finite alphabet $\mathcal A$.
\begin{enumerate}
\item Suppose the state $Z_j$ of a deterministic-state generator takes
at most $K_j$ values at each time $j$. Then, for every $t\ge0$ and
$T\ge1$,
\begin{equation}\label{eq:gen-finite-history}
\E D_{\mathrm{KL}}\!\left(
\mathcal L(Y_{t+1:t+T}\mid Y_{1:t})
\,\middle\Vert\,\mathsf Q_{t,Z_t}\right)
\ge2\sum_{j=t}^{t+T-1}\mathcal D_{K_j}(\mu_j).
\end{equation}
\item Suppose $Y$ is the restriction to positive times of a stationary
process $(Y_t)_{t\in\mathbb Z}$, and let $\mu$ be the law of its
infinite-past predictor
\begin{equation}\label{eq:gen-stationary-predictor}
V(y)=\Pp(Y_1=y\mid\sigma(Y_j:j\le0)),
\qquad y\in\mathcal A.
\end{equation}
Every $K$-state deterministic-state generator satisfying the expected
total-KL criterion \eqref{eq:gen-kl-criterion} obeys
\begin{equation}\label{eq:gen-quantization-converse}
\varepsilon\ge2T\mathcal D_K(\mu).
\end{equation}
\end{enumerate}
The uniform word criterion \eqref{eq:gen-criterion} implies
\eqref{eq:gen-kl-criterion}, so (ii) also holds under it. Neither part
requires ergodicity or stationarity of the machine.
\end{theorem}

The chain rule writes the total KL as
a sum of one-step losses along the source sequence, and at each time
the output laws of the at most $K_j$ states form a codebook for
$\Pi_j$. Pinsker's inequality bounds each one-step loss below by twice the squared total-variation distance from $\Pi_j$ to the output law of the current state, so its expectation is at least $2\mathcal D_{K_j}(\mu_j)$. For strictly positive decoders, \eqref{eq:common-kl} splits
each one-step loss into the information
$I(Y_{j+1};Y_{1:j}\mid Z_j)$ that the history carries about the next
symbol beyond the state, and the error of the decoder given the state.
Part (ii) follows by letting the observed past grow
(Appendix~\ref{app:hmm-converse}). In
\eqref{eq:gen-quantization-converse}, the horizon enters only through the
factor $T$, and the source only through the distortion $\mathcal D_K(\mu)$.

The upper bounds rest on the same accumulation. For
uniform word error, keeping the filter precision at $\varepsilon/T$
makes the successive conditional log errors sum to at most
$\varepsilon$. Theorem~\ref{thm:gen-hmm-upper} gives
\[
b_{\mathrm{gen}}(T,\varepsilon)
\le(m-1)\log_2(1+T/\varepsilon)+A_P.
\]

Total forward KL is quadratic in small one-step probability errors,
so belief precision of order $\sqrt{\varepsilon/T}$ suffices, as the
following construction shows.

\begin{theorem}[Square-precision generation for total KL]\label{thm:gen-kl-upper}
For the $m$-state hidden Markov model of Section~\ref{sec:hmm}, with any
fixed initial predictive belief, every $T\ge1$, and every
$\varepsilon>0$, set
\[
L_P=\log(p_{\max}/p_{\min}),\qquad
\xi=\log\!\left(1+\sqrt{\varepsilon/T}\right).
\]
If $L_P>0$, a time-homogeneous deterministic-state generator has
\begin{equation}\label{eq:gen-kl-state-upper}
K\le1+\left(1+
\left\lceil\frac{2L_P}{(1-\kappa)\xi}\right\rceil
\right)^{m-1}
\end{equation}
states and satisfies the stronger conditional bound
\begin{equation}\label{eq:gen-kl-conditional-upper}
\sup_{t\ge0}\ \sup_{h:\Pp(Y_{1:t}=h)>0}
D_{\mathrm{KL}}\!\left(
\mathcal L(Y_{t+1:t+T}\mid Y_{1:t}=h)
\,\middle\Vert\,\mathsf Q_{t,Z_t(h)}\right)\le\varepsilon.
\end{equation}
If $L_P=0$, two states suffice for exact generation, and one suffices
when the initial belief is uniform. For $0<\varepsilon/T\le1$,
\begin{equation}\label{eq:gen-kl-bit-upper}
b_{\mathrm{KL}}(T,\varepsilon)
\le\frac{m-1}{2}\log_2(T/\varepsilon)+B_P,
\end{equation}
where $B_P$ does not depend on $T$, $\varepsilon$, or the initial belief.
\end{theorem}

The estimate combines
$D_{\mathrm{KL}}(p\Vert q)\le\chi^2(p\Vert q)$ with the
mixture-ratio bound; Appendix~\ref{app:hmm-generation} gives the proof.

\subsection{Geometry of the output predictor}\label{sec:predictor-geometry}

By the converse, what matters is the law $\mu$ of the observable
predictor. For any stationary finite-alphabet
source, infinite support of $\mu$ already forces
$K_{\mathrm{KL}}(T,\varepsilon)$ and $K_{\mathrm{gen}}(T,\varepsilon)$
to diverge with $T$ at fixed $\varepsilon$
(Corollary~\ref{cor:gen-small-ball}). Quantitative rates depend on
how much predictor mass a small codebook can capture.

For a nondegenerate positive two-hidden-state model, two sufficiently long
constant observation blocks have separated filter images. Working
backward, a small target interval excludes one such block at each
scale. Positive emissions bound the excluded conditional mass away
from zero, so iteration gives a power-law small-ball estimate even
when the one-step filter images overlap. Together with
Corollary~\ref{cor:gen-small-ball}, this proves the state-count lower
bounds throughout the two-state class. The filter recursion is an
iterated function system with place-dependent probabilities, in the
framework of Barnsley et al.\ \cite{bdeg1988}. Its stationary law is a
Blackwell measure~\cite{blackwell1957}, whose singularity and dimension
are studied in~\cite{bps2012}.

Because the converse is stated through the distortion
\eqref{eq:gen-quantization}, the relevant notion of dimension is the
mean-square quantization dimension of $\mu$, which describes the decay
of $\mathcal D_K(\mu)$. Information dimension (fine-partition entropy)
and box dimension of the support (covering growth) enter mixed-state
memory~\cite{jurgens2021divergent,jurgens2021ambiguity}; the
quantization dimension can differ from both even for separated
self-similar measures~\cite{graf_luschgy2002}.

\subsection{Private hidden-state simulation}

The horizon cost disappears when the simulator can retain a hidden
regime. At a forecast request it draws that regime from the retained
belief $z$, then follows the original emission and transition laws.
Only initialization uses the approximate belief. This posterior
initialization mechanism~\cite{lohray2009} applies to the general
$m$-state HMM of Section~\ref{sec:hmm}; the finite recursive observer
supplies its approximate starting law.

\begin{proposition}[Initialization of a private-state generator]
\label{prop:gen-private-state}
For the same hidden Markov model and every $\varepsilon>0$, there
is a finite-state observer with the following property. At every
time and every positive-probability observed history, its state
initializes a randomized generator with $m$ private hidden states
whose future law satisfies the multiplicative bounds in
\eqref{eq:gen-criterion} simultaneously for all finite horizons.
The observer state count is the horizon-independent count of
Theorem~\ref{thm:hmm-memory}.
\end{proposition}

For a word $w=w_1\cdots w_T$, this mechanism gives probability
\[
zD_{w_1}P\cdots D_{w_T}P\one=zv_w.
\]
The mixture comparison controls this probability for every horizon
(Appendix~\ref{app:hmm-generation}). Observation and simulation reuse
storage, so $K$ observation labels and $m$ private labels cost at most
$K+m$ states. The simulated regime generally remains random even conditional
on the emitted word.

For the private-state generators of Section~\ref{sec:tasks}, let
$K_{\mathrm{priv}}(T,\varepsilon)$ be the smallest total number of
observation and simulation labels for which the generated future law
satisfies the multiplicative bounds of \eqref{eq:gen-criterion} after
every admissible history, for every word of length at most $T$. Write
$b_{\mathrm{priv}}=\lceil\log_2K_{\mathrm{priv}}\rceil$.

\Needspace{9\baselineskip}
\begin{corollary}[Optimal order of private-state memory]
\label{cor:gen-private-memory}
Consider a stationary hidden Markov model with two hidden states, a
finite observation alphabet, strictly positive transition and emission
probabilities, distinct transition rows and distinct emission laws.
There are model-dependent constants $c_p,C_p,d_p>0$ such that, for
every $T\ge1$ and $0<\varepsilon\le1$,
\begin{equation}\label{eq:gen-private-state-bounds}
c_p\varepsilon^{-d_p}
\le K_{\mathrm{priv}}(T,\varepsilon)
\le C_p(1+\varepsilon^{-1})+2.
\end{equation}
Consequently,
\begin{equation}\label{eq:gen-private-bit-law}
b_{\mathrm{priv}}(T,\varepsilon)
=\Theta\!\left(\log\frac1\varepsilon\right)
\qquad\text{as }\varepsilon\downarrow0,
\end{equation}
with constants uniform over $T\ge1$.
\end{corollary}

For the lower bound, average over the private randomness conditional
on an observation label. The first simulated symbol then has one of
finitely many marginal laws, to which the query lower-bound argument
applies. The finite observer followed by the two-state hidden
simulator attains the upper bound; both proofs are in
Appendix~\ref{app:hmm-generation}.

\subsection{Explicit state counts}

For the two-regime model of Section~\ref{sec:two-regime}, the hidden flip and emission-noise probabilities are
\begin{equation}\label{eq:gen-explicit-parameters}
r=\frac15,\qquad s=\frac1{10}.
\end{equation}
Table~\ref{tab:hmm-state-counts} gives the corresponding state counts at
$\varepsilon=0.1$. The query observer uses $29$
reachable states for every horizon. The private simulator adds two
hidden-state labels, for $29+2=31$ states, or five bits. For unifilar
generation, the reachable counts grow with $T$ under both losses, faster
for uniform word error than for total KL. The lower bounds come from separating six
observed histories and from exact quantization of finite-past
predictors; restricting each construction to its reachable states
gives the upper bounds. Appendix~\ref{app:hmm-illustrations}
states these bounds, and Appendices~\ref{app:hmm-constants}
and~\ref{app:hmm-finite-quantization} give the exact calculations.

\begin{table}[tb]
\caption{State bounds for the two-regime model at
$\varepsilon=0.1$. Event queries and uniform-word generation use uniform
log probability error; total-KL generation uses
\eqref{eq:gen-kl-criterion}. The lower and reachable columns give states,
with bits in parentheses. The full-grid column gives construction sizes
before pruning: log-odds grids with an exact initial label for event
queries and total KL, and the uniform belief grid of
Corollary~\ref{cor:gen-explicit} for uniform-word generation. Reachable
counts are upper bounds attained by these constructions.}
\label{tab:hmm-state-counts}
\centering
\small
\begin{tabular}{@{}llrrr@{}}
\toprule
Operation & Horizon & Lower & Full grid & Reachable\\
\midrule
Event queries & All $T$ jointly & $6\ (3)$ & $72$ & $29\ (5)$\\
Uniform-word generation & $T=100$ & $6\ (3)$ & $4{,}590$ & $257\ (9)$\\
Uniform-word generation & $T=10{,}000$ & $11\ (4)$ & $458{,}825$ & $3{,}139\ (12)$\\
Total-KL generation & $T=100$ & $4\ (2)$ & $225$ & $51\ (6)$\\
Total-KL generation & $T=10{,}000$ & $11\ (4)$ & $2{,}198$ & $183\ (8)$\\
\bottomrule
\end{tabular}
\end{table}

\begin{figure}[t]
\centering
\includegraphics[width=\textwidth]{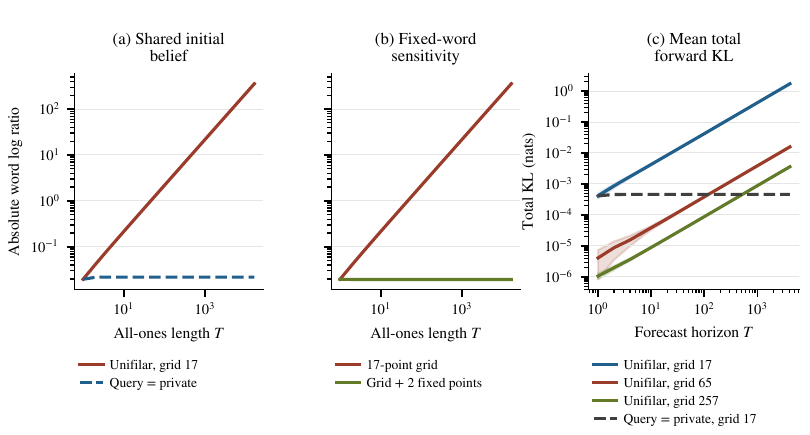}
\caption{Generation error in the two-regime model. (a) Absolute log error on the all-ones word of length $T$, from a shared initial belief, for the unifilar generator on a $17$-point grid and for event queries, which coincide with private-state simulation. (b) The same unifilar generator with and without the two filter fixed points added to its grid. (c) Estimated expected total forward KL up to horizon $T$ for unifilar generators on uniform grids of $17$, $65$ and $257$ points and for private-state simulation from the $17$-point observer; bands are $1.96$ standard errors.}
\label{fig:gen-hmm}
\end{figure}

Figure~\ref{fig:gen-hmm} shows where the extra cost of generation comes
from. In panels (a) and (b), the observed history is $0110100111$ and
every method starts from the same retained belief $z=0.7625$. Event
queries and private-state simulation from this belief keep the error
of the initial belief. The $17$-point unifilar generator rounds
after every symbol, and its error on the all-ones word grows linearly
in $T$. Adding the two exact filter fixed points to the grid gives $19$
points and $13$ reachable states instead of $9$, still four bits, and
stops the growth on this particular word. Averaged over source
histories and futures, however, the total KL of every unifilar grid in
panel (c) still grows linearly in $T$, consistent with the quantization
converse (Theorem~\ref{thm:gen-quantization}).
Appendix~\ref{app:gen-conditional-laws} gives the design.

\Needspace{10\baselineskip}
\section{Moving-window counts and rare-event forecasts}\label{sec:model}

A window estimator must form its first estimate from arriving observations
and then update it as older observations leave. Consider the count
\[
S_t=\sum_{i=t-n+1}^tX_i,\qquad t\ge n.
\]
For the source below, $S_t$ determines a conditional future-event
probability. We compare describing this count with forming and maintaining
an estimate by a finite-state recursion.

\subsection{A stream with delayed dependence}

Fix an integer $n\ge1$ and $0<\delta\le1/2$. Let $X_1,\ldots,X_n$ be independent fair bits. Independently, let $(N_t)_{t>n}$ be independent with
\[
\Pp(N_t=1)=\frac{1-\delta}{2}.
\]
For $t\ge n$, set
\begin{equation}\label{eq:source}
X_{t+1}=X_{t-n+1}\mathbin{\oplus}N_{t+1},
\end{equation}
where $\oplus$ denotes addition modulo two. The $n$ residue classes modulo
$n$ are independent stationary Markov chains with transition matrix
\[
\begin{pmatrix}(1+\delta)/2&(1-\delta)/2\\
(1-\delta)/2&(1+\delta)/2\end{pmatrix}.
\]
Their interleaving is stationary, and every consecutive block of at most
$n$ observations consists of independent fair bits. The dependence appears at lag $n$,
while each individual window has a product law.

To connect the count to prediction, define
\[
\mathcal A_t=\{X_{t+1}=\cdots=X_{t+n}=1\}.
\]
Write $p_+=(1+\delta)/2$, $p_-=(1-\delta)/2$, and $a=\log(p_+/p_-)$. Conditional on $X_{1:t}$, the next $n$ observations are independent noisy copies of the current window. Consequently,
\begin{equation}\label{eq:risk}
h_t:=\Pp(\mathcal A_t\mid X_{1:t})=p_+^{S_t}p_-^{n-S_t},
\qquad \ell_t:=\log h_t=n\log p_-+aS_t.
\end{equation}
In particular, $\Pp(\mathcal A_t)=2^{-n}$ and $S_t$ has distribution $\operatorname{Bin}(n,1/2)$ at every query time.

One-step prediction is inexpensive when the dependence is weak. For every
$t\ge n$ and every history,
\begin{align}\label{eq:one-step}
&D_{\mathrm{KL}}\bigl(\mathcal L(X_{t+1}\mid X_{1:t})\,\Vert\,\Ber(1/2)\bigr)\notag\\
&\hspace{15mm}=\frac{1+\delta}{2}\log(1+\delta)
+\frac{1-\delta}{2}\log(1-\delta)\le\delta^2.
\end{align}
The inequality follows from $D_{\mathrm{KL}}\le\chi^2$, and the divergence
is zero for $t<n$. A single-state fair predictor also has absolute log
probability error at most $-\log(1-\delta)\le2\delta$ for either next
symbol. The longer forecast in \eqref{eq:risk} depends on the current
count, and because $\ell_t$ is affine in $S_t$, an error in the count
becomes an error in the log forecast, multiplied by $a$.

\subsection{The cost of keeping the forecast current}

A $K$-state estimator has a state space $\mathcal Z$ of cardinality $K$, a fixed initial state $z_0$, and deterministic maps satisfying
\begin{equation}\label{eq:recursion}
\begin{gathered}
Z_0=z_0,\qquad Z_{t+1}=F_t(Z_t,X_{t+1}),\\
\widehat S_t=G_t(Z_t),\qquad
\widehat\ell_t=n\log p_-+a\widehat S_t.
\end{gathered}
\end{equation}
Define
\begin{equation}\label{eq:optimal-memory}
b_*(n,\delta,\varepsilon)
=\min\left\{\lceil\log_2K\rceil:
\sup_{t\ge n}\E(\widehat\ell_t-\ell_t)^2\le\varepsilon^2
\text{ for a recursion \eqref{eq:recursion}}\right\}.
\end{equation}
An exact window register makes the minimum finite. In count units, the
corresponding minimum for root mean-square tolerance $E$ is
\begin{equation}\label{eq:count-memory}
b_{\mathrm{count}}(n,\delta,E)=b_*(n,\delta,aE).
\end{equation}

The two scales are equivalent. Given a log-probability estimate, set
\begin{equation}\label{eq:normalized-error}
E=\frac{\varepsilon}{a},\qquad
\widehat S_t=\frac{\widehat\ell_t-n\log p_-}{a}.
\end{equation}
Then $\E(\widehat\ell_t-\ell_t)^2=a^2\E(\widehat S_t-S_t)^2$.
Clipping the count estimate to $[0,n]$ can only decrease its error and
uses the same state space.

\begin{theorem}[Optimal mean-square memory]\label{thm:main}
For every $c\in(0,1/4)$ there are constants $A_c,B_c>0$ and an integer $N_c$ such that, for all $n\ge N_c$, all $0<\delta\le1/2$, and all $\varepsilon\ge0$ with $E^2\le cn$,
\begin{equation}\label{eq:main}
A_c\left(\frac{n}{1+E^2}+\log_2n\right)
\le b_*(n,\delta,\varepsilon)
\le B_c\left(\frac{n}{1+E^2}+\log_2n\right).
\end{equation}
The constants and $N_c$ depend only on $c$.
\end{theorem}

The theorem makes precise the gap between describing a window and
maintaining its count~\cite{datar2002,bb}. At $E^2=cn$, a
single-window encoder needs only $O_c(1)$ labels
(Proposition~\ref{prop:offline}), while maintenance costs order $\log n$
bits. The restriction $c<1/4$ marks
the threshold below which a single-state forecast no longer suffices.

\begin{proposition}[The single-state threshold]\label{prop:zero}
One has $b_*(n,\delta,\varepsilon)=0$ if and only if $E^2\ge n/4$.
\end{proposition}

For the $n/(1+E^2)$ term, we read the state through its future updates. Given the state and the next window, the successive estimates recover the count of each block that is about to expire, with mean-square error at most $4E^2$. Conditional on the next window, the old coordinates remain independent with variance $v_\delta=(1-\delta^2)/4\ge3/16$, so the count of a block of $L$ of them keeps uncertainty of order $\sqrt L$. Once $L$ is a large multiple of $1+E^2$, each such block forces at least one bit of state, which gives the term $n/(1+E^2)$ (Proposition~\ref{prop:entropy}). The $\log n$ term
already arises at the first query, by the streaming variance theorem of
Braverman, Garg and Woodruff~\cite{bgw}
(Proposition~\ref{prop:log-lower}).

A fixed-capacity queue of centered block increments and a bounded
residual counter attain the upper bound. The residual stays exact during
overflow, so the estimator recovers as the queue refills; concentration
controls the loss at each query time. Only the independent fair
length-$n$ window marginals enter this construction
(Corollary~\ref{cor:window-marginal-upper}). The lower bound, by contrast, depends
on how successive windows are related. Periodically repeating an
independent fair block $U_1,\ldots,U_n$, with
$X_t=U_{1+((t-1)\bmod n)}$, gives a stationary stream with the same window
marginals and a constant window sum. Counting the first block and then
freezing is exact with $n+1$ states. Appendix~\ref{sec:appendix-window}
gives the proofs, the queue construction and the
constants in~\eqref{eq:main-constants}.

Mean-square loss permits large errors on rare histories. Requiring
accuracy on every history changes the dependence on $E$: let
$b_\infty(n,E)$ be the least memory of a recursion satisfying
\begin{equation}\label{eq:uniform-criterion}
|\widehat S_t-S_t|\le E
\quad\text{for every }t\ge n\text{ and every binary history}.
\end{equation}
All finite binary histories have positive probability under \eqref{eq:source}. Thus this criterion is also the uniform log probability criterion with tolerance $aE$.

\begin{proposition}[Uniform approximation]\label{prop:uniform}
If $0\le E<n/2$, let $L=\lfloor2E\rfloor+1$ and $B=\lfloor E\rfloor+1$. Then
\begin{equation}\label{eq:uniform-bounds}
\left\lfloor\frac nL\right\rfloor
\le b_\infty(n,E)
\le\min\left\{n,
\left\lceil\frac nB\right\rceil+\lceil\log_2(2B)\rceil\right\}.
\end{equation}
In particular, $b_\infty(n,E)=n$ for $E<1/2$.
\end{proposition}

The block lower bound and carry construction in Proposition~\ref{prop:uniform} adapt the arguments of Ben Basat et al.~\cite{bb} to our setting. For each fixed $c\in(0,1/4)$, in the common range $E^2\le cn$ and for sufficiently large $n$, they imply
\begin{align}\label{eq:uniform-mse-comparison}
b_\infty(n,E)&=\Theta_c\!\left(\frac{n}{1+E}\right),\notag\\
b_{\mathrm{count}}(n,\delta,E)
&=\Theta_c\!\left(\frac{n}{1+E^2}+\log_2n\right).
\end{align}
The counter cost $O(\log(1+E))$ is absorbed by $n/(1+E)$, and the
mean-square bound is uniform in $\delta$. At $E^2=cn$, uniform accuracy
therefore costs order $\sqrt n$ bits, compared with order $\log n$
under mean-square loss.

\subsection{A numerical example}\label{sec:window-example}

Take $n=512$ and $\delta=0.02$. A log probability RMS tolerance
$\varepsilon=0.1$ corresponds to count tolerance
$E=0.1/\log(51/49)\approx2.4997$. The completed-block estimator of
Proposition~\ref{prop:simple-mse} retains the latest $M=6$ completed
counts for blocks of length $B=99$, together with the current partial
count, and reads the block boundaries from the free clock. This uses
$K\le99\cdot100^6$ states, or at most $47$ bits. The queue used for
Theorem~\ref{thm:main} has a capacity bound of $1143$ bits at these
parameters, more than the $512$ bits of an exact register.
Estimating the oldest partial block from its complete count gives exact
phase-dependent count MSE $u(99-u)/(4\cdot99)$, and hence log probability
RMS error at most $0.0995121$ at every deterministic query time.

Requiring absolute log probability error at most $0.1$ on every history
instead needs at least $102$ bits; the uniform carry construction uses
at most $174$. Table~\ref{tab:example} lists these bounds
together with the error of the constant forecast.

\begin{figure}[t]
\centering
\includegraphics[width=\textwidth]{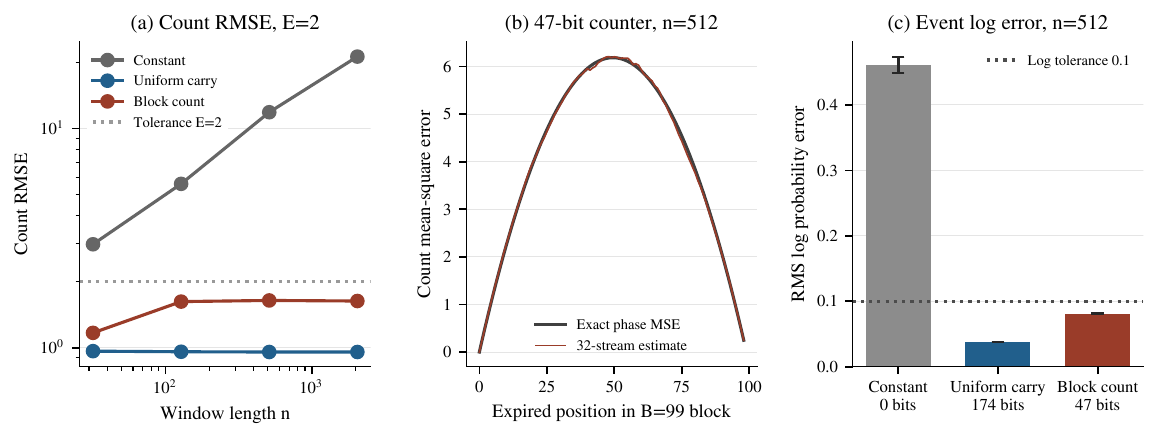}
\caption{Simulated window errors. (a)~Count RMS error of the constant estimate, uniform carry and completed-block counts against window length $n$, at $\delta=0.02$ and tolerance $E=2$ (dotted). (b)~Mean-square count error of the $47$-bit completed-block estimator ($n=512$, $B=99$) against the phase $u=(t-n)\bmod99$, with the exact curve $u(99-u)/(4\cdot99)$. (c)~Log probability RMS error of the methods in Table~\ref{tab:example} over $32$ streams, with the tolerance $0.1$ dotted. Appendix~\ref{app:window-finite-example} gives the design and the error bars.}
\label{fig:errors}
\end{figure}

Figure~\ref{fig:errors} compares empirical phase-binned mean-square
errors from independent realizations with the exact curve. Appendix~\ref{app:window-finite-example} gives the
implementation, experimental design, and numerical values.

\begin{table}[t]
\caption{Memory bounds for the window example with $n=512$, $\delta=0.02$ and count tolerance $E\approx2.4997$, which corresponds to log probability tolerance $\varepsilon=0.1$.}
\label{tab:example}
\centering
\small
\begin{tabular}{@{}>{\raggedright\arraybackslash}p{0.25\textwidth}r>{\raggedright\arraybackslash}p{0.25\textwidth}>{\raggedright\arraybackslash}p{0.25\textwidth}@{}}
\toprule
Method or bound & Bits & Count error & Log probability error\\
\midrule
Constant estimate & $0$ & RMS $\sqrt{128}$ & RMS $0.45260869\ldots$\\
Completed blocks, $B=99$ & $\le47$ & RMS $\le\sqrt{99}/4$ & RMS $\le0.0995121$\\
Uniform lower bound & $\ge102$ & Necessary for absolute error $\le E$ & Equivalent to absolute error $\le0.1$\\
Uniform carry, $B=3$ & $\le174$ & Absolute error $\le2$ & Absolute error $\le0.0800107$\\
\bottomrule
\end{tabular}
\end{table}

\subsection{Prediction with only a few states}\label{sec:critical-memory}

Near the single-state threshold, the relevant question is how much a
bounded state can gain over the constant forecast. Measure the gain by
$g=n/4-E^2$, and for an integer $b\ge0$ let
\begin{equation}\label{eq:critical-gain}
\Gamma_{n,\delta}(b)=
\sup_{\mathcal A:\,|\mathcal Z|\le2^b}
\inf_{t\ge n}
\left\{\frac n4-\E(S_t-\widehat S_t)^2\right\}
\end{equation}
be the largest gain available at every query time from a recursion
\eqref{eq:recursion} with at most $2^b$ states. Keeping the last
observation gains $1/4$. One bit can nearly double this: start at
$Z_0=0$ and alternate OR at odd times with AND at even times. This
program gains within $O(2^{-n})$ of $1/2$, while no one-bit recursion
gains $1/2$, so $1/2-\Gamma_{n,\delta}(1)=\Theta(2^{-n})$ uniformly in
$\delta$ (Theorem~\ref{thm:critical-one-bit}). One bit therefore
suffices for any fixed gain below $1/2$ once $n$ is large, and a gain
of $1/2$ needs two bits (Corollary~\ref{cor:critical-one-bit-threshold}).
The upper bound comes from an affine recursion for the Fisher
information of a two-state program and the conditional-score identity
\cite[Lemma~1]{bho20}.

A few more states let the gain grow with $n$. A six-state recursion
that runs the read-once Tribes computation \cite{bv10,bgz22} on one
block, and keeps its output while the next block is read, gains
$\gamma_w\sim(\log_2n)^2/(4(e-1))$ at every query time
(Theorem~\ref{thm:critical-tribes}). Read-once AND--OR trees of fixed
depth $2h$ \cite[full version, Section~5 and Appendix~C]{bv10} gain
$c_h(\log n)^{2h}$ with $\lceil\log_2(4h+2)\rceil$ bits
(Proposition~\ref{prop:critical-hierarchy}). Memory that does not grow
with $n$ thus buys any fixed power of $\log n$, whereas a fixed fraction of the variance $n/4$ needs
the growing memory of Theorem~\ref{thm:main}. For the Tribes gain the
optimal memory stays between two and three bits
(Corollary~\ref{cor:critical-nonuniformity}), so the logarithmic lower
bound in Theorem~\ref{thm:main} cannot hold uniformly as $c\uparrow1/4$.

These constructions use the free clock to locate block boundaries.
Storing the phase instead costs at most $\lceil\log_2m\rceil$ bits for
blocks of length $m$; in Theorem~\ref{thm:main} the corresponding
$\lceil\log_2B\rceil$ bits are absorbed by the logarithmic term. On
independent fair bits, a time-homogeneous one-bit update gains exactly
$1/4$, even with time-dependent decoding
(Proposition~\ref{prop:homogeneous-one-bit}).
Appendix~\ref{sec:appendix-small-memory} states and proves these
results; related one-bit computations and alternating Boolean functions
appear in \cite{vf91,khs13}.

\section{Discussion}\label{sec:scope}

In both settings, the loss decides which distinctions between histories the state must preserve, and the permitted state evolution decides what preserving them costs. We first relate the loss budget to perplexity and turn the generation bounds into precision requirements. Section~\ref{sec:sequence-models} then applies the generation converse to autoregressive sequence models, and Section~\ref{sec:open-questions} collects open questions.

\subsection{The role of the loss budget}

For generation, a fixed total error and a fixed average error lead to different precision requirements. Perplexity exponentiates average log loss: for a length-$T$ source law $P$ and model law $Q$, expected average log loss is $\{H(P)+D_{\mathrm{KL}}(P\Vert Q)\}/T$. A per-token excess-loss allowance $a$ therefore supplies a total KL budget $aT$. The converse becomes $a\ge2\mathcal D_K(\mu)$, and choosing $\xi=\log(1+\sqrt a)$ in the HMM construction gives a state count that does not depend on $T$. For the nondegenerate stationary sources of Theorem~\ref{thm:gen-binary-classification}, this gives
\begin{equation}\label{eq:per-token-bit-law}
b_{\mathrm{KL}}(T,aT)=\Theta\!\left(\log\frac1a\right)
\quad\text{as }a\downarrow0,
\end{equation}
with constants uniform over $T\ge1$. Each new symbol now brings its own allowance for approximation error, so the required memory can remain independent of the continuation length. Holding the total allowance fixed instead forces the predictor to become more accurate as the horizon grows.

\subsection{Precision requirements}

The generation bounds translate into precision requirements. For the $m$-hidden-state construction in Theorem~\ref{thm:gen-kl-upper}, $\tfrac12\log_2(T/\varepsilon)+O_P(1)$ bits per belief log-odds coordinate suffice for $0<\varepsilon/T\le1$. For the nondegenerate stationary two-state sources of Theorem~\ref{thm:gen-binary-classification}, the converse requires at least $\tfrac{d_*}{2}\log_2(T/\varepsilon)-O(1)$ bits for the whole state. The two coefficients measure different things. The factor $1/2$ is a precision per coordinate, and the factor $d_*/2$ comes from the nonconcentration of the stationary predictor.

For the two-regime model~\eqref{eq:gen-explicit-parameters}, the explicit constants give a lower coefficient of about $0.065$ for $\log_2(T/\varepsilon)$ under both criteria. The constructions give upper coefficients $1/2$ for total KL and $1$ for uniform word error.

\subsection{Autoregressive sequence models}\label{sec:sequence-models}

Fix a prompt $x$ and an autoregressive teacher with continuation law $\mathsf P_x$ on $\mathcal A^T$, for a finite vocabulary $\mathcal A$. The teacher need not be stationary or come from a hidden Markov model. A student generates by sampling from $q_{j,Z_j}$ and applying $Z_{j+1}=F_j(Z_j,Y_{j+1})$, starting from a state determined by $x$, and its maps may depend on $x$. Let $\mathsf Q_x$ be its continuation law, let $Z_j$ take at most $K_j$ values, and let $\mu_{j,x}$ be the law under $\mathsf P_x$ of the teacher's next-token distribution $\mathsf P_x(Y_{j+1}\in\cdot\mid Y_{1:j})$. Part (i) of Theorem~\ref{thm:gen-quantization} with $t=0$ gives
\begin{equation}\label{eq:prompt-bound}
D_{\mathrm{KL}}(\mathsf P_x\Vert\mathsf Q_x)
\ge2\sum_{j=0}^{T-1}\mathcal D_{K_j}(\mu_{j,x}).
\end{equation}
With fixed weights and deterministic updates, a student with $b$ bits of persistent state has $K_j\le2^b$, whatever its architecture or training, and a cache that grows with the context enters through $K_j$. The bits must cover everything that persists from one token to the next, including all recurrent layers, history-dependent buffers and scales, and attention caches. The student may sample its output tokens, because the bound only needs the state to be determined by the prompt and the emitted tokens. State-space students distilled from Transformers~\cite{bick2024} and linearized Transformers~\cite{zhang2025lolcats} satisfy this condition. A student whose state keeps randomness beyond its outputs falls outside the bound, and Corollary~\ref{cor:gen-private-memory} shows that such randomness can remove the horizon cost.

The chain rule~\eqref{eq:gen-kl-chain} evaluates the student's next-token law on prefixes drawn from the teacher, with the student's state computed by its own recursion along each prefix. Summed over the continuation, these one-step terms give the sequence-level forward KL exactly, so teacher forcing on teacher rollouts measures the forward KL of generation. Averaged over corpus prefixes instead, or computed with the teacher's cache supplied to the student, they no longer add up to this KL.

Across prompts, the bound depends on what the student can read. If a random prompt $X$ reaches the student only through an encoder $Z_0=g(X)$, the same argument applies to the source $(X,Y_1,\ldots,Y_T)$ with one codebook shared across prompts, and the information term in the split after Theorem~\ref{thm:gen-quantization} becomes $I(Y_{j+1};X,Y_{1:j}\mid Z_j)$, which includes what the encoder discards about $X$. If the decoder can reread the prompt, its codebook may depend on $X$, and the distortions must be conditioned on $X$ as in~\eqref{eq:prompt-bound}. Pooling them across prompts is then invalid. If $X$ is a fair bit and the teacher emits $Y_1=X$, a student that rereads $X$ is exact with a single internal state, while the pooled predictor law, uniform on two point masses, has $\mathcal D_1=1/4$ and would suggest a KL of at least $1/2$.

A memory law follows once the distortions are bounded below. Suppose that
\[
\frac1T\sum_{j=0}^{T-1}\mathcal D_K(\mu_{j,x})\ge cK^{-2/d}
\qquad\text{for all }K\le K_{\max}.
\]
Then a student with $b$ bits, $2^b\le K_{\max}$, and total KL at most $\varepsilon$ needs $b\ge\frac d2\log_2(2cT/\varepsilon)$, and $b\ge\frac d2\log_2(2c/a)$ under a per-token allowance $a$. The exponent $d$ is a quantization dimension of the teacher's next-token laws, as in Section~\ref{sec:predictor-geometry}, and so a property of the teacher's predictions rather than of its architecture. A teacher that itself runs in finite precision with bounded memory has finitely many next-token laws at every step. Then $K_{\max}$ is at most their number, and the scaling below it is an empirical question. Heuristically, $d$ should be large when the teacher's next-token laws vary along many directions of its state. For a randomly coupled linear system near instability, the number of principal directions needed to hold a fixed fraction of the stationary variance grows without bound while remaining a vanishing fraction of the dimension~\cite{feng2026covariance}.

\subsection{Open questions}\label{sec:open-questions}

The converse allows a new codebook at every time and ignores continuation consistency. A static quantizer of the current predictor may discard distinctions that later updates need, and the window results show that keeping a statistic current can cost far more than describing it. Matching the converse with a recursive machine therefore requires control of both the predictor geometry and the state evolution. For autoregressive teachers, the corresponding question is which teachers admit recursive students whose memory matches~\eqref{eq:prompt-bound}.

Beyond two hidden states, the recursive upper bounds still hold. The lower bounds are harder, because the emission map can collapse directions in the belief. Extending the matching classification requires quantitative nonconcentration of the observable predictor, which is the image of the belief under this map. For conformal systems with the strong open set condition, quantization theory for Gibbs measures~\cite{lindsay2002} gives finer distortion bounds; whether they can be attained recursively is open.

Randomized observation updates raise a further question. They help in finite-memory testing~\cite{hellman1971}; whether they change the memory laws here, when the loss is evaluated on the forecast law averaged over the observer's randomness, remains open. Private simulation already shows that randomness can replace some of the information a deterministic recursion must retain, and randomized observation would test how far this goes.

\section{Conclusion}\label{sec:conclusion}

A fixed-size state has to answer questions about the future and also remain a good starting point for later updates, and this paper measured what each requirement costs. For a known hidden Markov source with positive transitions, answering takes $O(\log(1/\varepsilon))$ bits at every horizon, because the filter forgets its rounding errors. Generating the future with a state determined by the emitted symbols costs more. Every step pays for approximating the next-symbol law with one of finitely many states, and for nondegenerate two-state sources the memory needed is of order $\log(T/\varepsilon)$ bits. Private randomness or a per-token error budget removes the dependence on $T$. For the moving-window counts of Section~\ref{sec:model}, the count takes $\log n$ bits to write down, and keeping it current to a mean-square error $E^2$ well below its variance takes order $n/(1+E^2)+\log n$ bits. For recurrent and state-space language models, the results on generation mean that a state sized for a fixed per-token loss can be too small to keep the total error of a long continuation bounded. Section~\ref{sec:sequence-models} expresses the required size through the teacher's next-token laws, and Section~\ref{sec:open-questions} asks when a recursive student can reach it.

\clearpage
\begin{appendix}
\section{Proofs for positive hidden Markov models}\label{app:hmm-proofs}

\subsection{Event queries and finite recursive filtering}\label{app:hmm-queries}

We use the model, filter and Hilbert metric of Section~\ref{sec:hmm}.
The contraction bound \eqref{eq:filter-contraction} applies to defined
updates, including when some emission probabilities vanish. On the common support $I_y=\{i:g_i(y)>0\}$, multiplication by $D_y$ preserves coordinate ratios and restriction can only reduce their oscillation. The positive rectangular matrix $P_{I_y,:}$ has projective diameter
at most $\Delta(P)$ and hence contraction coefficient at most
$\kappa$.

\begin{lemma}[Likelihood ratios of mixtures]\label{lem:mixture}
For positive probability vectors $u,w$ and any nonzero vector $v\ge0$,
\begin{equation}\label{eq:mixture-ratio}
\min_i\frac{u_i}{w_i}\le\frac{uv}{wv}\le\max_i\frac{u_i}{w_i},
\qquad
\left|\log\frac{uv}{wv}\right|\le d_H(u,w).
\end{equation}
In particular, the future laws arising from $u$ and $w$ have the same null events.
\end{lemma}

\begin{proof}[Proof of Lemma~\ref{lem:mixture}]
The ratio $uv/wv$ is a convex combination of the numbers $u_i/w_i$, with weights $w_iv_i/(wv)$. Since $u$ and $w$ are normalized, the smallest coordinate ratio is at most one and the largest is at least one. Taking logarithms proves the second inequality. For an event $B$, its probability vanishes precisely when $v_B$ is the zero vector.
\end{proof}

\begin{proof}[Proof of Theorem~\ref{thm:hmm-memory}]
Every $u\in\mathcal C$ has $u_i\in[p_{\min},p_{\max}]$. Its log-odds coordinates $\log(u_i/u_m)$, $1\le i<m$, lie in $[-L,L]^{m-1}$. Put $M=\lceil2L/\eta\rceil\ge1$ and choose the $M+1$ equally
spaced grid points $-L+2Lj/M$, $0\le j\le M$, in each coordinate.
Their spacing is at most $\eta$. Rounding each coordinate to the nearest grid point and normalizing back to a probability vector defines a quantizer $Q$ with
\[
d_H(Q(u),u)\le\eta,\qquad u\in\mathcal C.
\]
Each log-odds error is at most $\eta/2$. Normalization adds a common
constant to the log coordinate ratios, whose oscillation is therefore
at most $\eta$. The $1+\lceil2L/\eta\rceil$ grid points in each
coordinate give the codebook term in \eqref{eq:hmm-state-count}; the
remaining state will store the initial belief.

Use one additional state for the exact initial belief $u_0$. From any state $z$, upon observing a possible symbol $y$, update to $Q(\Psi_y(z))$. Assign any state to an impossible observation. For every initial
belief $u_0$, including a boundary point of the simplex, the first
defined update belongs to $\mathcal C$. The first rounding step
therefore has Hilbert error at most $\eta$. Subsequent steps satisfy,
by \eqref{eq:filter-contraction},
\[
e_{t+1}\le\kappa e_t+\eta,
\qquad e_t=d_H(Z_t,u_t).
\]
Iteration yields $e_t\le\eta/(1-\kappa)=\varepsilon$ for every
$t\ge1$, and the initial belief is exact. Applying
Lemma~\ref{lem:mixture} to the decoder \eqref{eq:hmm-decoder} proves
\eqref{eq:hmm-risk-bound} simultaneously for all event queries.

When $L=0$, every entry of $P$ equals $1/m$. Every update therefore gives the uniform predictive belief. Keeping this belief and, if needed, the initial belief gives the final assertion.
\end{proof}

\subsection{Saturation with the forecast horizon}\label{app:hmm-saturation}

\begin{proof}[Proof of Theorem~\ref{thm:saturation}]
For a possible prefix $r$ of length $T$, let $u_r,w_r$ be the predictive beliefs after observing it. Repeated contraction gives
\[
d_H(u_r,w_r)\le\kappa^Td_H(u,w).
\]
For an event $B$ in the entire future sigma-field, let $B_r$ be its
section after the prefix $r$. Applying Lemma~\ref{lem:mixture} to
the section, or noting that both sides vanish when it has probability
zero, gives
\[
e^{-\kappa^Td_H(u,w)}\mathsf P_{w_r}(B_r)
\le\mathsf P_{u_r}(B_r)
\le e^{\kappa^Td_H(u,w)}\mathsf P_{w_r}(B_r).
\]
The ratio
of prefix probabilities lies between $e^{-d_T(u,w)}$ and
$e^{d_T(u,w)}$. Multiplication followed by summation over the finitely
many prefixes gives
\begin{align*}
e^{-d_T(u,w)-\kappa^Td_H(u,w)}\mathsf P_w(B)
&\le\mathsf P_u(B)\\
&\le e^{d_T(u,w)+\kappa^Td_H(u,w)}\mathsf P_w(B).
\end{align*}
Taking the supremum over $B$ proves the upper bound in \eqref{eq:saturation}; inclusion of sigma-fields proves the lower bound. Finally, $d_T$ increases with $T$ and is bounded by $d_H(u,w)$. Letting $T$ tend to infinity proves the remaining assertions.
\end{proof}

\begin{corollary}[A fixed-word tail bound]\label{cor:hmm-fixed-word-tail}
Let $u,w$ be positive predictive beliefs, and let $a_1a_2\cdots$ be
an infinite observation word whose every finite prefix has positive
probability. Put
\[
\ell_T=\log\frac{\mathsf P_u(a_{1:T})}{\mathsf P_w(a_{1:T})}.
\]
Then $\ell_\infty=\lim_{T\to\infty}\ell_T$ exists, and
\begin{equation}\label{eq:hmm-fixed-word-tail}
|\ell_\infty-\ell_T|\le\kappa^Td_H(u,w),\qquad T\ge1.
\end{equation}
\end{corollary}

\begin{proof}[Proof of Corollary~\ref{cor:hmm-fixed-word-tail}]
For $N>T$, factor each length-$N$ probability into the probability of
its length-$T$ prefix and the conditional probability of its remaining
word. After the common prefix, the two predictive beliefs have Hilbert
distance at most $\kappa^Td_H(u,w)$. Lemma~\ref{lem:mixture}, applied
to the remaining word, therefore gives
$|\ell_N-\ell_T|\le\kappa^Td_H(u,w)$. This uniform bound makes
$(\ell_T)$ Cauchy; letting $N\to\infty$ proves the result.
\end{proof}

\subsection{The quantization converse and small-ball bounds}\label{app:hmm-converse}

The proof uses the generator and quantization distortion defined in
Section~\ref{sec:generation}, first for finite histories and then
for the stationary predictor.

\begin{proof}[Proof of Theorem~\ref{thm:gen-quantization}]
All relative entropies in this proof use natural logarithms. For
part (i), fix $t\ge0$ and $T\ge1$. The state is determined by the
observed symbols, so the relative-entropy chain rule gives
\begin{align}
&\E D_{\mathrm{KL}}\!\left(
\mathcal L(Y_{t+1:t+T}\mid Y_{1:t})
\,\middle\Vert\,\mathsf Q_{t,Z_t}
\right)\notag\\
&\hspace{2cm}=
\sum_{s=0}^{T-1}\E D_{\mathrm{KL}}
\left(\Pi_{t+s}\Vert q_{t+s,Z_{t+s}}\right).
\label{eq:gen-kl-chain}
\end{align}
On the right, the state follows \eqref{eq:gen-state-update} along the
actual source sequence. Pinsker's inequality yields
\[
D_{\mathrm{KL}}(v\Vert c)\ge2d_{\mathrm{TV}}(v,c)^2.
\]
At each fixed time $j$, the random vector $q_{j,Z_j}$ takes at most
$K_j$ values, allowing arbitrary clock dependence of the output map, so
$\E d_{\mathrm{TV}}(\Pi_j,q_{j,Z_j})^2\ge\mathcal D_{K_j}(\mu_j)$.
Summing over $j=t,\ldots,t+T-1$ proves \eqref{eq:gen-finite-history}.

For part (ii), apply part (i) with $K_j=K$. Under
\eqref{eq:gen-kl-criterion}, the left side of
\eqref{eq:gen-finite-history} is at most $\varepsilon$ for every start
time $t$, so
\begin{equation}\label{eq:gen-finite-past-bound}
\varepsilon\ge
2\sum_{s=0}^{T-1}\mathcal D_K(\mu_{t+s}).
\end{equation}

To pass to the stationary predictor, work on the two-sided source
probability space and put
\[
\widetilde\Pi_j(y)
=\Pp(Y_1=y\mid Y_{1-j:0}).
\]
For $j=0$ the conditioning sigma-field is trivial. Stationarity gives
$\mathcal L(\widetilde\Pi_j)=\mu_j$, and the martingale convergence
theorem, applied coordinatewise, gives
\[
\E d_{\mathrm{TV}}(\widetilde\Pi_j,V)\longrightarrow0.
\]
For any set of centers $\boldsymbol c=(c_1,\ldots,c_K)$, the function
\[
f_{\boldsymbol c}(v)
=\min_k d_{\mathrm{TV}}(v,c_k)^2
\]
is $2$-Lipschitz in total variation, uniformly over the centers. Indeed,
each distance belongs to $[0,1]$, and the triangle inequality bounds
the change of that distance by $d_{\mathrm{TV}}(v,v')$. Consequently,
\[
|\mathcal D_K(\mu_j)-\mathcal D_K(\mu)|
\le2\E d_{\mathrm{TV}}(\widetilde\Pi_j,V)
\longrightarrow0.
\]
Letting $t\to\infty$ in \eqref{eq:gen-finite-past-bound}, with $T$ and
the machine fixed, proves \eqref{eq:gen-quantization-converse}.
Under \eqref{eq:gen-criterion}, the conditional log likelihood ratio
$\log(P/Q)$ is at most $\varepsilon$ on every possible word.
Taking expectation under $P$ and then over histories gives
\eqref{eq:gen-kl-criterion}, proving the conclusion for uniform word
accuracy.
\end{proof}

For a binary alphabet, write $R=\Pp(Y_1=1\mid Y_{-\infty:0})$.
Then \eqref{eq:gen-quantization} reduces exactly to
\begin{equation}\label{eq:gen-scalar-quantization}
\mathcal D_K(\mathcal L(R))
=\inf_{c_1,\ldots,c_K\in[0,1]}
\E\min_j(R-c_j)^2.
\end{equation}
Theorem~\ref{thm:gen-quantization} also holds when an arbitrary
deterministic $K$-valued encoder supplies the initial forecast state,
and recursive updates are imposed during the forecast itself. At
each subsequent time the output distribution takes at most $K$
values, which is the property used in the proof.

\begin{corollary}[Small-ball bounds]\label{cor:gen-small-ball}
In the setting of part (ii) of Theorem~\ref{thm:gen-quantization}, suppose that, for
some $C\ge1$ and $d>0$,
\begin{equation}\label{eq:gen-small-ball}
\mu\{v:d_{\mathrm{TV}}(v,c)\le r\}\le Cr^d
\end{equation}
for every $c\in\Delta_{\mathcal A}$ and every $0<r\le1$. Then
\begin{equation}\label{eq:gen-small-ball-lower}
\mathcal D_K(\mu)\ge\frac12(2CK)^{-2/d},
\qquad
K_{\mathrm{gen}}(T,\varepsilon)
\ge K_{\mathrm{KL}}(T,\varepsilon)
\ge\frac1{2C}\left(\frac{T}{\varepsilon}\right)^{d/2}.
\end{equation}
If $\mu$ has infinite support, then for every fixed
$\varepsilon>0$,
\begin{equation}\label{eq:gen-divergence}
K_{\mathrm{KL}}(T,\varepsilon)\longrightarrow\infty,
\qquad K_{\mathrm{gen}}(T,\varepsilon)\longrightarrow\infty,
\qquad T\to\infty,
\end{equation}
without any small-ball assumption.
\end{corollary}

\begin{proof}[Proof of Corollary~\ref{cor:gen-small-ball}]
For arbitrary $K$ centers, let $r_K=(2CK)^{-1/d}\le1$.
By \eqref{eq:gen-small-ball}, the union of the $K$ closed balls of
radius $r_K$ has mass at most $1/2$. The distortion is therefore at
least $r_K^2/2$. Combining this with
Theorem~\ref{thm:gen-quantization} gives
\eqref{eq:gen-small-ball-lower}.

The center space $(\Delta_{\mathcal A})^K$ is compact and the
distortion is continuous in the centers, so the infimum in
\eqref{eq:gen-quantization} is attained. Its value is zero precisely
when $\mu$ is supported on at most $K$ points. Infinite support thus
gives $\mathcal D_K(\mu)>0$ for every finite $K$. For each such $K$,
\eqref{eq:gen-quantization-converse} fails once
$T>\varepsilon/(2\mathcal D_K(\mu))$, proving
\eqref{eq:gen-divergence}.
\end{proof}

\subsection{Generation upper bounds and initialization}\label{app:hmm-generation}

For the hidden Markov model of Section~\ref{sec:hmm}, assume that the
transition matrix $P$ is strictly positive and allow zero emission
probabilities. The initial predictive belief $u_0$ is fixed and known. Let $\kappa<1$ be the contraction coefficient in
\eqref{eq:birkhoff}, and put
\[
L_P=\log(p_{\max}/p_{\min}).
\]

\begin{theorem}[Quantized-filter generation]\label{thm:gen-hmm-upper}
For every $T\ge1$ and $\varepsilon>0$, there is a time-homogeneous
deterministic-state generator satisfying
\eqref{eq:gen-criterion}. If $L_P>0$, its state count obeys
\begin{equation}\label{eq:gen-hmm-state-upper}
K_{\mathrm{gen}}(T,\varepsilon)
\le1+\left(1+
\left\lceil\frac{2L_PT}{(1-\kappa)\varepsilon}\right\rceil
\right)^{m-1}.
\end{equation}
If $L_P=0$, two states suffice for exact generation, and one suffices
when $u_0$ is uniform. In particular, for a fixed model,
\begin{equation}\label{eq:gen-hmm-bit-upper}
b_{\mathrm{gen}}(T,\varepsilon)
\le(m-1)\log_2(1+T/\varepsilon)+A_P,
\end{equation}
where $A_P$ is independent of $T$, $\varepsilon$, and $u_0$.
\end{theorem}

\begin{proof}[Proof of Theorem~\ref{thm:gen-hmm-upper}]
Use the log-odds quantizer from the proof of
Theorem~\ref{thm:hmm-memory}, now with rounding radius
\[
\eta=(1-\kappa)\varepsilon/T.
\]
Keep the exact initial belief as an additional state. For a state $z$,
use output probabilities
\[
q_z(y)=zD_y\one
\]
and, after any possible symbol, update to $Q(\Psi_y(z))$.
The time-zero output distribution is exact. After any nonempty
positive-probability history, the true and retained predictive
beliefs are positive and satisfy
\begin{equation}\label{eq:gen-filter-error}
d_H(u_t,Z_t)\le\frac{\eta}{1-\kappa}
=\frac{\varepsilon}{T}.
\end{equation}
Lemma~\ref{lem:mixture}, applied to $D_y\one$, bounds each possible
next-symbol log likelihood error by $\varepsilon/T$.

Observation and generation use the same recursion, so the bound
holds after every possible extension of each observed history. Summing the conditional log likelihood
errors over a word of length at most $T$ proves
\eqref{eq:gen-criterion} on all positive-probability words.
The supports agree as well. The initial distributions coincide.
After any possible first observation, both predictive beliefs are
positive; a symbol then has zero probability precisely when every
hidden state assigns it zero emission probability. Induction gives
equal supports for all future words.

The codebook count is the one in
\eqref{eq:hmm-state-count} with the new value of $\eta$, proving
\eqref{eq:gen-hmm-state-upper}. If $L_P=0$, every row of $P$ is
uniform and the two-state construction in
Theorem~\ref{thm:hmm-memory} retains the exact predictive belief.
Taking logarithms and absorbing fixed constants gives
\eqref{eq:gen-hmm-bit-upper}.
\end{proof}

\begin{proof}[Proof of Theorem~\ref{thm:gen-kl-upper}]
Assume first that $L_P>0$. Use the same log-odds grid and exact initial
state as in Theorem~\ref{thm:gen-hmm-upper}, with rounding radius
$\eta=(1-\kappa)\xi$. Emit from $q_z(y)=zD_y\one$ and update to
$Q(\Psi_y(z))$. The error recursion gives
$d_H(u_t,Z_t)\le\xi$ after every nonempty positive-probability history;
the initial belief is exact. The true next-symbol distribution $p$ and
the machine distribution $q$ have equal supports. On their common
support, Lemma~\ref{lem:mixture} gives
$e^{-\xi}\le p(y)/q(y)\le e^\xi$.

The inequality $\log a\le a-1$ implies
\begin{align}
D_{\mathrm{KL}}(p\Vert q)
&\le\sum_{y:q(y)>0}\frac{(p(y)-q(y))^2}{q(y)}\notag\\
&\le(e^\xi-1)^2=\varepsilon/T.
\label{eq:gen-kl-local-upper}
\end{align}
The first inequality follows by summing
$p(y)\log(p(y)/q(y))\le p(y)(p(y)/q(y)-1)$ and using
$\sum_y p(y)=\sum_y q(y)=1$. The second uses the uniform ratio bound.
Common zero coordinates contribute zero, so the bound includes
models with zero emission probabilities.

Fix any observed history and start time. The same one-step bound holds
after every possible extension of that history. The conditional KL
chain rule sums at most $T$ such terms, proving
\eqref{eq:gen-kl-conditional-upper}. The grid has
$1+\lceil2L_P/((1-\kappa)\xi)\rceil$ points in each of its $m-1$
coordinates, plus the exact initial label. This proves
\eqref{eq:gen-kl-state-upper}. If $L_P=0$, the exact construction from
Theorem~\ref{thm:hmm-memory} applies.

For $0<\varepsilon/T\le1$, put $a=\sqrt{\varepsilon/T}\le1$.
Since $\log(1+a)\ge a/2$, the grid count is at most a
model-dependent constant times $(T/\varepsilon)^{(m-1)/2}$.
Taking the base-two logarithm and the ceiling proves
\eqref{eq:gen-kl-bit-upper}.
\end{proof}

For a single initial forecast, use the criterion
\eqref{eq:gen-initial-criterion} and the state count
$K_{\mathrm{init}}(T,\varepsilon)$ defined in
Section~\ref{sec:generation}.

\begin{proposition}[A single initial forecast]
\label{prop:gen-initial-forecast}
Suppose the positive hidden Markov model is initialized in its
stationary hidden-state distribution. Let $\mu$ be the law of its
stationary infinite-past output predictor, and put
\begin{equation}\label{eq:gen-initial-constant}
C_P=\frac{2L_P}{1-\kappa}.
\end{equation}
Every $K$-state deterministic-state generator with $D_{\mathrm{KL}}(\mathcal L(Y_{1:T})\Vert\mathsf Q_{0,z_\star})\le\varepsilon$, and in particular every generator satisfying \eqref{eq:gen-initial-criterion}, obeys
\begin{equation}\label{eq:gen-initial-converse}
\varepsilon+C_P\ge2T\mathcal D_K(\mu).
\end{equation}
\end{proposition}

\begin{proof}[Proof of Proposition~\ref{prop:gen-initial-forecast}]
Let $\pi$ be the stationary hidden-state distribution. On a
two-sided stationary realization, couple the finite-past filter
$u_j=\mathcal L(H_{j+1}\mid Y_{1:j})$, started from $u_0=\pi$,
with the stationary filter
$U_j=\mathcal L(H_{j+1}\mid\sigma(Y_i:i\le j))$.
Both initial beliefs belong to $\mathcal C$: for $\pi$ this
follows from $\pi=\pi P$, and for $U_0$ it follows from one
predictive transition. Their coordinates lie in
$[p_{\min},p_{\max}]$, so
\[
d_H(\pi,U_0)\le2L_P=:D.
\]
The common observation sequence and filter contraction give
\[
d_H(u_j,U_j)\le D\kappa^j,\qquad j\ge0.
\]

For positive probability vectors $u,v$, let
$\tau=d_{\mathrm{TV}}(u,v)$. Pinsker's inequality in both
directions and the definition of Hilbert distance give
\[
4\tau^2
\le D_{\mathrm{KL}}(u\Vert v)+D_{\mathrm{KL}}(v\Vert u)
=\sum_i(u_i-v_i)\log\frac{u_i}{v_i}
\le\tau\,d_H(u,v).
\]
For the last inequality, sum separately over the positive and
negative parts of $u-v$, each of which has total mass $\tau$.
Thus $d_{\mathrm{TV}}(u,v)\le d_H(u,v)/4$.
The emission kernel is nonexpansive in total variation. Thus the finite-past
output predictor $\Pi_j$ and the stationary output predictor induced
by $U_j$ couple $\mu_j$ and $\mu$ with expected total variation at
most $D\kappa^j/4$.

The uniform Lipschitz bound used in the proof of
Theorem~\ref{thm:gen-quantization} now gives, for every $K$,
\[
|\mathcal D_K(\mu_j)-\mathcal D_K(\mu)|
\le\frac D2\kappa^j.
\]
Part (i) of Theorem~\ref{thm:gen-quantization}, applied to the
initial forecast with $t=0$ and $K_j=K$, yields
\[
\varepsilon\ge2\sum_{j=0}^{T-1}\mathcal D_K(\mu_j)
\ge2T\mathcal D_K(\mu)-D\sum_{j=0}^{T-1}\kappa^j
\ge2T\mathcal D_K(\mu)-C_P.
\]
This proves \eqref{eq:gen-initial-converse}.
\end{proof}

\begin{proof}[Proof of Proposition~\ref{prop:gen-private-state}]
Let $z$ be the approximate predictive belief retained by
Theorem~\ref{thm:hmm-memory}. At the start of a forecast, draw a new
private hidden state $\widetilde H_1$ with law $z$. Generate the
first symbol according to $g_{\widetilde H_1}$, move the private
hidden state using $P$, and repeat. For every future event $B$,
the generated probability is $zv_B$. Theorem~\ref{thm:hmm-memory}
therefore gives the bounds simultaneously at every horizon.
\end{proof}

\begin{proof}[Proof of Corollary~\ref{cor:gen-private-memory}]
Let a private-state generator have $K_o$ observation labels. Its
observation state $Z_t$ is a deterministic function of $Y_{1:t}$.
For each label $z$ and time $t$, let $q_{t,z}$ be the distribution
of its first simulated symbol, averaging over all randomness used
to initialize and run the simulation. By the definition of a
private-state generator, $q_{t,z}$ depends on the observed history only
through $(t,z)$.
There are therefore at most $K_o$ such distributions at each time.

Write $\Pi_t=\mathcal L(Y_{t+1}\mid Y_{1:t})$ and
$\mu_t=\mathcal L(\Pi_t)$. Since $T\ge1$, the required
multiplicative guarantee applies to every one-symbol word. Summing
the resulting coordinate bounds gives
\[
d_{\mathrm{TV}}(\Pi_t,q_{t,Z_t})\le e^\varepsilon-1.
\]
It follows that
\[
\mathcal D_{K_o}(\mu_t)\le(e^\varepsilon-1)^2.
\]
The finite-past predictors converge to the stationary predictor as
in the proof of Theorem~\ref{thm:gen-quantization}; the same uniform
Lipschitz bound for squared distance to a codebook yields
$\mathcal D_{K_o}(\mu_t)\to\mathcal D_{K_o}(\mu)$.
For a nondegenerate model, the small-ball estimate in the proof of
Theorem~\ref{thm:gen-binary-classification} gives constants
$c_0,d_*>0$ with
\[
c_0K_o^{-2/d_*}\le\mathcal D_{K_o}(\mu)
\le(e^\varepsilon-1)^2.
\]
Using $e^\varepsilon-1\le(e-1)\varepsilon$ for
$0<\varepsilon\le1$ proves
$K_o\ge c_p\varepsilon^{-d_*}$. Since $K_{\mathrm{priv}}\ge K_o$,
the same lower bound holds for $K_{\mathrm{priv}}$.

For the upper bound, use the deterministic observer of
Theorem~\ref{thm:hmm-memory}, with at most
$C_p(1+\varepsilon^{-1})$ labels, and the initialization of
Proposition~\ref{prop:gen-private-state}. The two private hidden
labels used during simulation are disjoint from the observation
labels. Their total is at most $C_p(1+\varepsilon^{-1})+2$, and
the multiplicative guarantee holds for every horizon simultaneously.
This proves \eqref{eq:gen-private-state-bounds}, with $d_p=d_*$.
Taking logarithms proves \eqref{eq:gen-private-bit-law}, with constants
uniform in $T$.
\end{proof}

\subsection{Private states as projected outputs}\label{app:projected-output}

The private-state construction has a realization through auxiliary output symbols. From the retained belief $z$, emit an initialization label $h$ with probability $z_h$. At hidden label $h$, sample $(y,h')$ with probability $g_h(y)P_{hh'}$ and set the next retained label to $h'$. The state update is deterministic given the augmented output. Erasing the initialization label and projecting each pair onto $y$ yields the visible word probability
\[
zD_{y_1}P\cdots D_{y_T}P\one.
\]
The probability of a visible word is the sum over its augmented
paths. This gives the private hidden-state law of
Proposition~\ref{prop:gen-private-state}, using the finite-state
marginalization mechanism of \cite{nowak2024} for recurrent language
models with auxiliary symbols. The rollout retains the current hidden
label; an implementation that also stores an auxiliary transcript
must count that transcript as memory.

\subsection{Classification with two hidden states}\label{app:hmm-binary}

We prove Theorem~\ref{thm:gen-binary-classification} for \eqref{eq:gen-general-binary-parameters} by the backward-block argument outlined in Section~\ref{sec:predictor-geometry}. The main step is the power-law bound \eqref{eq:gen-general-interval-bound} for the stationary filter law, which holds even when the one-step filter images overlap. Corollary~\ref{cor:gen-small-ball} and the constructions of Appendix~\ref{app:hmm-generation} then give the state counts.

\begin{proof}[Proof of Theorem~\ref{thm:gen-binary-classification}]
If $\gamma=\alpha$, the stationary hidden chain is independent and
identically distributed, as are its conditionally independent
emissions. If $g_0=g_1$, every conditional emission law is the same
regardless of the hidden state, so the observations are again
independent and identically distributed. In either case the common
observation distribution defines an exact one-state generator.

Suppose now that $\gamma\ne\alpha$ and $g_0\ne g_1$. Put
\[
\Delta=\gamma-\alpha,\qquad
I=[\min(\alpha,\gamma),\max(\alpha,\gamma)]\subset(0,1).
\]
For $y\in\mathcal A$, the output probabilities and scalar filter maps
are
\begin{equation}\label{eq:gen-general-binary-filter}
p_y(u)=(1-u)g_0(y)+ug_1(y),\qquad
F_y(u)=\alpha+\Delta\frac{ug_1(y)}{p_y(u)}.
\end{equation}
Each $F_y$ maps $I$ into itself. Their derivatives have the same nonzero
sign and satisfy
\begin{equation}\label{eq:gen-general-binary-derivative}
F_y'(u)=\Delta\frac{g_0(y)g_1(y)}{p_y(u)^2}.
\end{equation}
Define
\begin{equation}\label{eq:gen-general-binary-lambda}
\lambda=\min\left\{\frac12,
\min_{y\in\mathcal A,\,u\in I}|F_y'(u)|\right\}>0,
\qquad \beta=\min_{i,y}g_i(y)>0.
\end{equation}
Thus $0<\lambda<1$, and a composition of $j$ filter maps is monotone
with absolute derivative at least $\lambda^j$.

Because $g_0$ and $g_1$ are different probability distributions with
positive coordinates, there are symbols $a_0,a_1$ with
$g_1(a_0)/g_0(a_0)\ne g_1(a_1)/g_0(a_1)$. Otherwise all these ratios
would be the same, and normalization would force $g_0=g_1$.
Hilbert contraction gives a unique fixed point for each $F_y$ in $I$,
and $F_y^k(I)$ converges uniformly to that point. The two selected
fixed points differ, since
\begin{equation}\label{eq:gen-general-fixed-points}
F_{a_1}(u)-F_{a_0}(u)
=\frac{\Delta u(1-u)
 [g_1(a_1)g_0(a_0)-g_1(a_0)g_0(a_1)]}
 {p_{a_1}(u)p_{a_0}(u)}\ne0,
\qquad u\in I.
\end{equation}
Consequently, for some integer $k\ge1$,
\begin{equation}\label{eq:gen-general-block-gap}
\psi_k=\operatorname{dist}(F_{a_0}^k(I),F_{a_1}^k(I))>0.
\end{equation}
Fix such a $k$, and define
\begin{equation}\label{eq:gen-general-small-ball-constants}
c_g=d_{\mathrm{TV}}(g_0,g_1),\qquad
G=c_g \psi_k,\qquad \theta=1-\beta^k,\qquad
d_* =\frac{\log\theta}{k\log\lambda}>0.
\end{equation}

Let $U_t=\Pp(H_{t+1}=1\mid\sigma(Y_i:i\le t))$ be the stationary
predictive filter, and let $\nu$ be its one-time law. It takes values
in $I$ and obeys $U_{t+1}=F_{Y_{t+1}}(U_t)$. We first prove, for every
interval $J\subset\mathbb R$ of length $h>0$, that
\begin{equation}\label{eq:gen-general-interval-bound}
\nu(J)\le(h/\psi_k)^{d_*}.
\end{equation}
For a word $v=v_1\cdots v_j$, write
$F_v=F_{v_j}\circ\cdots\circ F_{v_1}$. A word appended on the left
is older: $F_{bv}=F_v\circ F_b$. At depth $i$, call a word $v$ of
length $ik$ admissible if
\begin{equation}\label{eq:gen-admissible-node}
F_v(I)\cap J\ne\varnothing.
\end{equation}
Let $M_i$ be the sum of the probabilities of admissible words as the
most recent $ik$ observations. Then $M_0\le1$, and the filter recursion
implies $\nu(J)\le M_i$ at every depth: if $U_0\in J$, the actual
most recent word is admissible.

For an admissible word $v$ of length $ik$, the two older-block children
$a_0^kv$ and $a_1^kv$ have filter intervals separated by at least
$\lambda^{ik}\psi_k$. If
\begin{equation}\label{eq:gen-pruning-scale}
h<\psi_k\lambda^{ik},
\end{equation}
the target interval meets at most one of the two child intervals,
so at least one of these older blocks is inadmissible.

Every specified older block of length $k$, conditional on any more
recent observation word, has probability at least $\beta^k$. Indeed,
conditional independence of emissions given the hidden path gives
\begin{align}
\Pp(Y_{1:k}=b,\,Y_{k+1:k+j}=v)
&=\E\left[
\prod_{\ell=1}^k g_{H_\ell}(b_\ell)
\prod_{\ell=1}^j g_{H_{k+\ell}}(v_\ell)
\right]\notag\\
&\ge\beta^k\Pp(Y_{k+1:k+j}=v).
\label{eq:gen-backward-nonnull}
\end{align}
Stationarity translates this inequality to the preceding blocks in
the tree. For each admissible parent, at least $\beta^k$ of its
conditional child mass is excluded whenever
\eqref{eq:gen-pruning-scale} holds. Because the word events are disjoint even when their filter
intervals overlap, summing over parents yields
\begin{equation}\label{eq:gen-pruning-mass}
M_{i+1}\le\theta M_i.
\end{equation}
Since $F_b(I)\subset I$, every child of an inadmissible parent is
also inadmissible.

For $h\ge \psi_k$, \eqref{eq:gen-general-interval-bound} is immediate.
Otherwise choose $\ell\ge1$ with
$\psi_k\lambda^{\ell k}\le h<\psi_k\lambda^{(\ell-1)k}$. The pruning
condition holds for $i=0,\ldots,\ell-1$, and hence
\[
\nu(J)\le M_\ell\le\theta^\ell
=(\lambda^{\ell k})^{d_*}\le(h/\psi_k)^{d_*}.
\]
This proves \eqref{eq:gen-general-interval-bound}.

The stationary output predictor is the vector
$V=p(U_0)=(1-U_0)g_0+U_0g_1$. For any $u,v\in I$,
\begin{equation}\label{eq:gen-predictor-isometry}
d_{\mathrm{TV}}(p(u),p(v))=c_g|u-v|.
\end{equation}
The preimage under $p$ of a total-variation ball of radius $r$ has
diameter at most $2r/c_g$, and is contained in an interval of that
length. Thus its law $\mu$ satisfies
$\mu(B_{\mathrm{TV}}(c,r))\le(2r/G)^{d_*}$ for every center $c$.
Corollary~\ref{cor:gen-small-ball}, with $C=(2/G)^{d_*}$, gives
\begin{equation}\label{eq:gen-general-binary-lower}
\mathcal D_K(\mu)\ge\frac{G^2}{8}(2K)^{-2/d_*},
\qquad
K_{\mathrm{gen}}(T,\varepsilon)
\ge K_{\mathrm{KL}}(T,\varepsilon)
\ge\frac12\left(\frac{G^2T}{4\varepsilon}\right)^{d_*/2}.
\end{equation}
Theorem~\ref{thm:gen-hmm-upper}, with $m=2$, gives the upper bound
in \eqref{eq:gen-classification-state-bounds}. Theorem~\ref{thm:gen-kl-upper}
gives the upper bound in \eqref{eq:gen-kl-classification-state-bounds};
the same bound for $T/\varepsilon<1$ follows from a fixed grid since
$\log(1+\sqrt{\varepsilon/T})\ge\log2$. Taking logarithms proves
\eqref{eq:gen-general-binary-bit-law} and \eqref{eq:gen-kl-bit-law}.

For a single initial forecast, Proposition~\ref{prop:gen-initial-forecast}
gives
\[
K_{\mathrm{init}}(T,\varepsilon)
\ge\frac12\left(
\frac{G^2T}{4(\varepsilon+C_P)}\right)^{d_*/2}.
\]
Together with $K_{\mathrm{init}}\le K_{\mathrm{gen}}$, this proves
\eqref{eq:gen-general-binary-initial-law} for every fixed
$\varepsilon>0$.
\end{proof}

\subsection{The lower bound for event queries}\label{app:hmm-query-lower}

\begin{proof}[Proof of Theorem~\ref{thm:query-memory}]
In the independent cases identified in the proof of
Theorem~\ref{thm:gen-binary-classification}, the future law is the
same after every observed history. One state therefore answers all
queries.

In the nondegenerate case, choose a symbol $y_*$ with
$a_*:=|g_1(y_*)-g_0(y_*)|>0$. Use the constants $\psi_k,d_*$ and
the stationary hidden filter from Appendix~\ref{app:hmm-binary}.
The scalar predictor
$R=g_0(y_*)+(g_1(y_*)-g_0(y_*))U_0$ has a law $\mu_R$ satisfying
\[
\mu_R(J)\le\left(\frac{|J|}{a_*\psi_k}\right)^{d_*}
\]
for every interval of positive length. The same union-of-balls proof
as in Corollary~\ref{cor:gen-small-ball} therefore gives
\begin{equation}\label{eq:query-quantization-lower}
\mathcal D_K(\mu_R)\ge\frac{(a_*\psi_k)^2}{8}(2K)^{-2/d_*},
\end{equation}
where $\mathcal D_K$ here is scalar mean-square quantization.

Suppose a $K$-state observer has the required query accuracy. At a fixed
time $t$, let $\widehat R_t(z)$ be its answer for the event
$\{Y_{t+1}=y_*\}$, clipped to $[0,1]$ if necessary. Put
\[
R_t=\Pp(Y_{t+1}=y_*\mid Y_{1:t}),\qquad
\mu_{R,t}=\mathcal L(R_t).
\]
Every realized state satisfies
$|R_t-\widehat R_t(Z_t)|\le e^\varepsilon-1$. The $K$ answers at time $t$
form a codebook at that time, allowing arbitrary clock dependence.
Hence
\begin{equation}\label{eq:query-quantization-upper}
\mathcal D_K(\mu_{R,t})\le(e^\varepsilon-1)^2.
\end{equation}
Stationarity and coordinatewise martingale convergence give a coupling
under which the finite-past predictor tends to $R$ in mean absolute
error, as in the proof of Theorem~\ref{thm:gen-quantization}. Squared
distance to a codebook in $[0,1]$ is uniformly $2$-Lipschitz, so
$\mathcal D_K(\mu_{R,t})\to\mathcal D_K(\mu_R)$. Taking $t\to\infty$
in \eqref{eq:query-quantization-upper} and combining with
\eqref{eq:query-quantization-lower} gives
\[
K\ge\frac12\left(
\frac{(a_*\psi_k)^2}{8(e^\varepsilon-1)^2}
\right)^{d_*/2}.
\]
For $0<\varepsilon\le1$, convexity gives
$e^\varepsilon-1\le(e-1)\varepsilon$. This proves the lower bound
in \eqref{eq:query-state-bounds} with $d_q=d_*$. The upper bound is
Theorem~\ref{thm:hmm-memory} with $m=2$. Taking logarithms proves
\eqref{eq:query-bit-law}.
\end{proof}

\section{Proofs for moving-window memory}\label{sec:appendix-window}

\begin{proof}[Proof of Proposition~\ref{prop:zero}]
With one state, the output at each fixed time is deterministic. Since $S_t\sim\operatorname{Bin}(n,1/2)$, the minimum mean-square error among such outputs is $\Var(S_t)=n/4$, attained at $n/2$. The corresponding log probability estimate is $n\log p_-+an/2$.
\end{proof}

\subsection{Recovery of expiring block counts}\label{sec:lower}

Successive window estimates recover the counts of blocks that have left
the window. The following entropy bound relates the accuracy of this
recovery to the information retained in the state.

\begin{lemma}[Entropy and quadratic estimation error]\label{lem:dither}
Let $U$ be integer-valued and let $\widehat U$ be a function of a discrete random variable $W$. If $\E(U-\widehat U)^2\le D$, then
\[
H(U\mid W)\le\frac12\log\bigl(2\pi e(D+1/12)\bigr).
\]
\end{lemma}
\begin{proof}[Proof of Lemma~\ref{lem:dither}]
Let $V$ be independent and uniform on $[-1/2,1/2]$. Distinct integer
translates of this interval have disjoint interiors, so
$h(U+V\mid W)=H(U\mid W)$, where $h$ denotes differential entropy.
Conditional on $W$, the variance of $U+V$ is at most
$\E[(U-\widehat U)^2\mid W]+1/12$. The Gaussian entropy bound, followed
by Jensen's inequality, proves the claim.
\end{proof}

\begin{proposition}[Recovery of block counts]\label{prop:entropy}
Suppose a $K$-state recursion satisfies $\E(\widehat S_t-S_t)^2\le E^2$ at every $t\ge n$. Put $v=(1-\delta^2)/4$. For every integer $L\ge1$, with $m=\lfloor n/L\rfloor$,
\begin{equation}\label{eq:entropy-lower}
\log K\ge\frac m2\log\frac{4Lv}{\pi^2e(4E^2+1/12)}.
\end{equation}
In particular,
\begin{equation}\label{eq:block-bit-lower}
\log_2K\ge
\left\lfloor\frac{n}{\lceil1024(1+E^2)\rceil}\right\rfloor.
\end{equation}
\end{proposition}

\begin{proof}[Proof of Proposition~\ref{prop:entropy}]
Let $X=(X_1,\ldots,X_n)$, $Y=(X_{n+1},\ldots,X_{2n})$, and $Z=Z_n$. Conditional on $Y=y$, the coordinates of $X$ are independent, with
\[
\Pp(X_i=1\mid Y_i=1)=p_+,
\qquad \Pp(X_i=1\mid Y_i=0)=p_-.
\]
Each conditional variance is $v$. For $1\le j\le m$, define the old and new block counts
\[
U_j=\sum_{i=(j-1)L+1}^{jL}X_i,
\qquad V_j=\sum_{i=(j-1)L+1}^{jL}Y_i.
\]
They satisfy
\[
S_{n+jL}-S_{n+(j-1)L}=V_j-U_j.
\]
The deterministic recursion reproduces every subsequent state and output
from $Z$ and $Y$. Hence
\[
\widehat U_j=V_j-\bigl(\widehat S_{n+jL}-\widehat S_{n+(j-1)L}\bigr)
\]
is a function of $(Z,Y)$ and has mean square error at most $4E^2$. Lemma~\ref{lem:dither} yields
\begin{equation}\label{eq:block-upper-entropy}
H(U_j\mid Z,Y)\le\frac12\log\bigl(2\pi e(4E^2+1/12)\bigr).
\end{equation}

For the entropy conditional on the continuation alone, consider a sum
$U$ of $L$ independent Bernoulli variables, each with variance $v$.
Its characteristic function satisfies
\[
|\E e^{i\theta U}|\le\exp\{-2Lv\sin^2(\theta/2)\}.
\]
Indeed, each factor has squared modulus $1-4v\sin^2(\theta/2)$. Fourier inversion and $\sin(|\theta|/2)\ge|\theta|/\pi$ for $|\theta|\le\pi$ give
\begin{align*}
\max_k\Pp(U=k)
&\le\frac1{2\pi}\int_{-\pi}^{\pi}
e^{-2Lv\sin^2(\theta/2)}\,d\theta\\
&\le\frac1{2\pi}\int_{\mathbb R}e^{-2Lv\theta^2/\pi^2}\,d\theta
=\sqrt{\frac{\pi}{8Lv}}.
\end{align*}
Applying the resulting maximal-atom bound for each value of $Y$ gives
\begin{equation}\label{eq:block-lower-entropy}
H(U_j\mid Y)\ge\frac12\log\frac{8Lv}{\pi}.
\end{equation}
Since the $U_j$ are independent conditional on $Y$, conditional
subadditivity and \eqref{eq:block-upper-entropy}--\eqref{eq:block-lower-entropy}
give
\begin{align*}
\log K
&\ge I(U_1,\ldots,U_m;Z\mid Y)\\
&\ge\sum_{j=1}^m\{H(U_j\mid Y)-H(U_j\mid Z,Y)\}\\
&\ge\frac m2\log\frac{4Lv}{\pi^2e(4E^2+1/12)}.
\end{align*}
This proves \eqref{eq:entropy-lower}. Since $v\ge3/16$, choosing $L=\lceil1024(1+E^2)\rceil$ makes the ratio inside the logarithm at least $192/(\pi^2e)>4$. Each complete block therefore contributes at least one bit, proving \eqref{eq:block-bit-lower}.
\end{proof}

\subsection{The cost of the initial count}

The preceding argument uses the order of expiration. A further memory
cost arises when the recursion forms its first count, before any
observation has expired.

\begin{proposition}[The cost of the initial count]\label{prop:log-lower}
For every fixed $c\in(0,1/4)$ there are $\beta_c>0$ and $N_c'$ such that, for $n\ge N_c'$, any recursion initialized at a fixed state and satisfying $\E(\widehat S_n-S_n)^2\le cn$ has
\[
\log_2K\ge\beta_c\log_2n.
\]
\end{proposition}
\begin{proof}[Proof of Proposition~\ref{prop:log-lower}]
The inputs $X_1,\ldots,X_n$ are independent fair bits. Theorem~13 of the
full version of Braverman, Garg and Woodruff~\cite{bgw}, which allows an
arbitrary transition map at each time, gives the following implication: for each fixed $\eta>0$, there are $d_\eta>0$
and $N_\eta$ such that, for all $n\ge N_\eta$,
\[
\sum_{i=1}^n I(Z_i;X_{1:i})\le d_\eta n\log n
\quad\Longrightarrow\quad
\E\Var(S_n\mid Z_n)\ge(1/4-\eta)n.
\]
Take $\eta=(1/4-c)/2$. Each mutual information is at most $H(Z_i)\le\log K$. If $\log K\le d_\eta\log n$, the displayed implication forces the minimum mean square error based on $Z_n$ to exceed $cn$. Thus $\log K>d_\eta\log n$. Dividing both logarithms by $\log2$ proves the claim with $\beta_c=d_\eta$ and $N'_c=N_\eta$.
\end{proof}

\subsection{A finite-capacity estimator}\label{sec:upper}

A bounded counter retains the fluctuation within the current block,
while a fixed-capacity queue stores the increments of completed blocks.
We first bound the estimation error when the queue is full, then control
the loss during recovery from an overflow.

\begin{proposition}[Finite-capacity construction]\label{prop:upper}
Let $1\le E<\sqrt n/2$, and put
\begin{equation}\label{eq:construction-parameters}
B=\lfloor E^2\rfloor,\quad q=\lfloor E\rfloor,\quad
M=\left\lceil\frac nB\right\rceil,\quad
H=\lceil8M+64\log(2n)\rceil.
\end{equation}
There is a deterministic recursion, initialized before $X_1$, with
\begin{equation}\label{eq:upper-states}
K\le2^H(M+1)(q+2B)
\end{equation}
and $\E(\widehat S_t-S_t)^2\le E^2$ for every $t\ge n$.
\end{proposition}

\begin{proof}[Proof of Proposition~\ref{prop:upper}]
\smallskip
\noindent\emph{Centered carries.}

Let $C_t=\sum_{i=1}^tX_i$ and $D_t=2C_t-t$, with $C_0=D_0=0$. At block boundaries define
\[
Q_k=\left\lfloor\frac{D_{kB}}q\right\rfloor,
\qquad d_k=Q_k-Q_{k-1}.
\]
The retained residual is
\begin{equation}\label{eq:residual}
A_t=D_t-qQ_{\lfloor t/B\rfloor}.
\end{equation}
When the $t$th observation arrives, form $A_t^-=A_{t-1}+2X_t-1$.
If $t=kB$, emit $d_k=\lfloor A_t^-/q\rfloor$ and set
$A_t=A_t^--qd_k$; otherwise set $A_t=A_t^-$. This update uses only the
preceding residual and the new bit, and preserves the bound
\begin{equation}\label{eq:residual-range}
-(B-1)\le A_t\le q+B-2.
\end{equation}

Suppose first that the most recent $M$ increments are available. For a query at $t\ge n$, write
\[
k=\lfloor t/B\rfloor,\qquad
j=\lfloor(t-n)/B\rfloor,\qquad
u=t-n-jB.
\]
Then $0\le u<B$ and $k-j\le M$. Set
\begin{equation}\label{eq:ideal-estimate}
W_t=\sum_{r=j+1}^kd_r=Q_k-Q_j,
\qquad
\widetilde S_t=\clip\left\{
\frac n2+\frac{qW_t+A_t-(q-1)/2}{2}\right\}.
\end{equation}
To evaluate the error, decompose the expired cumulative sum as
\[
D_{t-n}=qQ_j+R_j+V_t,
\quad 0\le R_j\le q-1,
\quad V_t=D_{jB+u}-D_{jB}.
\]
The $u$ observations in this partial block are independent fair bits, so $\E V_t^2=u\le B$. Before clipping, the estimation error is
\[
\frac{R_j-(q-1)/2+V_t}{2}.
\]
Consequently,
\begin{equation}\label{eq:ideal-mse}
\E(\widetilde S_t-S_t)^2
\le\frac{(q-1)^2}{8}+\frac B2
\le\frac58E^2.
\end{equation}
The remaining error budget will cover periods when the queue is refilling.

\Needspace{5\baselineskip}
\smallskip
\noindent\emph{Encoding and recovery.}

Encode a signed integer $d$ by $|d|$ ones, a zero delimiter, and one sign bit; zero is assigned the nonnegative sign. These codewords are prefix-free and have length $|d|+2$. The queue is represented by an $H$-bit string, padded with zeros, together with its number $\nu$ of complete codewords, where $0\le\nu\le M$. The count $\nu$ determines where the padding begins.

Initially set $A_0=0$ and fill the queue with $M$ virtual zero increments. At each block boundary, remove the oldest codeword if the queue already contains $M$ words, and append the new increment. If it would exceed capacity $H$, clear the queue and set $\nu=0$. The residual $A_t$ keeps the value prescribed by \eqref{eq:residual}. Subsequent increments refill the queue. When $\nu=M$, use \eqref{eq:ideal-estimate}; when $\nu<M$, output $n/2$.

The virtual initial increments have indices at most zero. At every
query, $j\ge0$, so $W_t$ contains only observed increments. After an
overflow, the unchanged residual continues to produce the exact
increments $d_k$; once full, the queue again contains the required
suffix. The state consists of the padded bit string, the word count,
and the bounded residual. Their cardinalities, together with
\eqref{eq:residual-range}, give \eqref{eq:upper-states}.

\Needspace{5\baselineskip}
\smallskip
\noindent\emph{The overflow probability.}

Consider any $M$ consecutive completed blocks, with virtual zero blocks inserted when the interval extends before time zero. For an actual block let
\[
W_r=D_{rB}-D_{(r-1)B}.
\]
The increment obeys $|d_r|\le|W_r|/q+1$. The total code length of the $M$ blocks is therefore at most
\begin{equation}\label{eq:code-envelope}
\Phi=3M+\frac1q\sum_r|W_r|,
\end{equation}
where virtual blocks contribute zero to the sum. Since $B\le n$, each actual block is a sum of independent centered fair bits. Hence $\E W_r^2=B$ and
\begin{equation}\label{eq:mean-code-envelope}
\E\Phi\le3M+\frac{M\sqrt B}{q}\le5M.
\end{equation}

To account for dependence between blocks, group the observations by
their residue class modulo $n$. The interval in
\eqref{eq:code-envelope} has length $MB<n+B\le2n$, so the groups are
independent and each contains at most two observations. Replacing one
group changes $\sum_r|W_r|$ by at most four and $\Phi$ by at most
$4/q$. The bounded differences inequality~\cite{mcdiarmid} gives
\begin{equation}\label{eq:overflow-tail}
\Pp(\Phi>H)
\le\exp\left\{-\frac{(H-5M)^2q^2}{8n}\right\}
\le(2n)^{-12}.
\end{equation}
For the last step, $q^2\ge B/4$ and $n\le MB$ imply $q^2/n\ge1/(4M)$. If $s_n=\log(2n)$, then
\[
\frac{(H-5M)^2q^2}{8n}
\ge\frac{(3M+64s_n)^2}{32M}\ge12s_n.
\]

At an append operation, the candidate queue is a suffix of the latest
$M$ actual or initial virtual increments. An overflow therefore implies
the event in \eqref{eq:overflow-tail}. At a fixed time, an incomplete
queue implies an overflow at one of the last $M$ block boundaries.
A union bound yields
\begin{equation}\label{eq:invalid}
\Pp(\nu<M)\le M(2n)^{-12}.
\end{equation}
On this event the error of the constant output is at most $n/2$. Combining \eqref{eq:ideal-mse} and \eqref{eq:invalid} proves
\[
\E(\widehat S_t-S_t)^2
\le\frac58E^2+\frac{n^2}{4}M(2n)^{-12}
\le E^2.
\]
The last inequality uses $M\le n$, $E^2\ge1$, and $n\ge5$, all valid
in the stated regime.
\end{proof}

\begin{corollary}[An upper bound from window marginals]\label{cor:window-marginal-upper}
Suppose a binary stream has the independent fair distribution on every consecutive block of length $n$. For $1\le E<\sqrt n/2$, the construction and state bound in Proposition~\ref{prop:upper} still satisfy $\E(\widehat S_t-S_t)^2\le E^2$ at every $t\ge n$.
\end{corollary}
\begin{proof}[Proof of Corollary~\ref{cor:window-marginal-upper}]
The residual identity, the error bound for a full queue, and the mean
code length use only independent fair blocks of length at most $n$.
For the overflow probability, split the $M$ blocks in
\eqref{eq:code-envelope} into the first $M-1$ blocks and the last block.
Their observation intervals have lengths at most $(M-1)B<n$ and
$B\le n$, respectively; initial virtual blocks add no random
observations. Write
\[
T_1=\frac1q\sum_{r=1}^{M-1}|W_r|,\qquad
T_2=\frac1q|W_M|,
\]
with indices shifted to the interval under consideration. Each $T_i$ is a function of at most $n$ independent fair bits, and changing one of its bits changes its value by at most $2/q$. Thus, for $s>0$, bounded differences and a union bound give
\[
\Pp\{T_1+T_2-\E(T_1+T_2)>s\}
\le2\exp\{-s^2q^2/(8n)\}.
\]
The overflow and incomplete-queue probabilities in
\eqref{eq:overflow-tail} and \eqref{eq:invalid} are therefore at most
$2(2n)^{-12}$ and $2M(2n)^{-12}$. The mean-square bound becomes
\[
\E(\widehat S_t-S_t)^2
\le\frac58E^2+\frac{n^2}{2}M(2n)^{-12}\le E^2,
\]
since $M\le n$, $n\ge5$, and $E^2\ge1$.
\end{proof}

\subsection{Proof of the main memory theorem}

\begin{proof}[Proof of Theorem~\ref{thm:main}]
Put $R=n/(1+E^2)$ and take $n\ge\max\{5,N'_c\}$. For the upper bound, use the exact $n$-bit window when $E<1$, at cost at most $2R$ bits. When $E\ge1$, the condition $E^2\le cn<n/4$ permits Proposition~\ref{prop:upper}. Its parameters obey
\[
B\ge E^2/2,\qquad M\le4R+1,\qquad M\le n,
\qquad q+2B\le3E^2<n.
\]
By \eqref{eq:upper-states}, the number of bits needed by the construction is at most
\begin{align*}
H+\log_2(M+1)+\log_2(q+2B)+1
&\le32R+(64\log2+2)\log_2n+11+64\log2\\
&\le100(R+\log_2n).
\end{align*}
The last inequality holds for $n\ge5$. This proves the upper bound with $B_c=100$, uniformly in $\delta$ and $\varepsilon$.

For the lower bound, let $b=\lceil\log_2K\rceil$ be the memory of any admissible estimator. Proposition~\ref{prop:entropy} and
$\lceil1024(1+E^2)\rceil\le1025(1+E^2)$ give
\[
b\ge\left\lfloor\frac R{1025}\right\rfloor.
\]
Also $b\ge1$ by Proposition~\ref{prop:zero}, since $E^2<n/4$. Hence
$R<1025(b+1)\le2050b$. Proposition~\ref{prop:log-lower} gives
$\log_2n\le b/\beta_c$. Adding these inequalities yields
\[
b\ge(2050+\beta_c^{-1})^{-1}(R+\log_2n).
\]
Thus Theorem~\ref{thm:main} holds with
\begin{equation}\label{eq:main-constants}
A_c=(2050+\beta_c^{-1})^{-1},\qquad
B_c=100,\qquad N_c=\max\{5,N'_c\},
\end{equation}
where $\beta_c=d_\eta$ and $N'_c=N_\eta$ come from the streaming variance
theorem of~\cite{bgw} with $\eta=(1/4-c)/2$
(Proposition~\ref{prop:log-lower}).
\end{proof}

\subsection{Encoding and uniform-error comparisons}\label{sec:uniform}

We next separate the effects of recursion and loss. A single-window
encoder can choose its label after reading the entire window. For a
recursive estimator, requiring accuracy on every history gives a
different dependence on the tolerance.

\subsubsection{A finite encoding at one time}

\begin{proposition}[Single-window quantization]\label{prop:offline}
Fix $c\in(0,1/4)$. At any specified $t\ge n$, there is an encoding of $S_t$ into at most
\begin{equation}\label{eq:offline-labels}
1+\left\lceil\frac{2}{\sqrt c}
\sqrt{\frac12\log\frac4c}\right\rceil
\end{equation}
labels from which an estimate with mean square error at most $cn$ can be decoded.
\end{proposition}
\begin{proof}[Proof of Proposition~\ref{prop:offline}]
Let $W=(S_t-n/2)/\sqrt n$. The usual Bernoulli tail bound~\cite{hoeffding} gives $\Pp(|W|>x)\le2e^{-2x^2}$. It also follows directly from
\[
\E e^{\lambda W}
=\left(\cosh\frac{\lambda}{2\sqrt n}\right)^n
\le e^{\lambda^2/8}
\]
and Chernoff's inequality. For $R>0$, the squared error of clipping $W$ to $[-R,R]$ satisfies
\begin{align*}
\E(|W|-R)_+^2
&=\int_0^\infty2u\,\Pp(|W|>R+u)\,du\\
&\le4\int_0^\infty u e^{-2(R+u)^2}\,du
\le e^{-2R^2}.
\end{align*}
Choose $R=\sqrt{\tfrac12\log(4/c)}$ and divide $[-R,R]$ into $N=\lceil2R/\sqrt c\rceil$ equal intervals. Round the clipped value to the nearest endpoint. The rounding error is at most $\sqrt c/2$, so the total mean square error is at most $2e^{-2R^2}+c/2=c$. Rescaling and clipping the count estimate to $[0,n]$ proves the result with $N+1$ labels.
\end{proof}

\subsubsection{Uniform error over histories}

For the criterion in Section~\ref{sec:model}, the block argument and
carry construction of Ben Basat et al.~\cite{bb} take the following
form.

\begin{proof}[Proof of Proposition~\ref{prop:uniform}]
For the lower bound, put $m=\lfloor n/L\rfloor$ and consider the $2^m$ windows formed by $m$ constant blocks of length $L$, followed by $n-mL$ zeros. Suppose two such windows lead to the same state at time $n$. Take their rightmost unequal block and append enough zeros to discard all earlier blocks. The states remain identical, while the two window counts now differ by $L>2E$. A common output cannot satisfy \eqref{eq:uniform-criterion} for both histories. Hence the states reached at time $n$ after these $2^m$ windows must all be distinct.

For the upper bound, write $C_t=\sum_{i=1}^tX_i$, and define
\[
Q_k=\lfloor C_{kB}/B\rfloor,\qquad
d_k=Q_k-Q_{k-1}\in\{0,1\},\qquad
A_t=C_t-BQ_{\lfloor t/B\rfloor}.
\]
The residual satisfies $0\le A_t\le2B-2$. To update it, form $A_t^-=A_{t-1}+X_t$. At a block boundary $t=kB$, emit $d_k=\lfloor A_t^-/B\rfloor$ and set $A_t=A_t^--Bd_k$; otherwise set $A_t=A_t^-$. Store the most recent $M=\lceil n/B\rceil$ carry bits. Initially the residual and the queue are zero.

At time $t\ge n$, define
\[
k=\lfloor t/B\rfloor,\qquad
j=\lfloor(t-n)/B\rfloor,\qquad
u=t-n-jB,
\]
and let $W=\sum_{r=j+1}^kd_r$. Then $0\le u<B$ and $k-j\le M$. The expired cumulative count can be written
\[
C_{t-n}=BQ_j+R+V,\qquad
0\le R\le B-1,\quad0\le V\le u.
\]
Therefore
\[
\widehat S_t=\clip\left\{BW+A_t-\frac{B-1+u}{2}\right\}
\]
has error at most $(B-1+u)/2\le B-1\le E$. Since $j\ge0$, the queried suffix contains only actual carry bits. There are at most $2B\,2^M$ states. Exact storage of the window gives the alternative $n$-bit bound. When $E<1/2$, the lower bound has $L=1$ and equals $n$.
\end{proof}

The accuracy dependence in \eqref{eq:uniform-bounds} is $n/(1+E)$.
Under mean-square loss it becomes $n/(1+E^2)$, together with the
logarithmic cost of forming the initial count. Averaging over the
stationary law makes centered block fluctuations, and hence shorter
codes, sufficient for most windows. The fixed queue capacity in
Appendix~\ref{sec:upper} turns this saving into a bound on the total
number of states.

\section{Proofs for the small-memory regime}\label{sec:appendix-small-memory}

\subsection{Block transfer and the conditional score}

We obtain bounds on $\Gamma_{n,\delta}$ in \eqref{eq:critical-gain}
by relating window maintenance to a computation with one terminal query. Let
$\xi_1,\ldots,\xi_m$ be independent fair bits and $T_m=\sum_{i=1}^m\xi_i$.
Put
\begin{equation}\label{eq:prefix-gain}
V_m(b)=\max_{\mathcal P:\,|\mathcal Z|\le2^b}
\Var\bigl(\E[T_m\mid Z_m]\bigr),
\end{equation}
where $\mathcal P$ reads the $m$ bits once, starts from a fixed state, and may
use a different transition map at each position. For fixed $m$ and a
fixed finite state space there are finitely many transition sequences,
so the maximum is attained. The optimal decoder for each sequence is
the conditional mean.

\begin{proposition}[Block transfer]\label{prop:critical-transfer}
For $n\ge2$, $m=\lfloor n/2\rfloor$, and every integer $b\ge0$,
\begin{equation}\label{eq:critical-transfer}
V_m(\lfloor b/2\rfloor)
\le\Gamma_{n,\delta}(b)\le V_n(b).
\end{equation}
If a single-block program has $K$ internal states and a final output with
$L$ possible values, the variance gain of that output can be
maintained at all query times with at most $KL$ states.
\end{proposition}
\begin{proof}[Proof of Proposition~\ref{prop:critical-transfer}]
At the first query the recursion has processed exactly $n$ independent fair
bits. Conditional expectation minimizes squared error, so its gain is at
most $V_n(b)$.

For the lower bound, divide time into blocks of length $m$. Run a chosen
single-block program in a program register on the current block, and keep the final state of the preceding complete block in a storage register. The two registers use at most $K^2$ states. At a block boundary, copy the completed state into the storage register and
reset the program register. Both registers have fixed initial values, and
the clock specifies the block phase.

At $t=km+u$, $0\le u<m$, the preceding complete block consists of
positions $(k-1)m+1,\ldots,km$. It is contained in the current window since
$m+u\le2m-1\le n$. For $t\ge n$ this block has been fully observed.
Let $U$ be its count and $Z^{\mathrm{prev}}$ its retained state. The entire current window
consists of independent fair bits. Hence the decoder
\[
\widehat S_t=\frac{n-m}{2}+\E[U\mid Z^{\mathrm{prev}}]
\]
has mean square error
\[
\frac{n-m}{4}+\E\Var(U\mid Z^{\mathrm{prev}})
=\frac n4-\Var(\E[U\mid Z^{\mathrm{prev}}]).
\]
This proves the left inequality. If the decoder only uses an $L$-valued
function of the final state, retain that function instead, giving $KL$
states.
\end{proof}

The transfer depends only on the independent fair distribution within
each length-$n$ window, and therefore applies to all sources with these
window marginals.

The gain of a finite computation can be evaluated through its
sensitivity to the input bias. The score of a statistic is the
conditional expectation of the input score \cite[Lemma~1]{bho20}; for a
Bernoulli count, this gives the following identity.

\begin{lemma}[Fisher information of a finite computation]\label{lem:critical-fisher}
Fix a deterministic program on $m$ input bits. Keep its maps fixed and let
the input bits have law $\Ber(p)$. Write $\pi_z(p)=\Pp_p(Z_m=z)$ and let
\[
\mathcal I_{Z_m}(1/2)
=\sum_{z:\,\pi_z(1/2)>0}
\frac{\pi_z'(1/2)^2}{\pi_z(1/2)}.
\]
Then
\begin{equation}\label{eq:critical-fisher}
\mathcal I_{Z_m}(1/2)
=16\Var\bigl(\E[T_m\mid Z_m]\bigr).
\end{equation}
\end{lemma}
\begin{proof}[Proof of Lemma~\ref{lem:critical-fisher}]
The score of the input law at $p=1/2$ is $4(T_m-m/2)$. Differentiating
the finite sum defining $\pi_z$ gives
\[
\pi_z'(1/2)=4\pi_z(1/2)\E[T_m-m/2\mid Z_m=z].
\]
Substitution proves the identity. A state with zero probability at $1/2$
is unreachable and has zero probability for every $p\in(0,1)$.
\end{proof}

\subsection{Two-state information}

Venkatesh and Franklin~\cite{vf91} studied time-dependent one-bit
computation of Bernoulli streams. Alternating AND and OR also appears
in the analysis of nested canalizing functions~\cite{khs13}. For the
present squared-error criterion, we bound the state probability and its
derivative jointly to obtain the sharp limiting constant.

\begin{lemma}[Two-state information bound]\label{lem:critical-two-state}
For every deterministic two-state program on $m$ independent fair bits,
\begin{equation}\label{eq:critical-two-state}
\mathcal I_{Z_m}(1/2)\le8(1-2^{-m}).
\end{equation}
Consequently $V_m(1)\le(1-2^{-m})/2<1/2$ for every finite $m$.
\end{lemma}
\begin{proof}[Proof of Lemma~\ref{lem:critical-two-state}]
Encode the state by zero and one. At any position put
$q=\Pp_p(Z=1)|_{p=1/2}$ and
$r=\left.\frac{d}{dp}\Pp_p(Z=1)\right|_{p=1/2}$, and define
\[
v=\frac{r^2}{16q(1-q)}
\]
when $0<q<1$, with $v=0$ for a constant state. We show that each transition
satisfies
\begin{equation}\label{eq:critical-gain-recursion}
v_{\rm new}\le\frac14+\frac v2,
\end{equation}
provided $v\le1/2$. Initially $v=0$.

We consider the sixteen deterministic maps from a state and one input
bit to the next state. A map that ignores the input and preserves or
complements the state also preserves $v$, which is at most the right side of
\eqref{eq:critical-gain-recursion}. A constant map gives $v_{\rm new}=0$.
If both rows of the transition table are nonconstant functions of the
input, the new state has probability $1/2$ and its derivative is one of
$1,-1,1-2q,2q-1$. Thus $v_{\rm new}\le1/4$.

Each of the remaining eight maps has one constant row and one nonconstant row.
Complementing the old or new state reduces the calculation to
\[
q_{\rm new}=q/2,
\qquad |r_{\rm new}|\le q+|r|/2.
\]
For $0<q<1$, the Cauchy--Schwarz inequality gives
\begin{align*}
v_{\rm new}
&\le\frac{\bigl(\sqrt q+2\sqrt{v(1-q)}\bigr)^2}{4(2-q)}\\
&\le\frac{(q+2(1-q))(1+2v)}{4(2-q)}
=\frac14+\frac v2.
\end{align*}
From a constant old state, the new state is either constant or a function
of the new input bit, and the same inequality holds. Iterating
\eqref{eq:critical-gain-recursion} yields $v_m\le(1-2^{-m})/2$ and
preserves the hypothesis $v\le1/2$. The Bernoulli output has Fisher
information $16v_m$, so Lemma~\ref{lem:critical-fisher} completes the proof.
\end{proof}

\subsection{The alternating program at a fixed terminal time}

\begin{proposition}[Exact gain of the alternating program]\label{prop:alternating-prefix-gain}
Read $m\ge1$ independent fair bits $\xi_1,\ldots,\xi_m$ with the fixed initial state $Z_0=0$ and the updates
\[
Z_j=\begin{cases}
Z_{j-1}\vee\xi_j,&j\text{ odd},\\
Z_{j-1}\wedge\xi_j,&j\text{ even}.
\end{cases}
\]
For $T_m=\sum_{j=1}^m\xi_j$, the explained variance of this program is
\begin{equation}\label{eq:alternating-prefix-gain}
J_m^{\mathrm{alt}}:=\Var(\E[T_m\mid Z_m])
=\begin{cases}
\displaystyle\frac{2^m-1}{2^{m+1}+1},&m\text{ even},\\[2mm]
\displaystyle\frac{2^{m+1}-1}{2^{m+2}+4},&m\text{ odd}.
\end{cases}
\end{equation}
In particular,
\[
\frac12-J_m^{\mathrm{alt}}\sim\frac34\,2^{-m},
\qquad
J_m^{\mathrm{alt}}\le V_m(1)\le\frac12(1-2^{-m}).
\]
The program uses one state bit and the parity supplied by the free clock.
\end{proposition}
\begin{proof}[Proof of Proposition~\ref{prop:alternating-prefix-gain}]
Put $q_j=\Pp(Z_j=1)$ and $c_j=\operatorname{Cov}(T_j,Z_j)$, with
$q_0=c_0=0$. With the update maps fixed, let $f_j(p)$ be the success
probability under independent $\Ber(p)$ inputs. The Bernoulli score
identity gives $c_j=f_j'(1/2)/4$. Over two successive updates,
\[
f_{2k+1}(p)=p+(1-p)f_{2k}(p),\qquad
f_{2k+2}(p)=p^2+p(1-p)f_{2k}(p).
\]
Evaluation and differentiation at $p=1/2$ yield
\[
q_{2k+2}=\frac{1+q_{2k}}4,
\qquad
c_{2k+2}=\frac{1+c_{2k}}4.
\]
Therefore
\[
q_{2k}=c_{2k}=\frac{1-4^{-k}}3.
\]
At the intervening odd step,
\[
q_{2k+1}=\frac23-\frac{4^{-k}}6,
\qquad
c_{2k+1}=\frac{1-q_{2k}}4+\frac{c_{2k}}2
=\frac{q_{2k+1}}2.
\]
Since $Z_m$ is binary, its optimal decoder is
\[
\E[T_m\mid Z_m]
=\frac m2+\frac{c_m}{q_m(1-q_m)}(Z_m-q_m).
\]
Thus $J_m^{\mathrm{alt}}=c_m^2/[q_m(1-q_m)]$, which gives \eqref{eq:alternating-prefix-gain}. Its distance from $1/2$ is
$3/[2(2^{m+1}+1)]$ for even $m$ and $3/[4(2^m+1)]$ for odd $m$, proving the asymptotic formula. This particular program is admissible in \eqref{eq:prefix-gain}, giving the lower bound on $V_m(1)$; Lemma~\ref{lem:critical-two-state} gives its upper bound.
\end{proof}

\begin{remark}[Exact finite-length optimization]\label{rem:one-bit-finite-dp}
For every $1\le m\le22$, an exact dynamic program gives
$V_m(1)=J_m^{\mathrm{alt}}$. The optimization covers all deterministic
two-state programs, with a fixed initial state and arbitrary
position-dependent transitions, for one terminal query. At $m=22$ the
common value is $1398101/2796203$. The recurrence and its exact
verification are given in Appendix~\ref{app:one-bit-dp}.
\end{remark}

\subsection{One bit at every query time}

\begin{theorem}[One bit with a free clock]\label{thm:critical-one-bit}
With the free-clock convention in \eqref{eq:recursion}, for every $n\ge1$ and $0<\delta\le1/2$,
\begin{equation}\label{eq:critical-one-bit}
\max\left\{\frac14,
\frac12(1-2^{-n})^2-\frac94\,2^{-n}\right\}
\le\Gamma_{n,\delta}(1)\le\frac12(1-2^{-n}).
\end{equation}
In particular, $1/2-\Gamma_{n,\delta}(1)=\Theta(2^{-n})$ uniformly in $\delta$.
\end{theorem}

\begin{proof}[Proof of Theorem~\ref{thm:critical-one-bit}]
The upper bound follows by applying Lemma~\ref{lem:critical-two-state}
at the first query. Retaining $X_t$ and decoding by
$(n-1)/2+X_t$ gives the lower bound $1/4$.

For the sharper lower bound, start with $Z_0=0$ and use the recursion
\[
Z_t=
\begin{cases}
Z_{t-1}\vee X_t,&t\text{ odd},\\
Z_{t-1}\wedge X_t,&t\text{ even}.
\end{cases}
\]
Set
\[
q_t=\begin{cases}2/3,&t\text{ odd},\\1/3,&t\text{ even},\end{cases}
\qquad c_n=\frac32(1-2^{-n}),
\qquad
\widehat S_t=\frac n2+c_n(Z_t-q_t).
\]
These maps use only the state bit and the free clock.

Fix $t\ge n$. Under the delayed-copy law the actual state $Z_{t-n}$ can depend on the current window, so we compare the recursion with an auxiliary process. Initialize it at $\widetilde Z_{t-n}\sim\Ber(q_{t-n})$, independent of the current window, and apply the same updates to its $n$ observations. Under independent fair inputs the
alternating recursion sends $\Ber(1/3)$ to $\Ber(2/3)$ and back. Thus
$\Var(\widetilde Z_t)=2/9$. At each input position $i$ in the window,
\[
\operatorname{Cov}(X_i,\widetilde Z_i)=\frac16,
\qquad
\operatorname{Cov}(X_i,\widetilde Z_t)
=\frac16\,2^{-(t-i)}.
\]
The second identity follows because the conditional expectation of each
subsequent state is affine in the preceding state with slope $1/2$.
Consequently
\[
\operatorname{Cov}(S_t,\widetilde Z_t)
=\frac13(1-2^{-n}).
\]
The auxiliary decoder
$\widetilde S_t=n/2+c_n(\widetilde Z_t-q_t)$ is therefore
$\E[S_t\mid\widetilde Z_t]$, and
\begin{equation}\label{eq:critical-auxiliary-loss}
\E(S_t-\widetilde S_t)^2
=\frac n4-\frac12(1-2^{-n})^2.
\end{equation}

The actual and auxiliary trajectories coalesce whenever an OR step reads
one or an AND step reads zero. They can disagree at time $t$ only if the entire current
window is the unique alternating word with zeros at OR positions and
ones at AND positions. This event has probability $2^{-n}$. On this word,
$|S_t-n/2|\le1/2$, while $|\widehat S_t-n/2|\le1$. Hence
$(S_t-\widehat S_t)^2\le9/4$ there. Off this event the two estimates agree.
Combining this event bound with \eqref{eq:critical-auxiliary-loss} proves
the claim. The coupling is pathwise in the current window, so the dependence of $Z_{t-n}$ on that window does not enter.
\end{proof}

For $0\le g\le n/4$, write
\[
b_{\rm crit}(n,\delta,g)
=b_*(n,\delta,a\sqrt{n/4-g}).
\]
\begin{corollary}\label{cor:critical-one-bit-threshold}
With a free clock, for $0<g\le1/4$, one has $b_{\rm crit}(n,\delta,g)=1$.
For every fixed $g\in(0,1/2)$, this equality holds for all sufficiently
large $n$, uniformly in $\delta$. At $g=1/2$ and $n\ge2$,
\[
b_{\rm crit}(n,\delta,1/2)=2.
\]
\end{corollary}

\begin{proof}[Proof of Corollary~\ref{cor:critical-one-bit-threshold}]
A positive gain requires at least two states. The first two claims
follow from \eqref{eq:critical-one-bit}. Its strict upper bound makes
a gain of $1/2$ require more than one bit. Storing the last two observations and
adding $(n-2)/2$ to their sum attains this gain with four states.
\end{proof}

\subsection{The six-state construction}

\begin{theorem}[Six-state Tribes gain with a free clock]\label{thm:critical-tribes}
Let $n\ge4$, $m=\lfloor n/2\rfloor$, and let $w$ be the largest positive
integer with $w2^w\le m$. Put $k=2^w$ and $q=2^{-w}$. There is a
deterministic six-state recursion with a free clock, started from a fixed state, for which
at every $t\ge n$,
\begin{equation}\label{eq:critical-tribes-error}
\E(S_t-\widehat S_t)^2=\frac n4-\gamma_w,
\qquad
\gamma_w=\frac{w^2}{4}
\frac{(1-q)^{k-2}}{1-(1-q)^k}.
\end{equation}
Moreover,
\begin{equation}\label{eq:critical-tribes-asymptotic}
\gamma_w=\frac{w^2}{4(e-1)}\bigl(1+O(2^{-w})\bigr)
\sim\frac{(\log_2n)^2}{4(e-1)}.
\end{equation}
\end{theorem}

\begin{proof}[Proof of Theorem~\ref{thm:critical-tribes}]
In each block of $m$ observations, use its first $kw$ bits to form
\[
F=\bigvee_{j=1}^{k}\bigwedge_{i=1}^{w}\xi_{j,i}.
\]
For one such block let $U$ be its total count and
$p_F=\Pp(F=1)=1-(1-q)^k$. Under Bernoulli input parameter $p$, the
success probability is $1-(1-p^w)^k$. The score identity gives
\[
\operatorname{Cov}(U,F)
=\frac{kwq}{2}(1-q)^{k-1}
=\frac w2(1-q)^{k-1}.
\]
Since $F$ is binary,
\[
\Var(\E[U\mid F])
=\frac{\operatorname{Cov}(U,F)^2}{p_F(1-p_F)}
=\gamma_w.
\]
An explicit decoder for a window containing this block is
\[
\widehat S=\frac n2+
\frac{\operatorname{Cov}(U,F)}{p_F(1-p_F)}(F-p_F).
\]

The three program states record: no successful tribe with the current
tribe still all one; no successful tribe with a zero already in the
current tribe; and a successful tribe. The third state is
absorbing within the block. At each tribe boundary, which the clock identifies, the first state moves to success and the second returns to the first.
After the $kw$th input, the result is retained through the remaining
positions of the block.

Retain the current program state together with the preceding complete
block's Boolean result. At each block boundary, replace the Boolean result
by the completed output and restart the three-state program. This gives
$3\cdot2=6$ states and a fixed initialization. The preceding block is
inside every queried window, as in Proposition~\ref{prop:critical-transfer}.
Applying the displayed decoder to its result proves
\eqref{eq:critical-tribes-error} at every $t\ge n$.

Finally, maximality of $w$ gives
\[
w2^w\le\lfloor n/2\rfloor<(w+1)2^{w+1},
\]
so $w=\log_2n-\log_2\log_2n+O(1)$. With $q=2^{-w}$ and $k=1/q$, the expansion
$k\log(1-q)=-1-q/2+O(q^2)$ yields
\[
\frac{(1-q)^{k-2}}{1-(1-q)^k}
=\frac1{e-1}\bigl(1+O(q)\bigr).
\]
Substitution proves \eqref{eq:critical-tribes-asymptotic}.
\end{proof}

\begin{corollary}[Nonuniformity at the single-state threshold]\label{cor:critical-nonuniformity}
For $g_n=\gamma_w$ from Theorem~\ref{thm:critical-tribes},
\[
2\le b_{\rm crit}(n,\delta,g_n)\le3
\]
for all sufficiently large $n$, uniformly in $\delta$. Here $g_n\to\infty$ and $g_n/n\to0$, yet the required memory remains bounded.
The logarithmic lower bound in Theorem~\ref{thm:main} therefore cannot be
made uniform as its fixed constant $c$ approaches $1/4$.
\end{corollary}
\begin{proof}[Proof of Corollary~\ref{cor:critical-nonuniformity}]
The six-state construction gives the upper bound. Since
$\gamma_w>1/2$ for sufficiently large $n$, Theorem~\ref{thm:critical-one-bit} rules out one bit. Here $E_n^2=n/4-\gamma_w$, for which
$n/(1+E_n^2)+\log n$ diverges while three bits suffice.
\end{proof}

\subsection{A fixed-memory hierarchy}

\begin{proposition}[A fixed-memory hierarchy with a free clock]\label{prop:critical-hierarchy}
Fix an integer $h\ge1$ and let
\[
b_h=\left\lceil\log_2(4h+2)\right\rceil.
\]
There is a constant $c_h>0$ such that, for all sufficiently large $n$,
\begin{equation}\label{eq:critical-hierarchy}
\Gamma_{n,\delta}(b_h)\ge c_h(\log n)^{2h},
\end{equation}
uniformly in $\delta$. More explicitly, put $m=\lfloor n/2\rfloor$ and
$w=\lfloor\log_2m/(2h)\rfloor$. Whenever $w\ge3$, a recursion with at
most $4h+2$ states and a free clock attains a gain of at least
\begin{equation}\label{eq:critical-hierarchy-explicit}
\frac14\left(\frac{\log2}{4}w\right)^{2h}.
\end{equation}
\end{proposition}

\begin{proof}[Proof of Proposition~\ref{prop:critical-hierarchy}]
Start with $F_0$ equal to one input bit and $q_0=1/2$. Given a Boolean
function $F_{j-1}$ with success probability $q_{j-1}$ at fair input, set
\[
k_j=\left\lceil
\frac{\log(1/2)}{\log(1-q_{j-1}^w)}
\right\rceil.
\]
Define $F_j$ as the OR of $k_j$ groups, each the AND of $w$ independent
copies of $F_{j-1}$ on disjoint input coordinates. The numbers $k_j$ are
chosen at fair input and remain fixed as the input parameter varies.

The calibration keeps $q_j\in[1/2,3/4]$ at every level. Indeed, with
$\zeta_j=q_{j-1}^w$, the definition gives
\[
\frac12\le q_j=1-(1-\zeta_j)^{k_j}
<\frac12+\frac{\zeta_j}{2}
\le\frac12+\frac12\left(\frac34\right)^3<\frac34.
\]
Also
\[
k_j\le (\log2)2^w+1\le2^w.
\]
The number $N_h$ of input bits used by $F_h$ is therefore at most
$(w2^w)^h\le2^{2hw}\le m$; unused block positions are ignored.

Let $f_j(p)$ be the success probability of $F_j$ under independent
$\Ber(p)$ input and put $D_j=f_j'(1/2)$. Then $D_0=1$ and
\[
\frac{D_j}{D_{j-1}}
=k_jwq_{j-1}^{w-1}(1-\zeta_j)^{k_j-1}.
\]
Since $\zeta_j<1/2$,
\[
k_j\zeta_j\ge(\log2)(1-\zeta_j)\ge\frac{\log2}{2},
\qquad
(1-\zeta_j)^{k_j-1}>\frac12.
\]
These inequalities follow, respectively, from
\[
-\log(1-\zeta_j)\le\frac{\zeta_j}{1-\zeta_j},
\qquad
k_j-1<\frac{\log(1/2)}{\log(1-\zeta_j)}.
\]
Consequently
\[
D_h\ge\left(\frac{\log2}{4}w\right)^h.
\]
By Lemma~\ref{lem:critical-fisher}, the gain of the binary statistic $F_h$
for the whole block is
\[
\frac{D_h^2}{16q_h(1-q_h)}\ge\frac{D_h^2}{4}.
\]

The function $F_h$ is a read-once AND--OR tree of depth $2h$.
A tree of depth $d\ge1$ can be evaluated from left to right with
$d+1$ states and a free clock, by the construction of
Brody and Verbin~\cite[full version, Section~5]{bv10}.
For a single AND or OR gate, two states suffice. Inductively,
suppose that each child of the root can be evaluated with $d$ states.
Reuse these states for the successive children, together with one
additional state for the root's absorbing value: zero for AND and one
for OR. At the last input of a child, an absorbing output sends the
program to this additional state, where it remains for the rest of the
tree. A neutral output resets the child program to its fixed initial
state. The clock identifies these last inputs and the next child's
transition maps. After the last child, the additional state encodes the
absorbing output and the reset state encodes the neutral output.
This proves the state bound by induction.

Apply this construction to $F_h$ using $2h+1$ states. After its last
input, keep its output unchanged through the unused positions of the
length-$m$ block. Retain the preceding block's Boolean output in a
second register. At a block boundary, copy the completed output into
this register and reset the tree program, all within the transition
that reads the block's last input. The initialization is fixed, and
the pair of registers has at most $2(2h+1)=4h+2$ states.
Proposition~\ref{prop:critical-transfer}, applied with $K=2h+1$ and $L=2$,
now proves \eqref{eq:critical-hierarchy-explicit}.
For fixed $h$, $w$ is of order $\log n$, giving
\eqref{eq:critical-hierarchy}.
\end{proof}

\subsection{Time-homogeneous one-bit updates on independent inputs}

The free clock also affects one-bit performance. On an infinite
independent fair stream, using the same update map at every time yields
the following optimum, even when the decoder depends on time.

\begin{proposition}[One-bit gain without a time-dependent update]\label{prop:homogeneous-one-bit}
Let $X_1,X_2,\ldots$ be independent fair bits and let $S_t$ be their length-$n$ window count, with $n\ge1$. Among recursions
\[
Z_t=F(Z_{t-1},X_t),\qquad Z_0\in\{0,1\},
\]
with a fixed Boolean map $F$ and arbitrary decoders $G_t$, one has
\begin{equation}\label{eq:homogeneous-one-bit}
\sup_{F,Z_0,(G_t)}\inf_{t\ge n}
\left\{\frac n4-\E\bigl(S_t-G_t(Z_t)\bigr)^2\right\}
=\frac14.
\end{equation}
The same value holds when the decoder is required to be time-homogeneous.
\end{proposition}
\begin{proof}[Proof of Proposition~\ref{prop:homogeneous-one-bit}]
At each time, conditional expectation is the optimal decoder. If the state is nonconstant, its explained variance is
\[
J_t=\Var(\E[S_t\mid Z_t])
=\frac{\operatorname{Cov}(S_t,Z_t)^2}{\Var(Z_t)};
\]
put $J_t=0$ for a constant state. The guaranteed gain of any decoder sequence is at most $\inf_{t\ge n}J_t$, hence at most $\liminf_{t\to\infty}J_t$.

We again classify the sixteen Boolean transition tables. Six ignore the
state or the input: the four maps $0,1,z,1-z$ yield deterministic states
and zero gain; the two maps $x,1-x$ yield gain $1/4$ at every query time.
The parity maps $z\mathbin\oplus x$ and $1-(z\mathbin\oplus x)$ give a
state independent of the current window when $t>n$, hence zero gain.

The remaining eight tables have one row constant and the other equal to $x$ or $1-x$. Four have an absorbing state. Their probability of not yet reaching it is at most $2^{-t}$. Since the two conditional means of $S_t$ lie in $[0,n]$, their explained variance satisfies $J_t\le n^2 2^{-t}$ and tends to zero.

In each of the other four tables, the constant row sends its state to the opposite state. Under fair inputs the state chain is irreducible and has conditional-mean slope $-1/2$:
\[
\E[Z_{t+1}\mid Z_t]=\eta_F-\tfrac12 Z_t,
\qquad \eta_F\in\{1/2,1\}.
\]
Its stationary state probability is $1/3$ or $2/3$, with variance $2/9$. Write $z_\ast$ for the state whose row depends on the new bit; its stationary probability is $2/3$. Consequently
\[
\operatorname{Cov}(X_i,Z_i)
=\frac{\sigma}{4}\Pp(Z_{i-1}=z_\ast)
\longrightarrow \sigma/6,
\qquad \sigma\in\{-1,1\},
\]
where the sign is fixed by the input literal in that row. Subsequent independent inputs multiply this covariance by $-1/2$ at each step. For fixed $n$,
\[
\operatorname{Cov}(S_t,Z_t)
\longrightarrow\frac{\sigma}{6}\sum_{j=0}^{n-1}(-1/2)^j
=\frac{\sigma}{9}\bigl(1-(-1/2)^n\bigr).
\]
It follows that
\[
\lim_{t\to\infty}J_t
=\frac{\bigl(1-(-1/2)^n\bigr)^2}{18}
\le\frac18<\frac14.
\]
These cases exhaust the transition maps and prove the upper bound.
The update $Z_t=X_t$ with decoder $(n-1)/2+Z_t$ attains gain $1/4$
at every $t\ge n$, establishing both assertions.
\end{proof}

For a single terminal query, the homogeneous OR update started at zero
and applied to two independent bits has final-state probability $3/4$,
variance $3/16$, and covariance $1/4$ with their sum. Its explained
variance is $1/3$. Thus the value $1/4$ in
\eqref{eq:homogeneous-one-bit} depends on requiring the gain at every
query time.

\section{Explicit bounds and computational details}\label{sec:examples}\label{app:calculations}
This appendix collects the explicit constants and computations behind the examples. The hidden Markov examples come first, with the explicit constants of Appendix~\ref{app:hmm-constants}, followed by the window example and the finite-window Tribes values (Appendices~\ref{app:window-finite-example}--\ref{sec:window-calculations}). Appendix~\ref{app:verification} describes the code, and the remaining subsections give the details of individual computations.
\subsection{HMM examples and finite state bounds}
\label{app:hmm-illustrations}

\subsubsection{A fixed-word saturation calculation}

\begin{figure}[t]
\centering
\includegraphics[width=\textwidth]{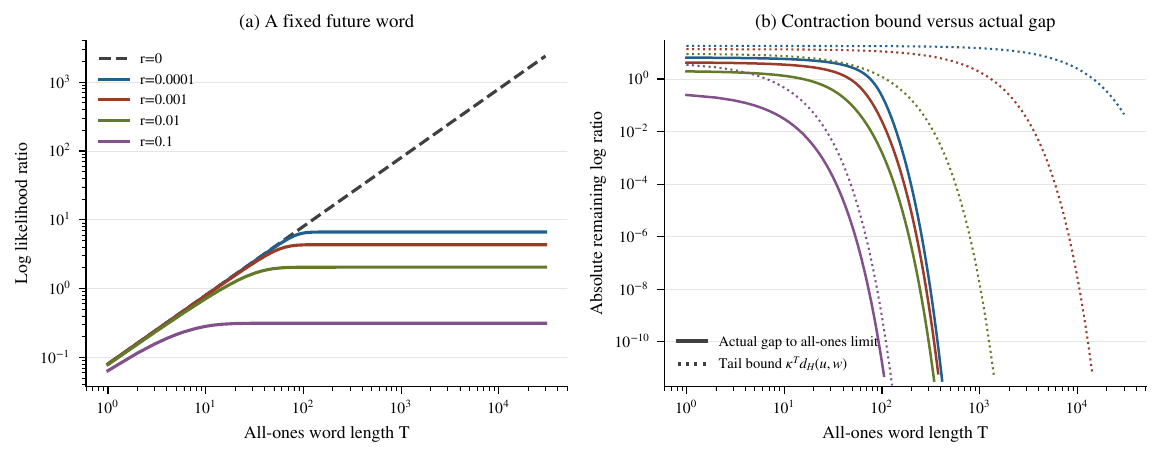}
\caption{Horizon saturation along the all-ones word, with alarm probabilities $0.52$ and $0.48$ and hidden flip probability $r$. (a)~Log likelihood ratio $\ell_T$ of $T$ future ones between the two present hidden states: linear in $T$ at $r=0$, bounded for $r>0$. (b)~Distance of $\ell_T$ from its limit, with the bound \eqref{eq:hmm-fixed-word-tail}.}
\label{fig:hmm}
\end{figure}

Figure~\ref{fig:hmm} examines horizon saturation with alarm
probabilities $0.52$ and $0.48$. The two-regime model of Section~\ref{sec:two-regime} uses $0.1$ and
$0.9$; the present comparison varies the hidden transition matrix
\[
P_r=\begin{pmatrix}1-r&r\\r&1-r\end{pmatrix},\qquad 0\le r<1.
\]
The comparison starts from each of the two possible \emph{present}
hidden states, so a transition precedes the first future emission.
With $D_1=\operatorname{diag}(0.52,0.48)$, its likelihood vector for
$T$ future ones is
\[
v_T=(P_rD_1)^T\one.
\]
At $r=0$ the log ratio is $T\log(0.52/0.48)$. For $0<r<1$, the
beliefs after the first transition are the positive rows of $P_r$;
the mixture comparison and contraction control the ratio as the
horizon grows. The curves use the exact two-dimensional matrix
recurrence, with the vector rescaled at each step.
Writing $\ell_T=\log(v_T(1)/v_T(2))$,
Corollary~\ref{cor:hmm-fixed-word-tail} gives, for $0<r<1$,
\[
|\ell_\infty-\ell_T|
\le\kappa^T d_H(P_r(1,:),P_r(2,:)).
\]
The second panel compares the fixed-word tail with its contraction
bound. At $r=10^{-4}$, the contraction timescale $1/(2r)=5000$ steps
exceeds the roughly $100$ steps needed by the all-ones word because
the bound covers every filtered continuation.

\subsubsection{Explicit state bounds and numerical comparisons}

For the two-regime model \eqref{eq:gen-explicit-parameters}, let
$d=\log(37/50)/\log(135/1369)>0$. The separated filter images give
an explicit lower bound, and a uniform belief grid gives the upper
bound below.

\begin{corollary}[A fully explicit two-state example]
\label{cor:gen-explicit}
For the source \eqref{eq:gen-explicit-parameters},
\begin{equation}\label{eq:gen-explicit-state-bounds}
\frac12\left(\frac{(12/65)^2T}{4\varepsilon}\right)^{d/2}
\le K_{\mathrm{gen}}(T,\varepsilon)
\le1+\left\lceil\frac{78}{17}\frac{T}{\varepsilon}\right\rceil,
\end{equation}
where $d$ is given in \eqref{eq:gen-explicit-constants}.
\end{corollary}

Appendix~\ref{app:hmm-constants} gives the constants and proof. The
lower exponent is $d/2\simeq0.065$, while the upper exponent is one.
At $\varepsilon=0.1$, the real-valued state lower bounds are about
$0.574700$ for $T=100$ and $0.775212$ for $T=10{,}000$.
Both values lie below one, so these bounds are vacuous here. Finite
histories give the following stronger bound.

\begin{proposition}[Finite-history lower bounds]
\label{prop:hmm-four-histories}\label{prop:hmm-finite-history}
For the two-regime model \eqref{eq:gen-explicit-parameters} at
$\varepsilon=0.1$, every observer satisfying the uniform event-query
criterion has at least six states. Every generator satisfying
\eqref{eq:gen-criterion} has at least six states for $T\ge2$ and at
least four states for $T=1$.
\end{proposition}

The six equal-length histories
$00000$, $00010$, $01110$, $00001$, $11101$ and $01011$ are separated
by next-two-symbol probabilities. Appendix~\ref{app:hmm-constants}
verifies this separation in exact rational arithmetic and proves the
four-history bound for $T=1$. The finite-past quantization bound of
Appendix~\ref{app:hmm-finite-quantization} separately
gives total-KL lower bounds of four states at $T=100$ and eleven at
$T=10{,}000$, as recorded in Table~\ref{tab:hmm-state-counts}.

Figure~\ref{fig:gen-hmm} in Section~\ref{sec:generation} compares event
queries, unifilar generation and private-state simulation on finite
grids. Additional parameters,
including overlapping images, a negative transition determinant and
both independent degeneracies, are examined in
Appendix~\ref{app:hmm-model-checks}.

\subsection{Explicit constants for symmetric hidden Markov models}\label{app:hmm-constants}

Let $H_t\in\{0,1\}$ be a stationary Markov chain with
\[
\Pp(H_{t+1}\ne H_t)=r,\qquad 0<r<\frac12,
\]
and stationary marginal $\operatorname{Ber}(1/2)$. Conditional on
the hidden states, the observations are independent and satisfy
\[
\Pp(Y_t\ne H_t\mid H_t)=s,\qquad 0<s<\frac12.
\]
Thus both the transition matrix and the emission matrix are
strictly positive. For a predictive hidden-state probability
$u\in I=[r,1-r]$, define
\begin{align}
p_1(u)&=s+(1-2s)u,\notag\\
F_1(u)&=r+(1-2r)\frac{(1-s)u}{s+(1-2s)u},\label{eq:gen-binary-filter}\\
F_0(u)&=r+(1-2r)\frac{su}{1-s-(1-2s)u}.\notag
\end{align}
Both filter maps preserve $I$ and are strictly increasing on $I$. The finite-past filter starts at $u_0=1/2$.

Theorem~\ref{thm:gen-binary-classification} gives the logarithmic
memory law for all $0<r,s<1/2$. Direct separation of the one-step
filter images gives sharper explicit constants on the open region
\begin{equation}\label{eq:gen-separated-parameters}
0<s<r<\frac12.
\end{equation}
Put
\begin{align}
a_H&=r+s-2rs,&\rho&=1-a_H,
&\lambda&=\frac{(1-2r)s(1-s)}{\rho^2},\label{eq:gen-family-constants-a}\\
g_F&=\frac{(1-2r)(r-s)}{a_H},
&g_V&=(1-2s)g_F,
&d&=\frac{\log\rho}{\log\lambda}.
\label{eq:gen-family-constants-b}
\end{align}
Here $0<a_H<1/2$, $1/2<\rho<1$, $0<\lambda<1$, and
$g_F,g_V,d$ are positive. The exponent $d$ follows from the derivative and cylinder-mass
estimates used below.

\begin{theorem}[Logarithmic memory on an open parameter family]
\label{thm:gen-binary-memory}
For the stationary hidden Markov source with parameters
\eqref{eq:gen-separated-parameters}, every $T\ge1$ and
$\varepsilon>0$ satisfy
\begin{equation}\label{eq:gen-family-state-bounds}
\frac12\left(\frac{g_V^2T}{4\varepsilon}\right)^{d/2}
\le K_{\mathrm{gen}}(T,\varepsilon)
\le 2+\left\lceil
\frac{\log((1-r)/r)}{r}\frac{T}{\varepsilon}
\right\rceil.
\end{equation}
In particular,
\begin{equation}\label{eq:gen-family-bit-law}
b_{\mathrm{gen}}(T,\varepsilon)
=\Theta_{r,s}\!\left(\log\frac{T}{\varepsilon}\right)
\qquad\text{as }T/\varepsilon\to\infty.
\end{equation}
The upper bound is achieved by a time-homogeneous machine; the lower
bound permits arbitrary time-dependent maps and a free clock.
\end{theorem}
\begin{proof}[Proof of Theorem~\ref{thm:gen-binary-memory}]
For $u\in I$, both symbol probabilities belong to $[a_H,\rho]$.
Differentiating \eqref{eq:gen-binary-filter} gives
\begin{equation}\label{eq:gen-binary-derivatives}
F_y'(u)=\frac{(1-2r)s(1-s)}{p_y(u)^2}
\ge\lambda>0.
\end{equation}
The numerator is less than $1/4$ and $\rho>1/2$, so
$\lambda<1$. Monotonicity and direct evaluation show that
\begin{equation}\label{eq:gen-binary-gap}
\operatorname{dist}(F_0(I),F_1(I))
=F_1(r)-F_0(1-r)=g_F>0.
\end{equation}

Let $U_t=\Pp(H_{t+1}=1\mid\sigma(Y_j:j\le t))$ be the
stationary predictive filter. For a word
$w=w_1\cdots w_\ell$, define the closed filter interval
\[
I_w=F_{w_\ell}\circ\cdots\circ F_{w_1}(I).
\]
Any two distinct length-$\ell$ intervals have distance at least
\begin{equation}\label{eq:gen-cylinder-separation}
g_F\lambda^{\ell-1}.
\end{equation}
To see this, remove the common outer maps. The first differing outer
maps place the intervals in $F_0(I)$ and $F_1(I)$, separated by
$g_F$. Restoring $j$ common outer maps multiplies the separation
by at least $\lambda^j$, by \eqref{eq:gen-binary-derivatives}. Since $0\le j\le\ell-1$ and
$\lambda<1$, \eqref{eq:gen-cylinder-separation} follows.

The recursion for $U_t$ implies that $U_0$ belongs to
$I_{Y_{-\ell+1:0}}$ almost surely. Since these intervals are
pairwise disjoint,
\begin{equation}\label{eq:gen-cylinder-mass}
\Pp(U_0\in I_w)
=\Pp(Y_{-\ell+1:0}=w)\le\rho^\ell.
\end{equation}
Successive conditioning on the infinite observation past gives the
last inequality, since each conditional symbol probability is at
most $\rho$.

Let $\mu$ be the law of the stationary output predictor
$V=p_1(U_0)$. The affine map $p_1$ multiplies all distances by
$1-2s$, so the corresponding predictor intervals have mutual
distance at least $g_V\lambda^{\ell-1}$ and mass at most
$\rho^\ell$. For any interval $J\subseteq\mathbb R$ of length
$h>0$, we claim
\begin{equation}\label{eq:gen-interval-bound}
\mu(J)\le(h/g_V)^d.
\end{equation}
For $h\ge g_V$ the right side is at least one. Otherwise choose
$\ell\ge1$ so that
\[
g_V\lambda^\ell\le h<g_V\lambda^{\ell-1}.
\]
The interval $J$ meets at most one level-$\ell$ predictor interval.
Because these intervals carry all the mass of $\mu$,
\[
\mu(J)\le\rho^\ell
=(\lambda^\ell)^d\le(h/g_V)^d,
\]
as claimed.

For the binary simplex, total variation is the distance between
the probabilities of the symbol one. A ball of radius $r_0$ is
contained in an interval of length $2r_0$. Thus
\eqref{eq:gen-interval-bound} gives
\eqref{eq:gen-small-ball} with $C=(2/g_V)^d$.
Corollary~\ref{cor:gen-small-ball} yields the lower bound in
\eqref{eq:gen-family-state-bounds}. Equivalently, it gives the
explicit quantization estimate
\begin{equation}\label{eq:gen-family-distortion}
\mathcal D_K(\mu)\ge\frac{g_V^2}{8}(2K)^{-2/d}.
\end{equation}

For this two-state transition matrix,
\[
\kappa=1-2r,\qquad L_P=\log((1-r)/r).
\]
Substitution into Theorem~\ref{thm:gen-hmm-upper} proves the upper
bound in \eqref{eq:gen-family-state-bounds}. Taking logarithms of
the two state-count bounds proves \eqref{eq:gen-family-bit-law}.
\end{proof}

Condition $s<r$ supplies disjoint one-step filter images and the
explicit concentration exponent $d$ in
\eqref{eq:gen-family-constants-b}. The block pruning argument in
Theorem~\ref{thm:gen-binary-classification} gives the same
logarithmic order throughout $0<r,s<1/2$. A positive lower bound on
the absolute derivative gives the separation needed for the lower
bounds; Hilbert contraction controls approximation in the upper bounds.

\begin{corollary}[Memory for one initial forecast]
\label{cor:gen-binary-initial}
For the parameter region \eqref{eq:gen-separated-parameters}, put
$C_r=\log((1-r)/r)/r$. Then
\begin{equation}\label{eq:gen-binary-initial-lower}
K_{\mathrm{init}}(T,\varepsilon)
\ge\frac12\left(
\frac{g_V^2T}{4(\varepsilon+C_r)}
\right)^{d/2}.
\end{equation}
Consequently, for every fixed $\varepsilon>0$,
\[
\left\lceil\log_2K_{\mathrm{init}}(T,\varepsilon)\right\rceil
=\Theta_{r,s,\varepsilon}(\log T)
\qquad\text{as }T\to\infty.
\]
\end{corollary}
\begin{proof}[Proof of Corollary~\ref{cor:gen-binary-initial}]
For the two-state model, the constant in
\eqref{eq:gen-initial-constant} is $C_P=C_r$.
Combine Proposition~\ref{prop:gen-initial-forecast} with
\eqref{eq:gen-family-distortion} to obtain
\eqref{eq:gen-binary-initial-lower}. The upper bound follows
from $K_{\mathrm{init}}\le K_{\mathrm{gen}}$ and
Theorem~\ref{thm:gen-binary-memory}.
\end{proof}

\subsubsection{Constants for the two-regime model}

For the parameters \eqref{eq:gen-explicit-parameters}, direct evaluation
gives

\begin{align}
I&=[1/5,4/5],
&F_0(I)&=[8/37,5/13],
&F_1(I)&=[8/13,29/37],\notag\\
g_F&=\frac3{13},
&g_V&=\frac{12}{65},
&\rho&=\frac{37}{50},\label{eq:gen-explicit-constants}\\
\lambda&=\frac{135}{1369},
&d&=\frac{\log(37/50)}{\log(135/1369)}.
\notag
\end{align}
The stationary predictor satisfies
\eqref{eq:gen-interval-bound} and
\eqref{eq:gen-family-distortion} with these constants.

\begin{proof}[Proof of Corollary~\ref{cor:gen-explicit}]
The lower bound is Theorem~\ref{thm:gen-binary-memory}. For the
upper bound, the explicit filter derivatives obey
\[
\frac{135}{1369}\le F_y'(u)\le\frac{135}{169}=:L_e<1,
\qquad u\in I.
\]
Choose an endpoint-inclusive uniform grid of $K\ge2$ points in
$I$, round the initial belief $1/2$ to the nearest grid point,
and round after every filter update. The maximum rounding error
is
\[
\eta=\frac{3}{10(K-1)}.
\]
For every finite word, the error $e_t$ between the true and stored
belief satisfies $e_0\le\eta$ and
\[
e_{t+1}\le L_e e_t+\eta,
\qquad e_t\le\frac{\eta}{1-L_e}.
\]
Both symbol probabilities are at least $13/50$, while their
absolute derivatives with respect to the belief equal $4/5$.
Hence
\[
|\log p_y(u)-\log p_y(v)|
\le\frac{40}{13}|u-v|.
\]
The log likelihood error of any word of length at most $T$,
from any observed history, is bounded by
\[
\frac{40T}{13}\frac{\eta}{1-L_e}
=\frac{78}{17}\frac{T}{K-1}.
\]
Taking $K=1+\lceil78T/(17\varepsilon)\rceil$ proves the upper
bound.
\end{proof}

\begin{proof}[Proof of Proposition~\ref{prop:hmm-finite-history}]
For a one-symbol forecast, start from the stationary belief
$u_0=1/2$ and apply the exact filter twice. The resulting probabilities
of a one after the indicated histories are
\begin{equation}\label{eq:hmm-four-predictions}
p_{00}=\frac{481}{1730},\qquad
p_{10}=\frac{289}{770},\qquad
p_{01}=\frac{481}{770},\qquad
p_{11}=\frac{1249}{1730}.
\end{equation}
They are strictly increasing in this order. Define
\[
d_{\mathrm B}(p,q)=
\max\!\left\{\left|\log\frac pq\right|,
\left|\log\frac{1-p}{1-q}\right|\right\}.
\]
For the first adjacent pair, $p_{10}/p_{00}>5/4$; for the middle pair,
$p_{01}/p_{10}>5/4$; and for the last pair,
$(1-p_{01})/(1-p_{11})=p_{10}/p_{00}>5/4$.
Since $\log(5/4)>1/5$, every adjacent pair has $d_{\mathrm B}>0.2$.
Monotonicity extends the separation to every other pair.

If two histories shared a state at time two, the observer would give
the same answer for each of the events $\{Y_3=0\}$ and $\{Y_3=1\}$.
Two multiplicative error bounds with tolerance $0.1$ would then give
$d_{\mathrm B}\le0.2$, a contradiction. All histories have positive
probability, so four distinct states are required. For a generator,
the length-one instances of \eqref{eq:gen-criterion} give the same
contradiction, for every $T\ge1$.

For the stronger bound, take the six histories in the order
\[
(h_1,h_2,h_3,h_4,h_5,h_6)
=(00000,00010,01110,00001,11101,01011).
\]
Their predictive hidden-state probabilities, obtained by five exact
filter updates from $1/2$, are
\begin{align*}
u_1&=\frac{716437}{3285890},&
u_2&=\frac{200893}{680450},&
u_3&=\frac{174973}{473090},\\
u_4&=\frac{781933}{1243010},&
u_5&=\frac{479557}{680450},&
u_6&=\frac{246013}{319490}.
\end{align*}
These beliefs are strictly increasing. At predictive hidden
probability $u$, the probabilities of the next two zeros and of the
next two ones are
\[
p_u(00)=\frac{333-320u}{500},\qquad
p_u(11)=\frac{13+320u}{500}.
\]
Substitution gives the exact adjacent-pair bound
\begin{equation}\label{eq:hmm-six-history-ratio}
\min\!\left\{
\min_{1\le i\le3}\frac{p_{u_{i+1}}(11)}{p_{u_i}(11)},
\min_{4\le i\le5}\frac{p_{u_i}(00)}{p_{u_{i+1}}(00)}
\right\}
=\frac{422842040885}{345978333749}=:R_*.
\end{equation}
The exponential series yields
\[
e^{1/5}<1+\frac15+\frac{1/50}{1-1/15}
=\frac{171}{140}<R_*.
\]
The final strict inequality follows from the integer identity
\[
140\cdot422842040885-171\cdot345978333749
=35590652821>0.
\]
Since $p_u(11)$ increases and $p_u(00)$ decreases with $u$, the
same separation holds for every nonadjacent pair: extending the
belief interval of any separated adjacent pair can only increase
the witnessing probability ratio. Thus each pair of histories has
probabilities differing by a factor greater than $e^{0.2}$ on at
least one of the words $00$ and $11$.

All six histories have length five and positive probability. If
two led to the same retained state, they would receive the same
answer for the witnessing event at time five. The two error
bounds with tolerance $0.1$ would force their true probability
ratio to be at most $e^{0.2}$, a contradiction. A query observer
therefore requires at least six states. For a generator, the same
argument uses the length-two instances of
\eqref{eq:gen-criterion}, which are available whenever $T\ge2$.
\end{proof}

\subsubsection{Reachable states and state counts}\label{app:hmm-reachability}

\begin{lemma}[Restriction to reachable states]\label{lem:hmm-reachable}
Let a time-homogeneous finite-state observer or generator have initial
state $z_\star$ and transitions $F_y$. If $\mathcal R$ contains
$z_\star$ and satisfies $F_y(\mathcal R)\subseteq\mathcal R$ for every
symbol, restriction to $\mathcal R$ preserves its state trajectory
after every input word and its probability for every generated word.
\end{lemma}

\begin{proof}[Proof of Lemma~\ref{lem:hmm-reachable}]
Induction on the word length shows that every visited state remains in
$\mathcal R$ and that the restricted and original trajectories agree.
The output maps are unchanged on these states. Their products in
\eqref{eq:gen-word-law}, and every event answer, therefore agree.
\end{proof}

For the query observer of the two-regime model at $\varepsilon=0.1$, the log-odds grid
is $x_j=-\log4+2j\log4/70$, $0\le j\le70$, with beliefs
$u_j=e^{x_j}/(1+e^{x_j})$. An additional initial label has belief $1/2$.
For the uniform-word generator, the grid is
$u_j=1/5+3j/[5(K-1)]$, $0\le j<K$, with
$K=1+\lceil78T/(17\varepsilon)\rceil$; the initial belief $1/2$ is
rounded to that grid. Ties are resolved toward the larger belief.
Breadth-first enumeration from the initial label gives the smallest
set closed under transitions. The generation transitions and their
midpoint comparisons are rational and are checked exactly. Enclosing
intervals decide the log-grid comparisons, with exact ties resolved
by the same rule.

More explicitly, put $\vartheta=2^{1/35}$. The query grid odds are
$\vartheta^{2j}/4$, and the rounding boundaries, at log-odds
midpoints, have odds $\vartheta^{2j+1}/4$. An integer $A$ satisfying
\[
A^{35}<2\cdot10^{2100}<(A+1)^{35}
\]
shows $A/10^{60}<\vartheta<(A+1)/10^{60}$. Raising these positive
rational endpoints to the required integer powers gives enclosing
intervals for every grid point and rounding boundary. The filter
update in odds $x=u/(1-u)$ is the rational map
\[
x\longmapsto\frac{1+4L_yx}{4+L_yx},\qquad L_1=9,\quad L_0=1/9.
\]
Exact interval propagation verifies that each reachable transition
lies strictly between the appropriate two rounding boundaries.
Floating-point arithmetic proposes candidate indices; rational
inequalities decide the final comparisons.

For the query observer the closure consists of $28$ grid labels and
the exact initial label. For the two uniform-word generation rows of
Table~\ref{tab:hmm-state-counts}, the closures have $257$ and $3{,}139$
states. These transition checks and Lemma~\ref{lem:hmm-reachable}
show that the reduced state sets keep the original probability
guarantees.

For the total-KL construction, put
\[
\xi_T=\log(1+\sqrt{0.1/T}),\qquad
M_T=\left\lceil\frac{2\log4}{(2/5)\xi_T}\right\rceil.
\]
The grid has log odds
$x_j=-\log4+2j\log4/M_T$, $0\le j\le M_T$, and an additional
exact initial label. Rational intervals for the square root and the
positive series
\[
\log(1+x)=2\sum_{k=0}^{\infty}
\frac{(x/(2+x))^{2k+1}}{2k+1},\qquad x>0,
\]
with its geometric tail bound give $M_{100}=223$ and
$M_{10{,}000}=2196$. Hence the nominal counts are $225$ and $2198$.

To verify the reachable sets, let $\theta_T=2^{1/M_T}$. Integer comparisons
\[
A^{M_T}<2\cdot10^{80M_T}<(A+1)^{M_T}
\]
enclose $\theta_T$ between $A/10^{80}$ and $(A+1)/10^{80}$. The grid odds and
rounding-boundary odds are respectively
$\theta_T^{4j}/4$ and $\theta_T^{4j+2}/4$. Outward-rounded rational
intervals propagate these enclosures through the two increasing
odds updates
\[
\Psi_1(x)=\frac{1+36x}{4+9x},\qquad
\Psi_0(x)=\frac{9+4x}{36+x}.
\]
Starting at the exact initial odds $1$, enumeration to closure
finds $50$ and $182$ reachable grid labels. With the initial label,
the totals are $51$ and $183$, requiring six and eight bits.
All $102$ and $366$ outgoing transitions are verified by strict
interval comparisons. For $T=10{,}000$ the grid
contains odds $1$ at a grid label unreachable after an observation.
The separate initial label is counted once.
Lemma~\ref{lem:hmm-reachable} therefore transfers the conditional
total-KL guarantee of Theorem~\ref{thm:gen-kl-upper} to both
reduced machines.

\subsubsection{Finite-past quantization bounds}
\label{app:hmm-finite-quantization}

Finite observation histories give an approximation to the stationary
predictor law with a computable error bound. For the two-regime model, write
$p(u)=1/10+4u/5$, $I=[1/5,4/5]$, and
$F_w=F_{w_j}\circ\cdots\circ F_{w_1}$ for a word $w$ of length $j$,
with $F_{\varnothing}$ the identity.
Let $\omega_w=\Pp(Y_{1:j}=w)$ and define
\begin{equation}\label{eq:hmm-finite-predictor-law}
\mu_j=\sum_{w\in\{0,1\}^j}\omega_w\,
\delta_{p(F_w(1/2))},\qquad
\zeta_j=2\sum_{w\in\{0,1\}^j}\omega_w\,
\operatorname{diam}p(F_w(I)).
\end{equation}
Here $\mu$ denotes the stationary infinite-past predictor law.

\begin{proposition}[Finite-past quantization bound]
\label{prop:hmm-finite-past-quantization}
For every $j\ge0$ and $K\ge1$,
\begin{equation}\label{eq:hmm-cylinder-distortion-gap}
\left|\mathcal D_K(\mu)-\mathcal D_K(\mu_j)\right|\le\zeta_j.
\end{equation}
In particular, if $L_{j,K}\le\mathcal D_K(\mu_j)$ and
$\zeta_j\le\overline\zeta_j$, then
\begin{equation}\label{eq:hmm-finite-quantization-test}
2T(L_{j,K}-\overline\zeta_j)>\varepsilon
\quad\Longrightarrow\quad
K_{\mathrm{KL}}(T,\varepsilon)\ge K+1.
\end{equation}
\end{proposition}

\begin{proof}
On the two-sided stationary source, let $W=Y_{-j+1:0}$ and let
$U_{-j}$ be the predictive hidden-state belief before those observations.
Both $U_{-j}$ and $1/2$ belong to $I$, so
$U_0=F_W(U_{-j})$ and $F_W(1/2)$ belong to the same interval $F_W(I)$.
The pair $p(U_0),p(F_W(1/2))$ couples $\mu$ and $\mu_j$, with
\[
\E\left|p(U_0)-p(F_W(1/2))\right|
\le\sum_w\omega_w\operatorname{diam}p(F_w(I)).
\]
For centers $c_1,\ldots,c_K\in[0,1]$, the function
$x\mapsto\min_\ell(x-c_\ell)^2$ is $2$-Lipschitz on $[0,1]$.
Applying the Lipschitz bound to the coupling and taking infima over
centers in both directions gives \eqref{eq:hmm-cylinder-distortion-gap}.
The implication \eqref{eq:hmm-finite-quantization-test} then follows
from Theorem~\ref{thm:gen-quantization}.
\end{proof}

For $j=16$, the finite law has $N=65{,}536$ atoms. The integer forward
recursion represents their weights and coordinates, after sorting, as
\[
\omega_i=\frac{A_i}{D},\qquad
x_i=\frac{B_i}{10A_i},\qquad D=2\cdot50^{16},\qquad A_i>0.
\]
Every adjacent comparison in the ordering is verified by integer
cross multiplication. For a consecutive block of indices, write
$A_{a:b}=\sum_{i=a}^b A_i$ and $B_{a:b}=\sum_{i=a}^b B_i$.
An optimal scalar squared-error codebook has consecutive cells, and
each cell center is its weighted mean. Consequently,
\begin{equation}\label{eq:hmm-quantization-dp}
\begin{split}
R_k(\ell)
&=\max_{k-1\le i<\ell}
\left\{R_{k-1}(i)+\frac{B_{i+1:\ell}^{\,2}}{A_{i+1:\ell}}\right\},\\
\mathcal D_K(\mu_{16})
&=\frac1{100D}\left\{\sum_{i=1}^{N}\frac{B_i^2}{A_i}-R_K(N)\right\}.
\end{split}
\end{equation}
The initialization is $R_0(0)=0$; infeasible entries are $-\infty$.

Let $C(a,b)$ be the minimum distortion of one cell containing the
atoms with indices $a,\ldots,b$. These interval costs are Monge.
For three adjacent blocks the means are ordered. The weighted-variance
merge identity then shows that merging both outer blocks into the
middle block costs at least the sum of the two separate merges.
Equivalently,
$C(a,c)+C(b,d)\le C(a,d)+C(b,c)$ for $a\le b\le c\le d$.
Thus the leftmost optimal split is nondecreasing in the endpoint,
which permits divide-and-conquer evaluation of
\eqref{eq:hmm-quantization-dp}. All objective comparisons use exact
integer cross products. The common second moment and the cylinder
sum are enclosed by outward-rounded rationals with denominator
$10^{50}$; the resulting interval width for each sum is at most
$65{,}536\cdot10^{-50}$.

The computation gives
\begin{align}
\zeta_{16}&<1.756943268\times10^{-10},\notag\\
\mathcal D_3(\mu)&>0.00122730555,
&\mathcal D_{10}(\mu)&>0.00000623283118.
\label{eq:hmm-distortion-bounds}
\end{align}
Substitution in \eqref{eq:hmm-finite-quantization-test} yields
\begin{equation}\label{eq:hmm-kl-count-bounds}
K_{\mathrm{KL}}(100,0.1)\ge4,\qquad
K_{\mathrm{KL}}(10{,}000,0.1)\ge11.
\end{equation}
The second count also bounds uniform-word generation, since
$K_{\mathrm{gen}}\ge K_{\mathrm{KL}}$.

The program \texttt{finite\_past\_quantization.py} carries out the
computation for $K=1,\ldots,10$.

\subsubsection{Model comparisons and empirical loss}\label{app:hmm-model-checks}

\begin{figure}[t]
\centering
\includegraphics[width=\textwidth]{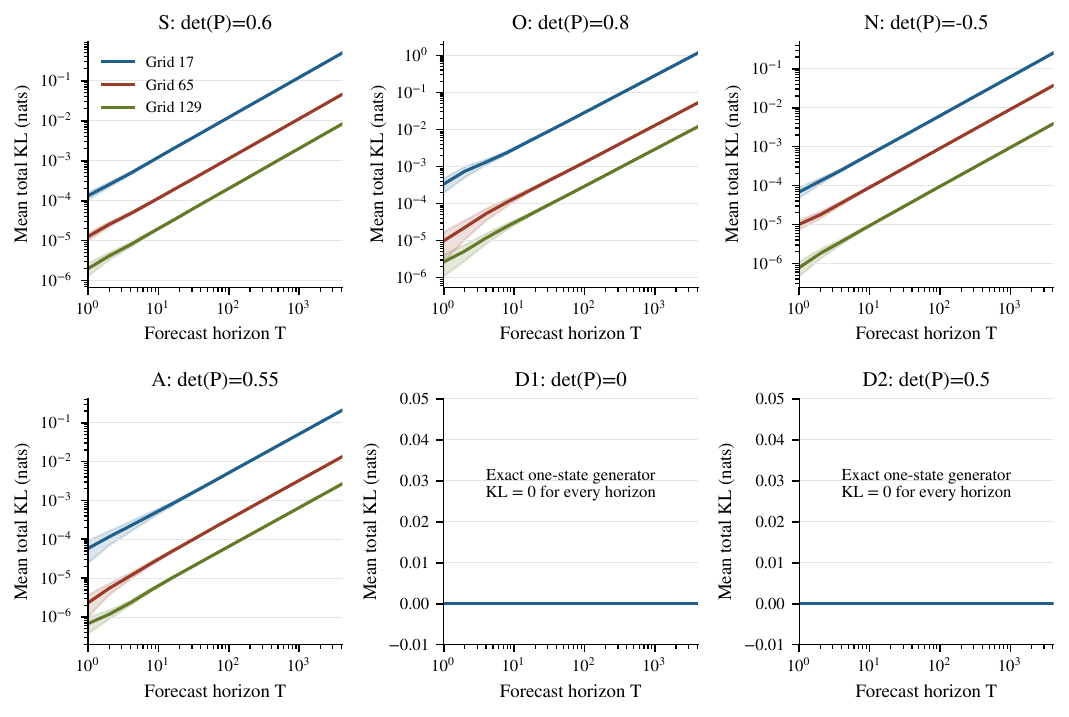}
\caption{Estimated expected total forward KL against horizon $T$ for the six models of Table~\ref{tab:hmm-models}, with log-odds grids of $17$, $65$ and $129$ points; bands are $1.96$ standard errors. In the independent models D1 and D2, a single state generates the source exactly.}
\label{fig:gen-models}
\end{figure}

Figure~\ref{fig:gen-models} compares the models in
Table~\ref{tab:hmm-models}, using the parameterization
\eqref{eq:gen-general-binary-parameters}. The two independent cases
have exact one-state output laws: Bernoulli$(0.34)$ for D1 and
Bernoulli$(0.4)$ for D2. These laws provide analytic baselines for
the belief-grid implementations.

\begin{table}[tb]
\caption{Binary two-state models covering the cases of Theorem~\ref{thm:gen-binary-classification}. All parameters lie strictly between $0$ and $1$.}
\label{tab:hmm-models}
\centering
\small
\begin{tabular}{@{}lrrrrl@{}}
\toprule
Model & $\alpha$ & $\gamma$ & $e_0$ & $e_1$ & Property\\
\midrule
S & $0.20$ & $0.80$ & $0.10$ & $0.90$ & Separated images\\
O & $0.10$ & $0.90$ & $0.20$ & $0.80$ & Overlapping images\\
N & $0.80$ & $0.30$ & $0.15$ & $0.85$ & Negative determinant\\
A & $0.15$ & $0.70$ & $0.35$ & $0.75$ & Asymmetric transitions\\
D1 & $0.30$ & $0.30$ & $0.10$ & $0.90$ & Coincident rows\\
D2 & $0.20$ & $0.70$ & $0.40$ & $0.40$ & Equal emissions\\
\bottomrule
\end{tabular}
\end{table}

For each model, requested log-odds grid sizes are $5$, $9$, $17$,
$33$, $65$, and $129$; model D1 has a degenerate belief interval and
uses one grid point. The implementations are checked on every
history of length at most four and every nonempty future word of
length at most eight. For the long-horizon losses, with grid sizes
$17$, $65$, and $129$, each of $32$ independent replicates
starts a stationary source, observes a history of length $256$, and
draws $32$ conditionally independent futures of length $4096$ from
its true predictive belief. Summing the conditional Bernoulli KL terms
along each future estimates the total forward KL without bias, by the
chain rule. Standard errors are computed across replicates, because
futures that share a history are dependent.

In the four nondegenerate models, the estimated total KL of every grid grows linearly in $T$, as the quantization converse requires for a fixed number of states, and a finer grid lowers it by a roughly constant factor. Each curve estimates the loss of one grid generator, and that loss bounds the optimal loss from above. Lower bounds come from the finite-past
calculation in Appendix~\ref{app:hmm-finite-quantization}, which
optimizes over all codebooks for the exactly weighted finite law and
bounds the error of passing to the stationary law, and asymptotically
from \eqref{eq:gen-general-binary-lower} and
\eqref{eq:gen-family-distortion}.

\subsection{The 512-observation window example}\label{app:window-finite-example}

For the example in Section~\ref{sec:window-example},
take
\[
n=512,\qquad \delta=0.02,\qquad
E=\frac{0.1}{\log(51/49)}.
\]
The count root mean-square tolerance is about $2.5$ observations,
equivalent under \eqref{eq:risk} to log probability tolerance
$\varepsilon=0.1$. More precisely,
\[
a=\log(51/49)=0.0400053346\ldots,\qquad
E=\frac{0.1}{\log(51/49)}=2.49966663\ldots.
\]
The constant count estimate $n/2$ has root mean-square count error $\sqrt{128}=11.313708498\ldots$. The equivalent root mean-square log probability error is
\[
a\sqrt{n/4}=0.45260869\ldots.
\]

Divide the stream into blocks of length $B$ and retain the counts of the
latest $M=\lceil n/B\rceil$ complete blocks and the current partial
block. The clock supplies the block phase. Complete blocks inside the
window contribute their exact counts. If a prefix of length $u$ has
expired from the oldest intersecting block, estimate its remaining
suffix by $(1-u/B)C$, where $C$ is the complete block count.

Only this oldest partial block contributes error. Since its $B$ observations are independent fair bits, the exact count mean-square error is
\[
\frac{u(B-u)}{4B}\le\frac B{16}.
\]
All other contributions cancel from the estimation error. The
calculation therefore depends only on the marginal law of this block.
Proposition~\ref{prop:simple-mse} gives the decoder and proof.

Choose
\[
B=\lfloor16E^2\rfloor=99,\qquad M=6.
\]
Each complete-block count has $100$ possible values. The current partial count has at most $99$ possible values because it is transferred to the queue at a block boundary. Thus
\[
K\le99\cdot100^6,\qquad \lceil\log_2K\rceil\le47.
\]
Unused queue entries are initialized to zero; all counts needed by the first query at $t=n$ have then been observed. The count root mean-square error is at most $\sqrt{99}/4=2.48746859\ldots$. At every deterministic query time, its log probability counterpart satisfies
\[
\sqrt{\E(\widehat\ell_t-\ell_t)^2}
\le a\frac{\sqrt{99}}4
=0.099512013394\ldots<0.1.
\]

For a uniform guarantee, requiring
$|\widehat\ell_t-\ell_t|\le0.1$ on every binary history gives, by Proposition~\ref{prop:uniform}, the lower bound
\[
\left\lfloor\frac{512}{\lfloor2E\rfloor+1}\right\rfloor=102
\]
bits. The uniform carry construction with block length $B=3$ uses at most
\[
\left\lceil\frac{512}{3}\right\rceil+\lceil\log_2 6\rceil=174
\]
bits and has absolute log probability error at most
$2a=0.08001066\ldots$.

Through \eqref{eq:risk}, the retained count also gives the log
probability of a future all-one cycle, whose unconditional probability
is $2^{-512}\simeq7.46\times10^{-155}$. The fair one-step predictor has
relative entropy $0.0002000133\ldots$ nats.

The source-averaged comparisons in Figure~\ref{fig:errors} use
\[
n\in\{32,128,512,2048\},\qquad
\delta\in\{0.02,0.1,0.5\},
\]
an independent fair-stream control, and the distinct tolerances in
$\{1,2,4,\sqrt{n/8}\}$ satisfying $E^2<n/4$. Each of these $52$ configurations uses $32$ realizations of length $9n$, with all $8n+1$ queries from $t=n$ retained. Panel~(a) displays the slice $\delta=0.02$, $E=2$.

\begin{samepage}
For the $512$-observation parameters above, $32$ streams each contain
$32\,769$ queried windows. The pooled count root mean-square errors are
$11.51999$, $0.95730$, and $2.02835$ for the constant estimate, uniform
carry, and completed-block counts, respectively. The corresponding log
probability errors are $0.46086$, $0.03830$, and $0.08114$. Panel~(b)
compares the block estimator's phase-binned error with the exact formula.
In panel~(c), the error bars run from $a\sqrt{m-1.96s}$ to
$a\sqrt{m+1.96s}$, where $m$ is the mean of the $32$ replicate
mean-square count errors and $s$ is its standard error.
Table~\ref{tab:example} gives the analytic bounds for these constructions.
\end{samepage}

\subsection{Finite-window Tribes gains}\label{app:window-tribes-example}

The width in Theorem~\ref{thm:critical-tribes} satisfies
\[
w=\log_2n-\log_2\log_2n+O(1).
\]
Retaining $w$ in the leading term captures the finite-window correction.
At $n=10^7$, one has $w=18$: the exact gain is
$47.140330668\ldots$, compared with $47.140113256\ldots$ from
$w^2/[4(e-1)]$ and $78.672354289\ldots$ from the leading expression in
$\log_2n$.

\begin{figure}[t]
\centering
\includegraphics[width=\textwidth]{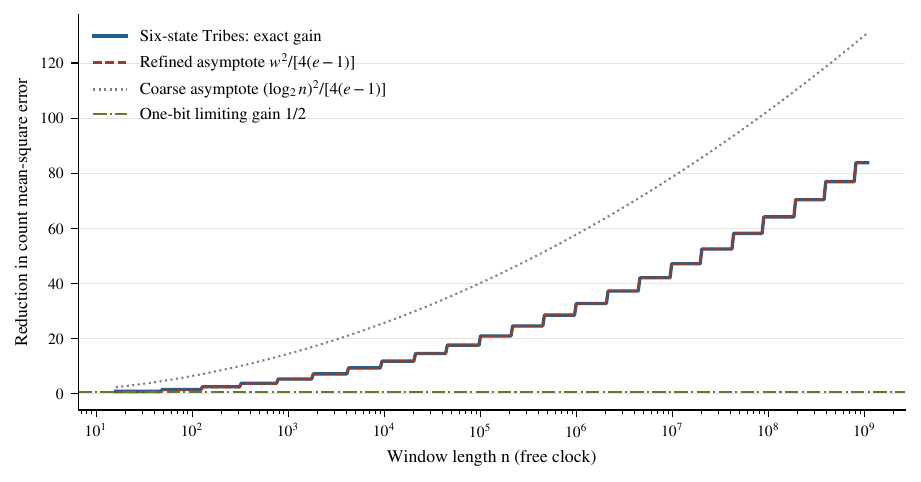}
\caption{Reduction in count mean-square error below the constant estimate's $n/4$ for the six-state Tribes recursion of Theorem~\ref{thm:critical-tribes}, against window length $n$, with the asymptotes $w^2/[4(e-1)]$ and $(\log_2n)^2/[4(e-1)]$. The horizontal line is the limiting one-bit gain $1/2$.}
\label{fig:critical-gain}
\end{figure}

\subsection{Exact error of completed-block counts}\label{sec:window-calculations}

The block estimator used in Table~\ref{tab:example} has an exact error
formula. Divide time into blocks of length $1\le B\le n$. Store the
counts of the latest $M=\lceil n/B\rceil$ completed blocks and the
count $P_t$ in the current partial block, with all counters and queue
entries initialized to zero. The number of states is at most
\begin{equation}\label{eq:simple-states}
K\le B(B+1)^M.
\end{equation}
Here the free clock gives the position within the current block, whose partial count has at most $B$ possible values.

At query time, let $k=\lfloor t/B\rfloor$, $j=\lfloor(t-n)/B\rfloor$, and $u=t-n-jB$. If $c_r$ is the count of completed block $r$, use
\begin{equation}\label{eq:simple-estimate}
\widehat S_t=P_t+\sum_{r=j+2}^kc_r+
\left(1-\frac uB\right)c_{j+1}.
\end{equation}
The inequalities $k-j\le M$ and $j\ge0$ ensure that every required
count is available. The estimate lies in $[0,n]$ and replaces the
expired prefix of the oldest block by its conditional mean given the
complete block count.

\begin{proposition}[Exact phase-dependent error]\label{prop:simple-mse}
The estimator \eqref{eq:simple-estimate} satisfies
\begin{equation}\label{eq:simple-mse}
\E(\widehat S_t-S_t)^2=\frac{u(B-u)}{4B}\le\frac B{16}.
\end{equation}
\end{proposition}
\begin{proof}[Proof of Proposition~\ref{prop:simple-mse}]
By its length-$B$ marginal law, the oldest block contains $B$ independent
fair bits. Let $U$ and $V$ be its prefix and suffix counts, over lengths
$u$ and $B-u$. The estimation error is
\[
U-\frac uB(U+V)=\left(1-\frac uB\right)U-\frac uB V.
\]
The two variables are independent, with variances $u/4$ and $(B-u)/4$. Substitution gives \eqref{eq:simple-mse}.
\end{proof}

Taking $B=\min\{n,\max\{1,\lfloor16E^2\rfloor\}\}$ meets the
mean-square tolerance; when $B=1$ the estimate is exact. Together with
\eqref{eq:simple-states}, the phase-dependent formula
\eqref{eq:simple-mse} gives the memory and error at each parameter value.

\subsection{Code and computations}\label{app:verification}

The ancillary files contain the manuscript sources and eight Python programs in the \texttt{anc} directory, together with a \texttt{README} that explains how to rebuild the paper and rerun the programs. Exact rational or integer arithmetic gives the reachable state sets and the computer-assisted lower bounds, which are computed in \texttt{exact\_checks.py} and \texttt{finite\_past\_quantization.py}. Each implementation is compared with direct computation on finite domains small enough to enumerate. The program \texttt{numerical\_experiments.py} produces the figures. Figure~\ref{fig:gen-hmm}(c), Figure~\ref{fig:errors} and Figure~\ref{fig:gen-models} come from simulation with independent replicates, with error bands of $1.96$ standard errors across replicates, and overlapping windows from one trajectory stay in the same replicate. Figure~\ref{fig:gen-hmm}(a,b), Figure~\ref{fig:hmm} and Figure~\ref{fig:critical-gain} evaluate recursions and formulas exactly. The guarantees stated in the paper rest on the proofs and the exact computations.

\subsection{Reachable states and numerical bounds}

The reachable sets in Appendix~\ref{app:hmm-reachability} are computed
with exact comparisons for uniform belief grids and rational interval
comparisons for log-odds grids. Uniform grids with $17$, $41$, $65$,
$257$, $4590$ and $458825$ points have reachable closures of $9$,
$23$, $23$, $55$, $257$ and $3139$ states; the last two give the nine-
and twelve-bit generators in Table~\ref{tab:hmm-state-counts}. The grid augmented by the two exact filter fixed points has $13$ reachable states. The construction sizes before pruning ($72$ observer states, $74$ labels including the two hidden states, and $4590$ and $458825$ generator grid points) and all bounds in Table~\ref{tab:example} are computed in rational arithmetic with $60$-digit logarithms.

\subsection{Conditional laws and sequential generation}\label{app:gen-conditional-laws}

For the two-regime model, all histories of length at most six and
all future words of length one to eight are enumerated for
$K=5,17,65$. On this domain the private-state word probabilities and the KL chain rule agree with direct computation to floating-point accuracy.

Panels (a) and (b) of Figure~\ref{fig:gen-hmm} use history
$0110100111$ and the all-ones continuation through $T=16384$,
starting from the retained belief $z=0.7625$. The final absolute log
errors are $355.553796$ for the $17$-point grid and $0.01943474$
after adding the two filter fixed points. If the augmented grid also
reads the observed history, it starts at $0.7819705149\ldots$ and the
error is $0.00226668$; the figure uses the common initial belief.

For panel (c), each of $32$ independent replicates draws a source
history of length $128$ and $64$ conditional futures through horizon
$4096$, and the conditional Bernoulli KL terms are summed along each
future. At $T=4096$, the uniform grids with $17$, $65$ and $257$
points have estimated total KL $1.712960$, $0.015759$ and $0.003601$;
private-state simulation from the same retained beliefs has
$0.0004605$, $0.00000448$ and $0.00000119$.

\subsection{Window estimators and clock phases}

The window simulations in Appendix~\ref{app:window-finite-example}
compare the constant estimate, the exact register, uniform carry, block
counts and the fixed-capacity queue, which falls back to the exact
register when the register needs fewer bits; independent fair streams
serve as a control. For $n=512$ the queue bound is $1143$ bits, so the
$512$-bit exact register is used, and the $47$-bit estimator is the
block-count construction of Proposition~\ref{prop:simple-mse}. The uniform-carry and fixed-capacity implementations are compared with direct computation on short streams, including streams that force the queue to overflow.

\subsection{Saturation and few-state calculations}

The limiting all-ones ratio in Figure~\ref{fig:hmm} is computed from
the positive Perron eigenvector of the all-ones matrix; gaps below
$2\times10^{-12}$ are omitted from the logarithmic plot.

For the few-state results, all sixteen deterministic one-bit transition tables are enumerated, and the score and covariance formulas agree with direct computation on small delayed-copy sources. The AND--OR evaluator of Proposition~\ref{prop:critical-hierarchy} agrees with direct Boolean evaluation.
Figure~\ref{fig:critical-gain} evaluates the exact Tribes formula and
its two asymptotic expressions; the finite-window values appear in
Appendix~\ref{app:window-tribes-example}.

\subsection{Exact finite-length optimization for one bit}\label{app:one-bit-dp}

The computation in Remark~\ref{rem:one-bit-finite-dp} optimizes a
two-state program on $m$ independent fair bits at one terminal time.
Write $N=2^m$. For a program with output $Z_m\in\{0,1\}$, define
\begin{align*}
n&=\#\{x\in\{0,1\}^m:Z_m(x)=1\},\\
L&=\sum_{x\in\{0,1\}^m}
\left(2\sum_{i=1}^m x_i-m\right)\mathbf1_{\{Z_m(x)=1\}}.
\end{align*}
For $0<n<N$, its explained variance is
\begin{equation}\label{eq:one-bit-dp-gain}
\Var(\E[T_m\mid Z_m])=\frac{L^2}{4n(N-n)}.
\end{equation}
Constant outputs have gain zero. Let $D_m(n)$ be the largest attainable
$|L|$ at output count $n$. Complementing the output shows that
$D_m(n)=D_m(N-n)$. Complementing every input bit preserves $n$ and
reverses the sign of $L$, so both extremal signs are attainable.

To derive a recurrence, consider all sixteen possible next-state maps.
Suppose the current input length is $m-1$, put $M=2^{m-1}$, and write
$D=D_{m-1}$. For a transition $f:\{0,1\}^2\to\{0,1\}$, set
$a_z=f(z,0)+f(z,1)$ and $c_z=f(z,1)-f(z,0)$. An old output count $n$
and score $L$ give
\begin{align*}
n'&=a_0(M-n)+a_1n,\\
L'&=(a_1-a_0)L+c_0(M-n)+c_1n.
\end{align*}
Each row of $f$ is constant zero, constant one, the input bit, or its
complement. If exactly one row is nonconstant, the largest score at
each new count is
\[
A_m(k)=
\begin{cases}
D(k)+k,&0\le k\le M,\\
D(k-M)+2M-k,&M\le k\le2M.
\end{cases}
\]
The two expressions agree at $k=M$, where they equal $M$. The other
maps either preserve or complement the old state, give a constant,
or have both rows nonconstant. Preservation and complementation give
$2D(k/2)$ at even $k$. Constants give zero at the endpoints, and two
nonconstant rows give at most $M$ at $k=M$. Consequently the exact
recurrence is
\begin{equation}\label{eq:one-bit-dp-recurrence}
D_m(k)=
\begin{cases}
\max\{A_m(k),\,2D_{m-1}(k/2)\},&k\text{ even},\\
A_m(k),&k\text{ odd}.
\end{cases}
\end{equation}
The initial values $D_0(0)=D_0(1)=0$ correspond to the two possible
labels for a fixed initial state. Every candidate is attained by one of
the transitions and sign choices above, so the recurrence gives the
exact optimum over the program class.

Evaluating \eqref{eq:one-bit-dp-recurrence} through $m=22$ in exact
integer arithmetic shows that every ratio in \eqref{eq:one-bit-dp-gain}
is at most $J_m^{\mathrm{alt}}$, with equality for at least one output
count; by output complementation it suffices to take
$1\le n\le2^{m-1}$.

\section{Conditional risk and recursive consistency}\label{app:common-risk}

\begin{proof}[Proof of Proposition~\ref{prop:common-risk}]
Write
\[
W_t-a_t(Z_t)=(W_t-\bar W_t)+(\bar W_t-a_t(Z_t)).
\]
The first term has conditional mean zero given $Z_t$, so the two terms
are orthogonal in $L^2$. Expanding the square yields
\eqref{eq:common-variance}. The same orthogonal decomposition
conditional on $(Z_t,C)$ gives \eqref{eq:common-probe}.

For log loss, split the likelihood ratio at the conditional-mean predictor:
\[
\sum_y\Pi_t(y)\log\frac{\Pi_t(y)}{q_t(Z_t)(y)}
=
\sum_y\Pi_t(y)\log\frac{\Pi_t(y)}{\bar\Pi_t(y)}
+
\sum_y\Pi_t(y)\log\frac{\bar\Pi_t(y)}{q_t(Z_t)(y)}.
\]
If $\bar\Pi_t(y)=0$, then $\Pi_t(y)=0$ almost surely conditional on
$Z_t$, and that coordinate contributes zero. In the last sum the
logarithm is measurable with respect to $Z_t$; conditional expectation
therefore replaces $\Pi_t(y)$ by $\bar\Pi_t(y)$. Taking expectations
gives \eqref{eq:common-kl}.

Finally, suppose that $Z_t(h)=Z_t(h')$. Both histories have the same
length, so the first symbol of $w$ is processed by the same map $F_t$
from the same state. The resulting states at time $t+1$ are equal.
Repeating the argument for the remaining symbols proves
\eqref{eq:common-continuation}, including for time-dependent maps.
\end{proof}

\end{appendix}

\bibliographystyle{spr-ims-number}
\bibliography{references}
\end{document}